\documentclass[sigconf, natbib=false]{acmart}

\usepackage{dirtytalk}
\usepackage{tabularx}
\usepackage{graphicx}
\usepackage{pdfpages}

\AtBeginDocument{%
  }

\copyrightyear{2025}
\acmYear{2025}
\setcopyright{cc}
\setcctype{by}
\acmConference[CHI '25]{CHI Conference on Human Factors in Computing Systems}{April 26-May 1, 2025}{Yokohama, Japan}
\acmBooktitle{CHI Conference on Human Factors in Computing Systems (CHI '25), April 26-May 1, 2025, Yokohama, Japan}\acmDOI{10.1145/3706598.3713759}
\acmISBN{979-8-4007-1394-1/25/04}

\RequirePackage[
  datamodel=acmdatamodel,
  style=acmnumeric,
  ]{biblatex}

\begin{document}

\title[Let’s Talk Futures: A Literature Review of HCI’s Future-Orientation]{Let’s Talk Futures: A Literature Review of HCI’s Future-Orientation}

\author{Camilo Sanchez}
\orcid{0000-0002-8486-031X}
\email{camilo.sanchez@aalto.fi}
\affiliation{%
  \institution{Aalto University}
  \city{Espoo}
  \country{Finland}
}

\author{Sui Wang}
\orcid{0009-0009-3748-3417}
\email{suiwang@usc.edu}
\affiliation{%
  \institution{University of Southern California}
  \city{Los Angeles}
  \country{United States of America}}

\author{Kaisa Savolainen}
\orcid{0000-0003-2907-6036}
\email{kaisa.savolainen@aalto.fi}
\affiliation{%
  \institution{Aalto University}
  \city{Espoo}
  \country{Finland}
}

\author{Felix Anand Epp}
\orcid{0000-0001-6252-7244}
\email{mail@felix.science}
\affiliation{%
  \institution{Aalto University}
  \city{Espoo}
  \country{Finland /} 
}
\affiliation{%
  \institution{University of Helsinki}
  \city{Helsinki}
  \country{Finland}
}

\author{Antti Salovaara}
\email{antti.salovaara@aalto.fi}
\orcid{0000-0001-7260-8670}
\affiliation{%
  \institution{Aalto University}
  \city{Espoo}
  \country{Finland}
}

\renewcommand{\shortauthors}{Sanchez et al.}

\begin{abstract}
HCI is future-oriented by nature:
it explores new human--technology interactions and applies the findings to promote and shape vital visions of society. Still, the visions of futures in HCI publications seem largely implicit, techno-deterministic, narrow, and lacking in roadmaps and attention to uncertainties. A literature review centered on this problem examined futuring and its forms in the ACM Digital Library's most frequently cited HCI publications. This analysis entailed developing the four-category framework SPIN, informed by futures studies literature. The results confirm that, while technology indeed drives futuring in HCI, a growing body of HCI research is coming to challenge techno-centric visions. Emerging foci of HCI futuring demonstrate active exploration of uncertainty, a focus on human experience, and contestation of dominant narratives. The paper concludes with insight illuminating factors behind techno-centrism's continued dominance of HCI discourse, as grounding for five opportunities for the field to expand
its contribution to futures and anticipation research.

\end{abstract}

\ccsdesc[500]{Human-centered computing~HCI theory, concepts and models}

\keywords{Literature Review, Futures, Futures Studies}
\received{12 September 2024}
\received[revised]{12 November 2024}
\received[accepted]{16 January 2025}


\maketitle

\begin{refsection}
\section{Introduction}
Human--computer interaction (HCI) is a future-oriented enterprise. Interactive prototypes, technical demonstrations and new design methods materialize future visions and shape our ideas of what futures can be \cite{bell_yesterdays_2007, kinsleyFuturesMakingPractices2012, salovaaraEvaluationPrototypesProblem2017, lindley_implications_2017}. They do so by introducing imaginaries of advances wrought by technological development \cite{bell_yesterdays_2007, lindley_implications_2017}.
However, because the HCI's field primary concerns are technology development and implications, it risks overlooking other matters of importance, such as systemic implications and long-term consequences\cite{nathan_envisioning_2008, mankoffLookingYesterdayTomorrow2013a, salovaaraEvaluationPrototypesProblem2017}. 
This leads us to ask whether the future-orientation so central to our discipline needs refining to support more meaningful research inquiry. 

For example, while scenario-building is commonly found in the HCI toolbox, nearly always wielded for envisioning future contexts of technology use \cite{carroll_making_2003}, detailed instructions for constructing the scenarios and on how HCI scholars should carry out their future-building \cite{nathan_envisioning_2008, pargmanSustainabilityImaginedFuture2017} are scarce within HCI, and could benefit from critical attention. The intricacies of how societies and perspectives get portrayed in such scenarios and the visions further illustrate this problem. Seemingly neutral narratives that incorporate certain visions may readily simplify the complex weave wherein interactive technologies exert influence over day-to-day life \cite{bell_yesterdays_2007, nathan_envisioning_2008, ashby_fourth-wave_2019}. If nothing else, visions for HCI often express interests of actors invested in the technologies \cite{bell_yesterdays_2007, kinsleyFuturesMakingPractices2012, harrington_eliciting_2021} and perpetuate socio-technological foci that may leave the agencies of more-than-human systems to the side \cite{forlanoPosthumanismDesign2017} or neglect underprivileged and vulnerable stakeholders. 

Enhancing HCI's future-savviness could safeguard against superficial engagement with futures and help us identify pressing questions, alongside rich opportunities \cite{nathan_envisioning_2008}. 
Interest in futuring is growing in design-oriented HCI studies especially, as approaches such as design fiction \cite{bleecker_design_2009, sterling_design_2009}, critical design \cite{dunne_hertzian_2006, bardzell_critical_2012}, speculative design \cite{dunneSpeculativeEverythingDesign2013a, wongSpeculativeDesignHCI2018}, and more-than-human design \cite{forlanoPosthumanismDesign2017, giaccardiTechnologyMoreThanHumanDesign2020, wakkary2021things,lu2024ecological,sanchezPeeringThroughTime2025} attest, yet the level of exploration and application of futures knowledge remains uneven across our field. Some future-oriented studies are far more comprehensive than others. 

This paper contributes to the solid development of future-oriented research in HCI by examining the prevalence of short-sighted futuring in HCI. We conducted in-depth investigation informed by works that have called for stronger engagement between HCI and futures studies \cite{mankoffLookingYesterdayTomorrow2013a, pargmanSustainabilityImaginedFuture2017, epp2022reinventing, moesgenDesigningUncertainFutures2023}, to construct a comprehensive picture of HCI's status as a future-oriented research discipline. Specifically, we analyzed how the field takes futures into account and formulated ways to undertake more reflective \textit{futuring}, where we use the latter term to describe a dual process: firstly, striving to understand current events deeply and secondly, using that understanding to inform future-oriented actions by exploring potential possibilities \cite{bellWhatWeMean1996}. 

To address these goals, we conducted a literature review assessing the extent to which HCI scholarship engages in futuring and carried out analysis directed at the following research questions: 

\begin{description}
    \item{RQ1:} \textit{How has the HCI field's future-orientation developed over the last 15 years?} 
    \item{RQ2:} \textit{What traits characterize the field's comprehensive futuring?}
\end{description}

We began by applying keywords and citation metrics to identify influential future-oriented papers from 11 high-profile journals and conferences in the ACM Digital Library (DL). 
Proceeding from this initial set of articles, we devised selection criteria that afforded manually inspecting the papers' Introduction, Discussion, and Conclusion sections and, accordingly, grouping the future-oriented publications into those attending to futures only fleetingly and the ones whose futuring could be considered comprehensive. For the 205 articles that appeared comprehensive in their future-orientation, we performed full-text qualitative analysis. To support our analysis, we developed a futures studies-informed analysis framework encompassing four categories of futuring: epistemic stance, contingency perceptions, systemic integration, and narrative. Via the outputs from this process, we are able to describe the timelines followed by the future-orientation work behind the most influential HCI papers, over the years and by publication channel; describe how the most future-savvy papers have handled futuring; and articulate the opportunities available to HCI researchers today, so as to deepen futures inquiry in our field. 

\section{Related Research} \label{sec:related-research}

Ever since the discipline's inception as a response to technological advancement \cite{baecker_timelines_2008}, HCI has been evolving, in waves \cite{harrison_making_2011, bodkerWhenSecondWave2006, bodker_third-wave_2015, frauenbergerEntanglementHCINext2019}, with its approach to futures adapting accordingly.  
Recent development suggests that the field is starting to take a more active and political stance to the futures addressed \cite{ashby_fourth-wave_2019}. 
Increasingly often, HCI researchers step beyond user studies, into co-creation of futures with overlooked communities \cite{harrington_eliciting_2021, freeman_rediscovering_2022} and non-human actors \cite{liu_design_2018, reddy_making_2021, heitlingerAlgorithmicFoodJustice2021}.
Below, we review the vast expanse of both conventional and emergent approaches to considering the future in HCI. Against this backdrop, we can paint a picture of dovetailing and divergences between HCI and futures studies, the branch of social science most directly tackling systematic explorations of alternative futures~\cite{bellWhatWeMean1996}.

\subsection{Conventional Approaches to Examining the Future in HCI}
To study futures, one needs to study possibilities. Scholars of HCI tackle this in various ways, among which are prototyping \cite{limAnatomyPrototypesPrototypes2008}, scenario-building (either by the researchers themselves or with possible future users \cite{carroll_making_2003, stromberg2004interactive, epp2022reinventing}), observing (potentially opportunity-exposing) ways in which users make use of technologies \cite{norman2008workarounds,norman2013incremental}, and reflecting on implications for future technologies' development on the basis of user studies \cite{luria_re-embodiment_2019}. 

Despite the work's future-orientation, HCI researchers rarely reflect comprehensively on the complexity of developing futures. This leaves room for improvement in several areas. One issue that rises to prominence is \emph{techno-centrism}: HCI scenarios often are informed by the goal of suggesting new technologies or uses, and the studies typically present prototypes without deeper considerations of how their adoption might develop or of various contingencies' potential influences \cite{lindley_implications_2017,pargmanSustainabilityImaginedFuture2017}. A few scholars have acknowledged the pressing need to consider futures more broadly. Among the most prominent has been \citeauthor{carroll_making_2003}, whose book \emph{Making Use} \cite{carroll_making_2003} discusses the history of scenario-building by referencing the work of \citeauthor{kahn1962thinking} \cite{kahn1962thinking}, a pioneer of scenario-based methods' use in strategic planning, head-on; however, even \citeauthor{carroll_making_2003} narrowed his attention to possible human interactions with technologies, as opposed to broader socio-political considerations or environmental contexts.

While HCI studies may give superficial attention to ethics factors, many fail to integrate corresponding perspectives, political sensitivity, or justice considerations meaningfully into the design of futures. For example, surveillance issues often get sidelined via brief mentions of the pervasive rise of technologies that monitor/track user behavior \cite{zuboff2015bigother}. Likewise, disparities in access to future technologies and the costs of unintended consequences remain under examined \cite{benjamin2019raceaftertechnology}. As automation and AI-based systems threaten to reshape entire industries, often at the expense of vulnerable workers \cite{eubanks2018automatinginequality}, the shadow of labor displacement has taken on especially large dimensions. Yet the ethics implications of adopting new technologies at scale -- from environmental and sustainability domains to mechanisms for data governance and prevention of misuse -- often receive only surface-level treatment, if any at all \cite{jasanoff2016dreamscapes}.

Even in work attuned to ``far-future'' scenarios and active honing of visions, the scenarios' plausibility is seldom scrutinized. In contrast, researchers often approach long-time-horizon studies as thought experiments, seeds for speculative responses to the present \cite{coulton2017design,wong2016when,blythe_research_2014}. In such cases, the visions become \emph{leaps into the future}, in that the work glosses over the intermediate steps needed for their realization. Neglecting the pathways to the future leaves gulfs in understanding of how such futures might materialize.      

In summary, the promise of rapid, productive strides in optimizing and enhancing human interaction with technological systems has brought HCI research a double-edged sword. The methods developed have drawn the field's future-orientation to immediate events and concrete examples of technology use while broader, speculative engagements with possible futures, or possible routes to distant futures, often suffer from neglect. These shortcomings spotlight the need for an expanded approach to futures, wherein HCI scholars embrace diverse explorations that are scalable and context-aware.

\subsection{Speculative Futures in HCI Studies} \label{sec:speculative-futures-HCI} 
In the last decade, the field's traditional approaches to considering the future have been joined by techniques that place greater focus on innovative approaches whereby speculation and foresight aid in examining possible futures. One distinguishing feature of these emerging practices is their stress on envisioning alternative futures -- e.g., on futures that diverge from the most probable scenarios, challenge prevailing assumptions about potential futures, and could help us deepen our understanding of technology design by probing alternative outcomes \cite{kozubaev_expanding_2020}.  

The evolving HCI landscape has witnessed sprouting of speculative design, 
critical design 
and design fiction, 
as fruit of a rich tradition of critical theory and artistic critique. Emerging as responses to technology-design practitioners' conventional orientation toward utilitarian and market-driven outcomes, they are directed instead to challenging dominant narratives and inspiring profound reflection on societal impacts. Rather than content themselves with familiar tools such as prediction and forecasting, proponents of these methods foreground the role of design in crafting scenarios that provoke discussion, question current assumptions, and reimagine worldviews ``otherwise'' \cite{blythe_research_2014}. They embrace a plurality of futuring \cite{howell2021calling}, thus inviting a rich array of speculative visions that extend the sphere of inquiry beyond linear, deterministic projections.

Speculative design often applies the tools of artifact- and scenario-creation to stimulate discussion and, thereby, exlore possible futures \cite{wong2018speculative, dunneSpeculativeEverythingDesign2013a}. Where older methods focus solely on likely or desirable futures, speculative design thrives on ``what if...?'' scenarios to bring in also those futures that could yield insight even if they remain improbable. By letting designers and other participants examine the implications of today's decisions, technology trends, and developments in society, speculative design broadens views, for greater attention to possibilities that may challenge the \emph{status quo}, and homes in on avenues for critical thinking about future technologies.

Critical design goes further still, by explicitly focusing on generating controversy and debate \cite{bardzell_critical_2012, bardzell_reading_2014, bardzell_what_2013}. Often, the critique gets embodied in artifacts that challenge presumptions about the role of technology in society by pointing to the flaws and unfulfilled potential of current technology-use practices. It champions the principle of anti-solutionism \cite{blythe_anti-solutionist_2016}, contesting the supposition that technology must always offer solutions. Positing that deeper understanding and acknowledgment of problems can drive more meaningful developments, critical design creates criticism-imbued artifacts that can help society expose and interrogate the values, practices, and power structures enmeshed in everyday technologies. Thus it encourages a conscious, reflective approach to design.

Design fiction, in turn, is a strongly storytelling-centered approach that crafts narratives instrumented to probe and manifest alternative futures \cite{bleecker_design_2009, sterling_design_2009, tanenbaum_design_2014}. Particularly effective at rendering abstract scenarios tangible and relatable, it displays power to bridge the gap between scientific exploration of concepts and emotional engagement with an audience \cite{coulton2017design}. Design-fiction narratives situate hypothetical technologies within detail-populated social, cultural, and ethics-bearing contexts. These stories anchor potential technologies in fictional interactions between technologies and intended users. 

All these speculative methods echo ideas from critical future studies \cite{slaughterKnowledgeBaseFutures1998}, especially in their emphasis on sensitization to uncertainty 
and recognizing 
the open nature of future outcomes \cite{howell2021calling, poli_anticipation_2019, gallHowVisualiseFutures2022}. Together, they constitute a richly woven tapestry for grappling with possibilities that transcend technological determinism. Speculative methods contribute to futures studies via tools and frameworks that play with and strategize around uncertainty. By harnessing it as a resource to enrich discourse about the futures we choose to pursue \cite{epp_uncertainties_2024}, they work alongside other methods to flesh out the picture. 

\subsection{Possible Futures in HCI} \label{sec:possible-futures-HCI}

\citeauthor{weiser1991computer}’s vision of computing for the 21st century \cite{weiser1991computer} strongly influenced HCI research. That narrow view of the future reproduced mostly middle-class US societal ideals~\cite{bell_yesterdays_2007}, 
but reliance on \emph{any} monocultural imaginary of future computing is bound to lead to conceiving of futures principally on near horizons without reflecting on technologies at systemic scales \cite{nathan_envisioning_2008, mankoffLookingYesterdayTomorrow2013a}. 
The gap has gained increasing recognition in calls for longer-term perspectives and more holism in approaches to possible futures \cite{nathan_envisioning_2008, mankoffLookingYesterdayTomorrow2013a, salovaaraEvaluationPrototypesProblem2017, epp2022reinventing, elsden_speculative_2017, pargmanSustainabilityImaginedFuture2017, lightCollaborativeSpeculationAnticipation2021a}.

For interaction designers to think about their interactive technologies' evolution, consider the consequences 5--20 years down the line, and identify long-term implications \cite{nathan_envisioning_2008, mankoffLookingYesterdayTomorrow2013a}, they must be encouraged to attend to externalities, such as how the technology might affect the environmental, societal, ethics, or economic domain \cite{nathan_envisioning_2008, mankoffLookingYesterdayTomorrow2013a, pargmanSustainabilityImaginedFuture2017, epp2022reinventing, moesgenDesigningUncertainFutures2023}. Such consequences may evolve slowly and stay unforeseeable within short timeframes \cite{pargmanSustainabilityImaginedFuture2017}.

Another major challenge for futuring arises from future-contin\-gent factors' interdependencies. 
Since, for example, environmental and economic forces can interact in complex and
unexpected ways, teasing out the uncertainties necessitates a systemic, open-ended angle on the future 
\cite{jouvenelArtConjecture1967, adamFutureMattersAction2007, poli_anticipation_2019}, such as a perspective acknowledging that multiple futures may warrant exploration and that diverse trajectories have to be considered \cite{pargmanSustainabilityImaginedFuture2017, salovaaraEvaluationPrototypesProblem2017, epp2022reinventing, moesgenDesigningUncertainFutures2023}.

Inherent biases too warrant contemplation. Similarly to how a techno-optimistic approach to interaction design can leave one misapprehending stakeholder needs \cite{nathan_envisioning_2008}, so can insufficient awareness of one's assumptions about the alternative futures \cite{bell_yesterdays_2007, nathan_envisioning_2008, kinsleyFuturesMakingPractices2012, mankoffLookingYesterdayTomorrow2013a}. Some of the contributions to HCI mentioned above are grounded in the principles of value-sensitive design \cite{nathan_envisioning_2008} or participatory design \cite{elsden_speculative_2017, epp2022reinventing}.
Such grounding promotes increased attentiveness to different perspectives when alternative futures come into play, whether through user scenarios \cite{nathan_envisioning_2008}, futures workshops \cite{epp2022reinventing}, or field trials \cite{elsden_speculative_2017, odom_fieldwork_2012}.  

Thanks to the breadth of HCI research's spread of topics and methods, we have numerous approaches at our disposal for looking at futures. While the predominant pattern might still be to envision linear trajectories from short-term visions of emerging technologies, such methods as design fiction and speculative design have expanded the field's understanding of how technology, its futures included, may be examined.

\subsection{HCI's Future-Orientation in Light of Futures Studies}

HCI scholars are not the only ones interested in studying possible futures and exerting effects on them. Perhaps more than for any other field of research, this enterprise is the territory of futures studies, a branch of the social sciences devoted to forecasting, foresight, and anticipation of probable and possible futures alike \cite{bellWhatWeMean1996}. Correspondingly, it employs both probabilistic forecasting (e.g., trend analyses) and qualitative methods that explore plausible futures (e.g., Delphi studies, which involve distributed expert-based scenario-building \cite{roweDelphiTechniqueForecasting1999} and various scenario-development processes \cite{vanderheijdenScenariosArtStrategic2005}) \cite{bell_purpose_2009}. 

Futures studies has gone through various phases and turns, just as HCI has. An outgrowth from strategic planning~\cite{bellWhatWeMean1996}, it later experienced a critical turn~\cite{slaughterKnowledgeBaseFutures1998}. By shifting its focus to how systems connect with social reality, this continuously developing 
field highlighted how futures get differentially built conceptually and acted, varying with cultures, ways of knowing, and how people care for others~\cite{sardar_colonizing_1993, adam_futures_2011, sardar_postnormal_2015}. 
Current work in futures studies concentrates on \textit{futures literacy}~\cite{millerFuturesLiteracyTransforming2018} and \textit{futures consciousness}~\cite{ahvenharjuFiveDimensionsFutures2018}, in aims of illuminating how individuals and groups, respectively, engage in futuring. This encompasses their ways of seizing active agency in shaping the futures that affect them. Other streams of futures studies focus on the concept of \textit{postnormality} ~\cite{funtowicz_science_1993, sardar_postnormal_2015}, bound up with how volatility, uncertainty, complexity, and ambiguity are pressing us toward urgent decisions with pivotal consequences amid a marked lack of knowledge \cite{sardar_postnormal_2015, slaughter_farewell_2020}. 

The stress on examining possible futures does not imply that futures studies confines itself to studying the future. To explore alternative futures, futures studies researchers also take into account possible pasts~\cite{sardar_colonizing_1993, bendorLookingBackwardFuture2021}, the present~\cite{poliIntroductionAnticipationStudies2017, millerSensingMakingsenseFutures2018, bellWhatWeMean1996}, and the manifold futures that can open out from the exploration. These examinations require approaching the future as mutable and open-ended \cite{jouvenelArtConjecture1967, adamFutureMattersAction2007, poli_anticipation_2019}. Futures studies scholars arrive at knowledge by considering the complexity-rife arena in which several alternative futures could unfold \cite{poliIntroductionAnticipationStudies2017}. 

One commonplace conceptual technique that futures studies employs to understand the web of relations and consequences among numerous factors in the face of alternative futures is use of the STEEPLE framework (the acronym refers to social, technological, environmental, economic, political, legal, and ethics factors) \cite{aguilarScanningBusinessEnvironment1967, saritasMappingIssuesEnvisaging2012}. Digging into future alternatives unveils uncertainty patterns that have escaped our notice or that we have been consciously ignoring, and it assists in tending to the consequences of the decisions made in the present \cite{poliIntroductionAnticipationStudies2017, millerSensingMakingsenseFutures2018, adamFutureMattersAction2007}.

In spite of obvious thematic linkages due to their interest in futures, HCI and futures studies have not interacted extensively with each other. They have reached their closest in the area of scenario-based methods, with HCI studies having applied such techniques from futures studies as Delphi scenario-building \cite {mankoffLookingYesterdayTomorrow2013a} and the Futures Wheel \cite{epp2022reinventing}, a lightweight workshop-based method wherein a seed scenario stimulates envisioning of waves of first-order, second-order, and further consequences. Overlap is evident from the other side also: \citeauthor{dunneSpeculativeEverythingDesign2013a} introduced the futures cone \cite{taylorAlternativeWorldScenarios1993} in the HCI field as a visualization of how the present can develop toward preferred, probable, plausible, and possible futures. 
Beyond these arenas and envisioning (as employed in user studies), futures studies and HCI research have engaged in little collaboration, notwithstanding suggestions for joint work on speculative studies of interaction for in-the-wild studies \cite{elsden_speculative_2017,salovaaraEvaluationPrototypesProblem2017} and VR simulations \cite{simeone2022immersive}. 

Although the outline above points to several valuable contributions to HCI, we have pinpointed the field's most future-oriented publications for discussion. 
In contrast, the vast majority of HCI research has left the opportunities for futuring underexplored, even though this seems to be an inherent aspect of HCI. One reason might be that discussions of the future often hide in the bowels of design and evaluation of emerging technologies while HCI research typically gets judged not by the rigor of its futures thinking but by what opportunities it accords individuals and communities to use technology for purposes of control, reconfiguring day-to-day life, and living that life \cite{bodkerWhenSecondWave2006}.

Design fiction, speculative design, foresight-informed studies, and other investigation routes cited above have started to introduce new methods from which HCI can draw. This growing interest does not, however, shed light on the extent to which the field has actualized its inherent capacity for future-oriented thinking. 
For fuller understanding of how HCI engages with futures, also in contributions that are not explicitly future-oriented, we carried out a literature review. The next section presents its implementation.

\section{A Literature-Based Examination of Futuring in HCI} \label{sec:literature-review} 

As we discussed above, future-orientation is an integral part of HCI. Our discipline envisions and evaluates technologies that shape our imaginaries the future and engages individuals in these visions through interactive prototypes and technological demonstrations. 
However, the extent to which HCI consciously leverages this potential in HCI seems underexplored.
Although the techno-centric focus and short-termism in HCI have been acknowledged \cite{nathan_envisioning_2008, mankoffLookingYesterdayTomorrow2013a, pargmanSustainabilityImaginedFuture2017, salovaaraEvaluationPrototypesProblem2017}, these critiques remain on a general level.
However, we stated in \nameref{sec:related-research} that HCI has potential to dedicate its research on futures more reflectively. The following review seeks to find out how much that is already done in our field's most influential papers.

\subsection{Stage 1: Identification}
\label{sec:identification}

To honor the guiding proposition that HCI is inherently future-oriented, we could not limit our study's scope to explicitly future-oriented publications alone. A larger corpus also assists in reaching meaningful conclusions.

We took papers available in the ACM DL as our starting point for the endeavor on account of the ACM being the most important publisher of research in this field. We chose to narrow our investigation to a set of venues selected to represent the field's most respected research outlets -- CHI,\footnote{~Conference on Human Factors in Computing Systems events.} DIS,\footnote{~The Designing Interactive Systems Conference.} UIST,\footnote{~The Symposium on User Interface Software and Technology.} IUI,\footnote{~The Conference on Intelligent User Interfaces series.} CSCW,\footnote{~The Conference on Computer Supported Cooperative Work.} PACMHCI\footnote{~The journal \emph{Proceedings of the ACM on Human--Computer Interaction}.} IMWUT,\footnote{~\emph{Proceedings of the ACM on Interactive, Mobile, Wearable and Ubiquitous Technologies}.} TSC,\footnote{~\emph{Transactions on Social Computing}.} TIIS,\footnote{~\emph{ACM Transactions on Interactive Intelligent Systems}.} TOCHI,\footnote{~\emph{ACM Transactions on Computer--Human Interaction}.} and the \emph{Interactions} magazine. 
Although we could surmise that some of these attract papers less positioned for futuring than others, we did not wish to risk unwarranted assumptions, 
since our interest cohered around the HCI field as a whole.

\begin{figure}[tb]
    \centering\includegraphics[width=\linewidth]{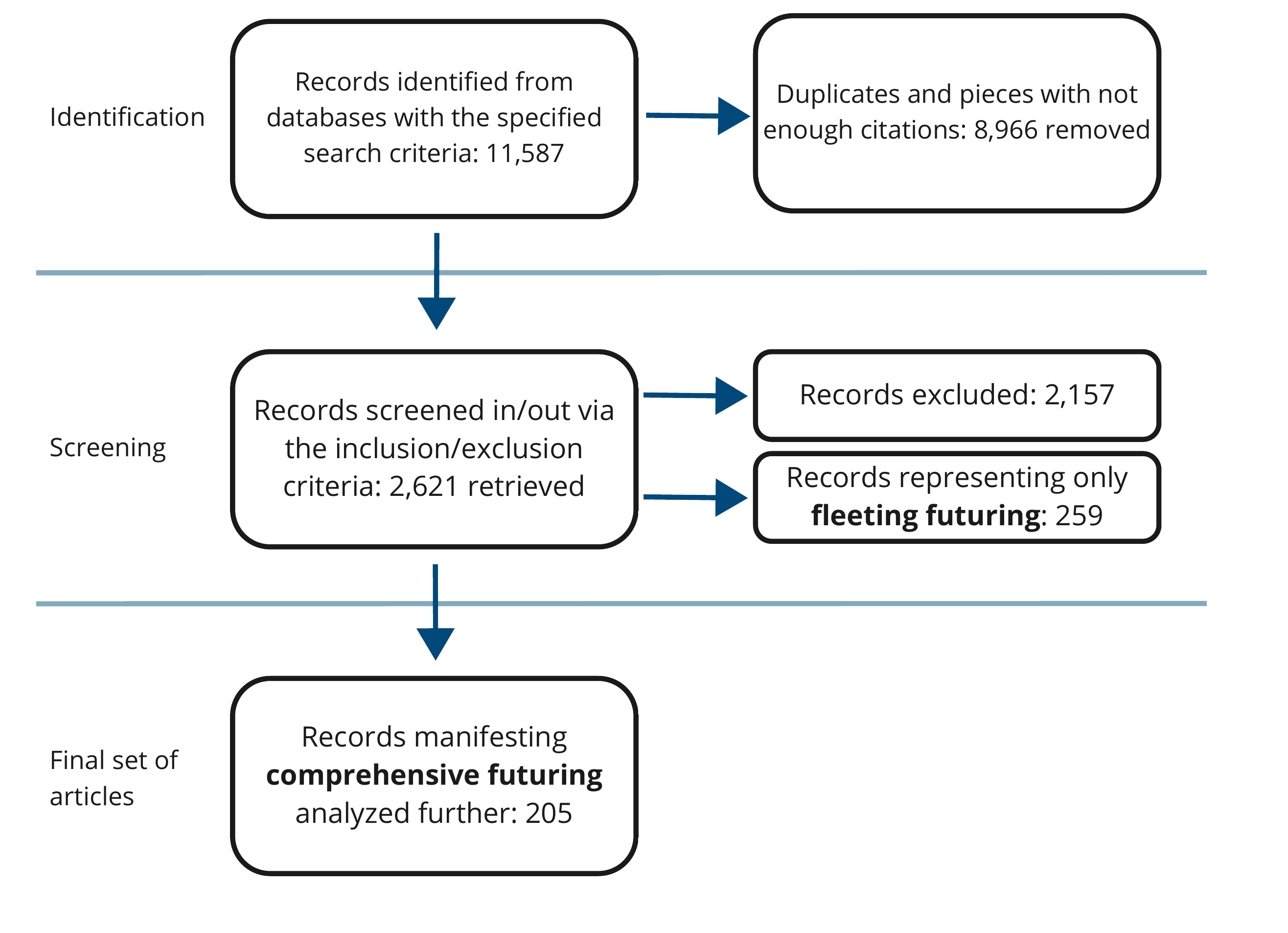}
  \caption{A PRISMA diagram of the article-screening process.}
  \Description{A flow diagram outlining the literature review inclusion/exclusion process, divided into three stages: Identification, Screening, and Final set of articles. 
  The identification stage comprised 11587 articles, out of which 8966 were removed after finding duplicates and pieces under the citation count criteria. 
  The Screening stage presents 2621 articles that were reviewed under the inclusion/exclusion criteria. This process excluded 2157 records and identified 259 articles as ``fleeting futuring''.
  The last stage presents the final set of articles. These 205 articles were categorized as comprehensive futuring and selected for qualitative full-text analysis.}
    \label{fig:prisma} 
\end{figure}

Figure~\ref{fig:prisma} presents a PRISMA-style flow diagram \cite{page2021prisma} for our review process. The first stage -- dubbed ``Identification'' -- posed the challenge of defining search terms that supported our objective of examining the HCI field as a whole yet offered high enough specificity to rule out the bulk of the false positives. We wrestled in particular with the fact that research articles may contain future-related terms for various reasons not arising from futuring. For example, multitudes of authors describe ideas for follow-up research under the rubric of limitations and ``future work.'' 

We could not safely resort to the common strategy of filtering papers on the basis of titles, keywords, and abstracts. While that provides for efficient keyword-based identification, it would have restricted our sample to papers that are explicitly futuring-oriented. To explore the role of futuring across the full gamut of HCI research, we exploited the DL's \textit{full-text} search functionality. 
Accordingly, we iteratively searched for workable \textit{keywords} suited to the ACM DL search interface. From the vantage point of full-text search mode, we explored varied keyword combinations with searches of the venues listed above. 

We also limited our search to ACM Digital Library's filter ``Article Type'' to ``Research Article''. This filter enabled systematic exclusion of such content types as short papers, extended abstracts, posters, and editorials. 
Since the latter decision necessitated ruling out records from before 
2008, when the DL applied its ``Research Article'' designation, our identification procedure retrieved records from then to 2023 (the most recent year with a complete set of articles at the time of our study, in 2024).
To avoid false positives stemming from musings on further studies (``future research''), we specified that ``the future'' or ``futures'' had to receive mention. We also searched for other future-related terms: ``visions,'' ``tomorrow,'' the stem ``specula*,'' etc. Finally, we added the requirement that the paper contain at least one of the terms attached to STEEPLE categories \cite{aguilarScanningBusinessEnvironment1967, saritasMappingIssuesEnvisaging2012}. 
Table~\ref{tab:search-terms} reproduces our final list of search keywords.

With most keyword combinations probed, our searches yielded 10,000+ articles, an infeasible volume for the next stage: manual analysis. However, we recognized that our well-honed search query returning 11,587 papers would be hard to improve upon. Therefore, we introduced two \textit{citation-based filters}, to reflect our interest in the most influential HCI papers. Referring to the ACM DL's paper-specific citation counts, we looked at various citation-metric cut-offs, ultimately limiting the corpus to articles that had been cited at least 50 times in all or were among the 15\% most cited in the year in question across the selected outlets. 
Excluding papers from 2024 was consistent with this rationale, in that they could not have amassed many citations.
Removal of duplicates left us with a final sample of 2,621 articles. 

\begin{table*}[t]
  \caption{The search terms used for the literature review}
  \label{tab:search-terms}
  \def\arraystretch{1.3}%
  \begin{tabularx}{\textwidth}{p{4cm}X} 
    \toprule
    \textbf{Term} & \textbf{Rationale}\\
    \midrule
    "the future" & Several notions serve to address the future: futures, futuring, future, etc. While these might refer to the same thing, they present an ontological stance to how futures are experienced (e.g., via singular, plural, or active expression). \citeauthor{sardar_namesake_2010} offers an overview of the challenge of defining futures studies \cite{sardar_namesake_2010}.\\
    "futures" & Talking about the future in the plural highlights the possibility of there being more than one future — the possible, the probable, the preferable, etc. -- and opens exploration to a panoply of futures \cite{bell_purpose_2009}. \\
    "visions" & Visions can refer to expressions of what the future might be or ideals we hold for the future \cite{vanderhelmVisionPhenomenonTheoretical2009}. \\
    envision & This element captures the process of constructing visions of the future \cite{vanderhelmVisionPhenomenonTheoretical2009}.\\ 
    tomorrow*& ``Tomorrow'' or ``tomorrows'' often functions as a synonym for the future \cite[e.g., ][]{bell_yesterdays_2007, kinsleyPractisingTomorrowsUbiquitous2010, sardarThreeTomorrowsPostnormal2016, mankoffLookingYesterdayTomorrow2013a}. \\
    specula*& Speculation has seen widespread adoption in HCI work. Building on the contribution of \citeauthor{dunneSpeculativeEverythingDesign2013a} \cite{dunneSpeculativeEverythingDesign2013a}, this method has served the field's critical examination of the future visions that new technologies represent, in particular \cite{wongSpeculativeDesignHCI2018}. \\
    imaginar*& ``Imaginary'' or ``imaginaries'' can allude to future shared imaginings of technology -- in other words, visions of the future that are constructed socially by a community, not just by a single individual \cite{fujimuraChapterFutureImaginaries2019}.  \\
    "design fiction"& Design fiction \cite{bleecker_design_2009, sterling_design_2009} is a method commonly followed to explore possible HCI futures by developing narratives without having to develop a new technology. \\
    societ* OR technolog* OR ethic* OR environmental* OR politic* OR legal* OR econom* & STEEPLE is a tool often used in futures studies to consider interrelationships among distinct factors associated with a future \cite{aguilarScanningBusinessEnvironment1967, saritasMappingIssuesEnvisaging2012}. \\ 
    \hline
    \textbf{Example queries} & [[Full text: "the future"] OR [full text: "futures"] OR [full text: "visions"] OR [Full text: envision*] OR [full text: tomorrow*] OR [full text: specula*] OR [Full text: imaginar*] OR [full text: "design fiction"]] AND [full text: societ*] \\
     & [[Full text: "the future"] OR [Full text: "futures"] OR [full text: "visions"] OR [full text: envision*] OR [Full text: tomorrow*] OR [full text: specula*] OR [full text: imaginar*] OR [Full text: "design fiction"]] AND [full text: ethic*] \\
    \bottomrule
    \end{tabularx}
\end{table*}

\subsection{Stage 2: Screening for Fleeting and Comprehensive Futuring}

Our analyses in the second stage allowed us to answer RQ1, pertaining to how HCI's future-orientation has developed in the last 15 years. 
Manual inspection of each paper pinpointed the works meriting full-text analysis in the final stage. Because experience had pointed to the paper's motivation or discussion/summaries of the findings as the strongest indicator of futuring, we directed our manual analysis to the corresponding sections of the articles 
(their introduction, discussion, and conclusion). 
We scripted downloads of the papers (full-text PDF versions of all 2,621) and programmatically highlighted the search terms' appearances within them, thus making the important parts easier to spot by eye. 

Collaborative classification of the papers and exclusion of those that do not express futuring 
required robust inclusion/exclusion criteria (see Appendix \ref{sec:inclusion-exclusion}) that the four researchers involved could reliably follow independently. This task was implemented in seven rounds, each with its own randomly selected set of 20 papers. After all individuals had worked through the set, we met and refined the inclusion/exclusion criteria. 

In the first decisions, from the round 1 meeting, our individual analyses of the paper sections' content led us to start categorizing the articles into three groups: 1) ones that provide comprehensive description of some future (``comprehensive futuring'' articles); 2) those not applying the key terms in the Introduction, Discussion, or Conclusion section or that refer to futures only in the context of proposed further work (``exclude'' articles); and 3) papers mentioning futures only superficially (``fleeting futuring''). At this point, we agreed to retain papers focused on futuring methods (e.g., speculative design) or with problematization of futuring in HCI even when explicit future visions were absent. Subsequent rounds saw us temporarily add a ``maybe'' category, for papers whose relevant sections promised that other sections would address futuring, and respecify the criteria for ``fleeting futuring'' by clarifying that the paper keeps the future technology uses or situations divorced from broader context (round 2); illustrate the definitions via excerpts from papers in each category (round 3); decide to limit our inspection to only those paragraphs in which search terms appear, rather than the entire section, and also to exclude papers that, while declaring that the work's contributions herald improvements, neglect to reflect on such implications for the future (round 4); remove the ``maybe'' category because it had become unnecessary and simultaneously exclude papers whose hypothetical use scenarios were not futuristic at the time of writing (round 5); decide to include (formerly ``maybe'') papers pointing to elaboration on futuring in sections other than the three target ones (round 6); and further explicate and fine tune the inclusion/exclusion criteria (round 7). In each round, we compared the team members' categorizations and computed the inter-rater agreement (via Fleiss kappa scores), which improved steadily, from 0.51 (in round 2) to 0.75 (in round 7), which lies near the top of the ``substantial agreement'' band (0.60--0.80) \cite{landis1977measurement}.

Upon reaching this level of reliability, we were ready to start the full-scale parallel screening. 
That entailed each of the sub-team's four researchers independently reviewing $4 \times 620$ articles not subjected to screening in the seven-round process described above. In the end, we had 205 articles screened in as manifesting comprehensive futuring, 2,157 excluded articles, and 259 in the ``fleeting futuring'' category.

\subsection{Stage 3: Qualitative Analysis Covering Comprehensive-Futuring Work}
\label{sec:qual-analysis}

We subjected the 205 comprehensive-futuring articles to in-depth reading, employing both inductive and abductive approaches. The goal for this stage of analysis was to answer RQ2, compassing the traits of comprehensive futuring in HCI. 
To analyze those characteristics, three of us began by reviewing typologies found in the futures studies literature \cite{borjesonScenarioTypesTechniques2006, ahvenharjuFiveDimensionsFutures2018, millerFuturesLiteracyTransforming2018, minkkinenSixForesightFrames2019, bergmanTruthClaimsExplanatory2010, vannottenUpdatedScenarioTypology2003, poli_anticipation_2019}.
Paying special heed to futuring models that might facilitate examining HCI's future-orientation, we selected those that articulate concepts suited to examining the degree or purpose of futuring in the literature review's sample \cite{borjesonScenarioTypesTechniques2006, minkkinenSixForesightFrames2019, millerFuturesLiteracyTransforming2018, ahvenharjuFiveDimensionsFutures2018}.
Then, the team synthesized these typologies into a four-category framework designed to inform analysis of the comprehensive-futuring papers. Table~\ref{tab:analytical-framework} presents the framework, denoted as ``SPIN'' (for its categories: Epistemic \textbf{S}tance, Contingency \textbf{P}erceptions, Systemic \textbf{I}ntegration, and \textbf{N}arrative).

\begin{table*}[t]
\caption{The SPIN framework's typology of HCI literature's futuring approaches and their possible manifestations} 
  \label{tab:analytical-framework}
\def\arraystretch{1.3}%
\begin{tabularx}{\textwidth}{l X}
\toprule
\textbf{Category} and manifestations & \textbf{Description} \\
  \midrule
  \textbf{S -- Epistemic \underline{S}tance} &
    \textit{What stance does the futuring take?} This communicates the kind of knowledge sought and the actions pursued through the futuring process. \\
  Normative &
    Promoting a certain vision (or multiple visions) of the future as desirable or pursuable \\ 
  Explorative &
    Exploring several scenarios for the future, to gain knowledge about the future, while not actively engaging with any particular vision as pursuable \\
  Predictive &
    Presenting some future vision(s) as likely to materialize or inevitable \\
  \hline
  \textbf{P -- Contingency \underline{P}erceptions} &
    \textit{How does the futuring account for the uncertainty arising from timescale, complexity, and complex relations among trajectories?} This reveals whether the futuring process allows for uncertainty vs. focusing on only narrow visions of the future. \\
  Narrow in approach &  
    Presenting a prediction / a vision of the future that features little or no uncertainty, space for other factors to divert its course, etc. \\
  Open in approach &
    In the vision(s) of the future, accounting for possible changes, factors that are not yet known, and/or relations between/among possible visions \\
  Short-term &
    Relying heavily on the past/present, with foreseeable implications \\
  Long-term &
    Acknowledging that complexity develops through time and that effects are largely unforeseeable \\
  Envisioning-strategy-based &
     Constructing the vision by using such methods as these: a linear trajectory (past to present), alternative pasts, a prediction model, current trends, weak signals, speculation, and questioning of the \emph{status quo} \\
  \hline
  \textbf{I -- Systemic \underline{I}ntegration} &
    \textit{How and to what extent are additional factors -- social, cultural, ethics, environmental, etc. -- considered?} This elucidates what factors beyond technological ones the vision treats and how they tie in with issues of scale. \\
  Oriented toward STEEPLE (with extensions) &
    Discussing factors of some/all of the following types: social, technological, economic, environmental, political, legal, ethics, more-than-human, personal-experience, community-related, and cultural \\
  Interconnection-focused &
    Considering interdependencies between/among factors and across scales \\
  \hline
  \textbf{N -- \underline{N}arrative} &
  \textit{What is the explicit and any 
underlying narrative, view, or belief about the future?} This reveals the aspects that the authors have highlighted to present or legitimize a given vision. \\
  Rationale-aligned &
    Using particular key elements to justify a future (e.g., technological development, climate crisis, or dystopia) \\  
  Centered on viewpoints/beliefs &
    Presenting certain assumptions and worldviews in the paper\\
    \bottomrule
\end{tabularx}
\end{table*}

The first category zeroes in on \textit{epistemic stance}, examined in \citeauthor{minkkinenSixForesightFrames2019}'s analysis of futures scenarios \cite{minkkinenSixForesightFrames2019} via the lens of the levels of change pursued.
Actors' stance hinges on their futures literacy \cite{millerSensingMakingsenseFutures2018}: they might link the knowledge/epistemology to concrete actions and goals (e.g., normative strivings toward particular futures), to more explorative investigations of what \textit{could} happen, or to predictive statements intended to capture the most likely trajectories \cite{borjesonScenarioTypesTechniques2006}. 
We saw parallels with researchers' possible epistemic stances, which could prove fruitful in distinguishing among types of HCI publications -- namely, those aligned for pursuing normative technological advances, speculating about possible futures through design-fiction explorations etc., and presenting predictable implications of technology-mediated activities.

The second category, \textit{contingency perceptions}, likewise draws from \citeauthor{minkkinenSixForesightFrames2019}'s analysis -- specifically, of futures scenarios' levels of perceived unpredictability. 
In addition to unpredictability and uncertainty, which are central to every futures studies investigation, we incorporated scope-related and temporal manifestations into this category, in line with the work of both \citeauthor{ahvenharjuFiveDimensionsFutures2018} \cite{ahvenharjuFiveDimensionsFutures2018} and \citeauthor{millerFuturesLiteracyTransforming2018} \cite{millerFuturesLiteracyTransforming2018}. A scenario might be short-term or longer-term but also could be narrow or more open. 
A narrow approach can be useful when it is reasonable to assume that the future will not be affected by major external factors \cite{millerSensingMakingsenseFutures2018}. 
Researchers who adopt an open approach, in contrast, can embrace complexity and uncertainty as resources for awareness-building \cite{millerSensingMakingsenseFutures2018}. 
For crystallizing temporalities, 
we adapted \citeauthor{ahvenharjuFiveDimensionsFutures2018}'s 
\cite{ahvenharjuFiveDimensionsFutures2018} 
subjective timescales to the horizons typical of HCI scholarship: we mapped short-term visions to incremental changes (as in iterative design) and long-term considerations to accounts of emerging/unforeseeable effects, alongside unpredictable consequences \cite{epp2022reinventing}. 
Our final insight here is that analysis might benefit from bearing in mind differences in envisioning strategies: 
HCI use-scenario methods tend to be unidirectional, progressing linearly from past to future \cite{nathan_envisioning_2008}, while HCI scholars' design fiction and speculative design may contemplate alternative-future trajectories with criss-crossing, parallel stretches, and several possible starting points \cite{blythe_research_2014}.

\textit{Systemic integration} captures the differences in futuring's angles of analysis. This category covers both the traditional STEEPLE \cite{aguilarScanningBusinessEnvironment1967, saritasMappingIssuesEnvisaging2012} 
and extensions that observations from stage 2 had led us to regard as fundamental: the screening process revealed that many HCI papers attend to individual-level and community experiences, cultural factors, and more-than-human futures -- none of which the STEEPLE nomenclature directly addresses. 
The importance of these factors' interconnections prompted us to conclude that they deserve consideration as a factor in their own right; our conceptualization here adhered to \citeauthor{ahvenharjuFiveDimensionsFutures2018}'s \cite{ahvenharjuFiveDimensionsFutures2018} systems-perception dimension and \citeauthor{saritasMappingIssuesEnvisaging2012}'s \cite{saritasMappingIssuesEnvisaging2012} suggestions for scenario-building. 
This category offered particular value for problematizing our initial perception of HCI work as populated largely by a techno-centric set of envisioned futures, whereby STEEPLE's technology factor eclipses others in the futuring. We concluded that zooming in on the other possible factors should help us gauge the truth behind that assumption. 

The final SPIN category, pertaining to \textit{narratives}, contends with the explanations and views communicated in the scenarios (for example, the rationales and worldviews behind them). For this, we were informed by \citeauthor{datorAlternativeFuturesManoa2019}'s scenario archetypes and \citeauthor{bazzaniFuturesActionExpectations2023}'s narrative dimension of the future.
Via reduction (distilling the complex) and appeals to emotion (evoking affective responses), narratives shape the future visions' appearance and how they -- and the underlying beliefs -- are legitimized for audiences~\cite{datorAlternativeFuturesManoa2019, bazzaniFuturesActionExpectations2023}. 
We delineated this category for assistance in identifying the article authors'
underlying sociopolitical motivations, as well as the archetypes, assumptions, and beliefs \cite{selinSociologyFutureTracing2008} in HCI's future visions \cite{datorAlternativeFuturesManoa2019}, whether overt or implicit. 

This four-category framework provided solid theory-underpinned grounding for the full-text qualitative analysis that followed.
From our sample of 205 comprehensive-futuring papers, we began by selecting 20 at random and distributed them across all five researchers for in-depth reading. Each paper was read by three of us. 
In our analysis of these 20 papers (12 articles per researcher), we observed that many HCI studies have approached the future via the lens of individuals' experiences as opposed to system-level relations -- a dimension overlooked by our framework's synthesis of futuring typologies. 
Accordingly we returned to the SPIN categories, refining them to incorporate manifestations of personal experience, cultural factors, and more-than-human approaches, all of which seemed relevant on the basis of our prior knowledge of HCI futures discourse. 

Next, we randomly distributed a further 90 papers for reading, this time across three researchers (30 papers each).
We left space both for further tuning of the framework and for flexibility beyond this or any other specific tool.
For instance, we remained free to probe facets not visible in the categories of the SPIN framework, such as commonplace methods, key stakeholders, and the foci of the study. Our analysis was platform-agnostic too: we did not limit ourselves to any analysis application (e.g., ATLAS.ti) or template (e.g., the MS Word file portraying the framework). 
This supported balanced analysis: we were guided by the four-category framing but also encouraged to abductively identify unforeseen futuring-related interpretations with a bearing on the research questions. In practice, 1) our coding highlighted the distinct manifestations of each SPIN category in every comprehensive-futuring article while, 
in addition, 2) any researcher who noticed an important mention of the/a future that did not fit the categories could augment the evaluation accordingly. 
Each researcher concluded the paper-specific analysis with a brief qualitative summation describing the paper's position relative to the dimensions expressed by the SPIN categories. 
To ensure equitable distribution of tasks within the team, only two authors analyzed the remaining papers. 
We developed our findings in full awareness that our corpus, being limited (partly for manageability's sake) to the papers likely to have exerted the greatest influence within the HCI field, is not representative of all futuring research in the HCI domain. The earlier stages in our analysis, on the other hand, fleshed out the broader picture of the discipline considerably.

\section{Findings} \label{sec:findings}

The careful analysis presented above supplied answers to both RQs, which we address in the following two subsections.

\subsection{How HCI's Future-Orientation Has Developed (RQ1)}

Owing to the breadth of RQ1, covering the landscape of HCI futuring practice across the board, we sought to uncover the path taken thus far in HCI's future-orientation in a manner that did not necessitate full-text examination of all our large corpus.

The screening stage and its inclusion/exclusion process (see Figure~\ref{fig:prisma} and Appendix~\ref{sec:inclusion-exclusion}, respectively) provided a suitable starting point for considering RQ1. That stage's decision-making, fruit of the authors' manual analyses of Introduction, Discussion, and Conclusion sections as described above and harmonized via seven iterations, 

yielded a classification that suits this aim well: dividing the publications into three categories: those for exclusion, ``fleeting futuring'' papers, and ``comprehensive futuring.'' 

We will now characterize these groupings, which served as ample foundations for answering the first research question. In the course of our overviews of the two main categories, we show how they represent the evolution of HCI scholarship's future-orientation over time. 

The \textbf{fleeting-futuring} papers' distinguishing property is content touching 
only briefly upon the future, often in passing or as a minor aspect of the topic under discussion. These fleeting references typically lack depth and seldom reflect the main focus of the paper. The piece might mention potential for further studies
or speculate vaguely about future trends but avoid committing to detailed explorations. The following extracts exemplify these patterns. Text included before and after each (italicized) fleeting reference to some future sets these references in context, thus demonstrating that their mention of a future does not constitute part of extensive elaboration. 

\begin{quote}
    \textit{We envision StickEar to be an empowering personal device that anyone would carry and use every day to augment objects and spaces.} 
    [end of paper] \cite[p. 226]{yeo_stickear_2013}
\end{quote}
\begin{quote}
    \textit{The high maintenance and environmental cost of batteries lead to concerns about the wireless sensor networks domain. This sustainability issue has led to battery-free embedded devices powered by ambient energy sources (vibrations, radio frequency transmissions, and light). \textit{These battery-free devices are likely to form the future of physical-computing devices and the Internet Of Things due to being maintenance-free and enhancing long-term deployment.} Recent battery-less device demonstrations include phones, satellites in space, implantables, devices conducting machine learning, handheld gaming consoles, and even underwater sensing}\cite[p. 2]{kraemer_battery-free_2022}
\end{quote}
As illustrated above, the future visions presented were often condensed in one or two sentences, usually imagining the positive influence of the technology discussed.

In \textbf{comprehensive futuring}, the
papers discussed futures in a more holistic manner. These papers did not only speculate on future technologies but also considered the socio-cultural, ethical, or practical implications of these technologies. Such comprehensive treatments often involved scenarios that were richly detailed, offering both broad visions and specific predictions to construct meaningful future worlds. Awareness of the future is typically woven throughout the content of papers in this category, or a whole section might be dedicated to discussing futuristic scenarios, inclusive of thorough and more sustained engagements with futuring. We provide multiple examples of this category in the next subsection in conjunction with the characteristics we pinpointed as constitutive of the last 15 years' comprehensive futuring in HCI. 

\begin{figure*}[tb]
    \centering\includegraphics[width=0.95\linewidth]{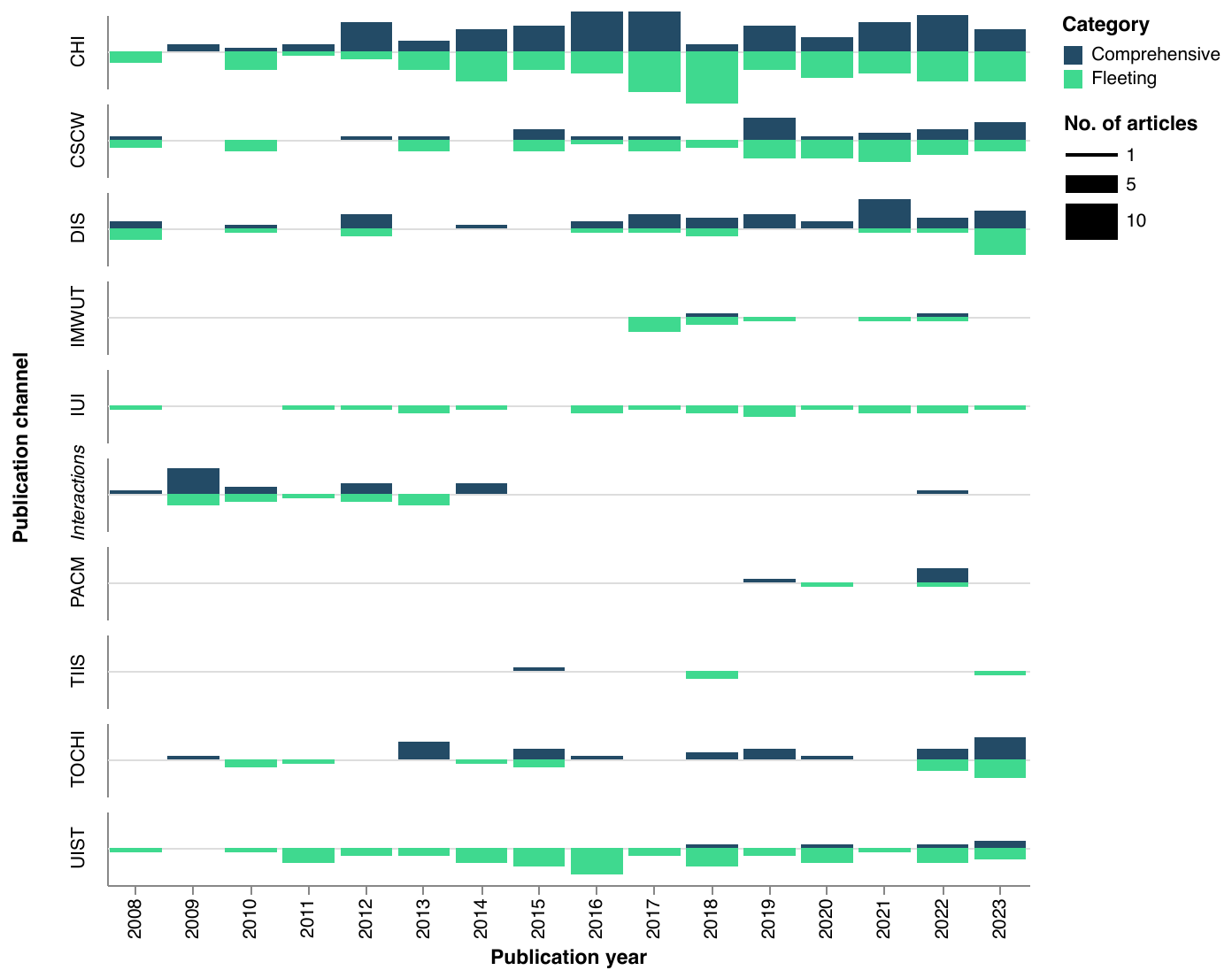}
  \caption{A histogram presenting the articles in our comprehensive-futuring sample (blue) and in the fleeting-futuring sample (green) by year (on the \emph{x}-axis) and by publication venue (on the \emph{y}-axis). Bar length corresponds to the number of articles in each category.
  Empty space indicates that no relevant articles were published in the relevant year and venue (CSCW material published in PACM is included (only) under the CSCW category). 
  }
  \Description{A histogram representing the amount of articles that were in the comprehensive futuring sample, represented by blue bars, and the fleeting sample, represented by green bars, and in 
  Both categories are distributed by year, from 2008 to 2023, in the x-axis and by publication series in the y-axis.
  Bar length offers an approximation of the number of articles in each category.
Empty space indicates no articles in that year and venue. 
  The histogram highlights a general upward trend in both categories across most series, particularly in CHI, CSCW and DIS. The comprehensive futuring contributions, while smaller in number, have increased noticeably in recent years.
  }
    \label{fig:heatmap}
\end{figure*}

Graphically depicting the development of futuring within HCI over the years (e.g., visually answering RQ1), Figure~\ref{fig:heatmap} presents a timeline of how both categories of future-oriented pieces in the venues examined have developed. 
Note that this graph covers the counts only of fleeting- and comprehensive-futuring papers that 
were above our citation cut-off (see Subsection~\ref{sec:identification}).
While Figure~\ref{fig:heatmap} hence does not represent the total volume of future-oriented papers in each venue, category, or year, it does offer a revealing overview of the most cited future-oriented papers in ACM's HCI venues over the last 15 years.

A trend toward future-oriented articles began to emerge in most of the venues in 2012 (TIIS and \emph{Interactions} constituted exceptions).\footnote{~Note, however, that not all of the publication venues were active throughout the span of time considered. For instance, TIIS was launched in 2011 while PACMHCI and IMWUT entered print only in 2017. 
Likewise, CSCW being biennial until 2012 and DIS until 2016 explains gaps in the corresponding timelines.}
This trend has accelerated since. With regard to such venues as CHI, DIS, CSCW, and TOCHI, we can interpret the growing number of comprehensive-futuring articles as evidence of a tendency to conceptually approach the future in an increasingly holistic manner. Growth in comprehensive futuring peaked in 2017, though the paper count is still showing considerable growth, in the 2020s. 

The fleeting-futuring category too shows growth over the years, particularly from 2011 onward. While the rise in this class of futuring is less strong, even the more technical areas of HCI research clearly demonstrate a growing inclination to extrapolate technologies to visions of the future. 
Overall, fleeting futuring seems to dominate the HCI domain, but there is an evident shift to engaging in comprehensive futuring, particularly in the venues where the social sciences and design practitioners exert a strong influence. And that trend is gaining pace.

To render the temporal development of HCI's future-orientation more comprehensible, it is worth considering 
the fundamental nature of this orientation: how it established itself in the first place. 
The ``fleeting futuring'' category's dominance in Figure~\ref{fig:heatmap} might be a consequence of HCI research’s inherently technology-oriented nature. 
In this light, it may not be surprising that technology forms the main lens through which visions of futures get articulated. 
The operationalization we developed for the fleeting mentions echoes this -- we found brief sketches of potential future applications that depend on the technology introduced in the paper. 
We noticed that fleeting visions of the future typically point to technologies as the main driver of change, therefore 
suggesting leaps or linear progression from certain current applications to future possibilities. These visions also often stress the more positive opportunities created by emerging technologies. 
The extracts below, all from the ``fleeting futuring'' category, illustrate the extrapolation of technology to visions of the future. 
Such extrapolation existed also in comprehensive-futuring papers, but space considerations preclude quoting from the latter, since their descriptions of the future generally extend through two or more paragraphs \cite[e.g., ][]{mueller_jogging_2015, su_dolls_2019, jhaver_designing_2022}.

\begin{quote}
    \textit{We envision eye trackers in the future to be integrated with consumer devices (laptops, mobile phones, displays), hence allowing the user’s gaze to be analyzed and used as input for interactive applications.} \cite[p. 267]{alt2014eye}
\end{quote}
\begin{quote}
\textit{In the future, it is possible that many people who are not already able to drive and are considering learning will have the option of buying an AV} [automated vehicle] \textit{instead.} \cite[p. 526]{hewitt_assessing_2019}
\end{quote} 

Emerging technologies were often treated as springboards that allow for elaboration on potential futures. This pattern reveals a propensity to envision futures through the lens of technological capabilities or implications, thereby confining the imagination of the future to the proximity of certain technologies' existence. 
For example, the authors of many papers in each of the categories focused predominantly on technological factors when envisioning the future. That said, there are notable exceptions: researchers whose envisioning of the future has encompassed non-technological factors \cite{tomlinson_collapse_2013, oogjes_designing_2018, vines_age-old_2015}.
The reliance on technology as the underpinnings to futuring in both categories of papers reveals a degree of techno-centrism in HCI. Indeed, studies' prominent focus on incremental advances in technology and on persuasion to use technology is called out in several papers addressing sustainable HCI \cite{pierce_beyond_2012, brynjarsdottir_sustainably_2012, chopra_negotiating_2022, crivellaro_contesting_2015}. Their authors have called for further engagement with the societal, ethics, and environmental aspects of technologies. 

\noindent Narrowing our examination of techno-centrist tendencies to papers in the comprehensive-futuring category uncovers greater variation in how scholars regard technologies. 
These works often expand the discussion to include broader future scenarios for technology use, specifically involving individual-level experiences \cite{devendorf2016dontwear, luria_re-embodiment_2019, sanchez-cortes_mood_2015, su_dolls_2019}, as alluded to above. For example, one project explored how wearables could mediate attention and contribute to stress management: 

\begin{quote}
    \textit{Even though she imagined activating her scarf with her phone, she felt that her scarf interface would ameliorate stress by providing a slowly shifting dedicated information stream that would allow her to direct her attention to the world, interacting with others, rather than frequently checking her phone.} \cite[p. 6036]{devendorf2016dontwear}
\end{quote}
Similarly, \citeauthor{luria_re-embodiment_2019} investigated how social presence in the auto\-nomous-vehicle space could foster user trust and understanding of technical capabilities:
\begin{quote}
    \textit{The driving social presence can also try to assure the user that it can handle multiple tasks safely; it is possible that more open communication about the car’s “cognitive” effort would have helped participants to better understand its technical abilities.} \cite[p. 642]{luria_re-embodiment_2019}
\end{quote} 

Researchers engaging in comprehensive futuring often delved into the societal implications of the various technologies, posing questions regarding how they might reshape societal norms and interactions. 
Finally, some studies in the comprehensive-futuring category elicited reflections on the values inscribed in the design of technologies (e.g., \citeyear{wong_eliciting_2017}'s from \citeauthor{wong_eliciting_2017} \cite{wong_eliciting_2017}, work by \citeauthor{sondergaard_intimate_2018} from \citeyear{sondergaard_intimate_2018} \cite{sondergaard_intimate_2018}, \citeyear{ballard2019judgment}'s paper by \citeauthor{ballard2019judgment} \cite{ballard_judgment_2019}, and \citeyear{suresh_beyond_2021}'s by \citeauthor{suresh_beyond_2021} \cite{suresh_beyond_2021}) rather than solely improving the technological aspects of design. For instance, \citeauthor{wong_eliciting_2017} approached design artifacts not as end products but as tools to provoke discussion about multiple potential futures and prompt professionals to reflect on underlying social values during product development \cite{wong_eliciting_2017}. With design fiction at their disposal for eliciting values, they shifted the focus from simply refining technologies to understanding the societal context in which these operate and incorporating ethics deliberations into design processes. 

\subsection{Characteristics of Comprehensive Futuring in the HCI Field (RQ2)}
    
To address RQ2, we purposefully performed full-text qualitative analysis of all papers in the ``comprehensive futuring'' category. Accordingly, the findings presented in this subsection are derived from the third stage of the paper-screening process (shown at the bottom in Figure~\ref{fig:prisma}). 
Through the abductive process described in Subsection~\ref{sec:qual-analysis}, we developed and employed the framework presented in Table~\ref{tab:analytical-framework}, which guided the team in mapping common ways of envisioning the future onto the future-orientation of the sample. Informed by the SPIN framework, we identified four main properties of the comprehensive futuring, denoted as ``exploration of uncertainty,'' ``prevailing short-term time horizons,'' ``focus on human experience,'' and ``contestation of dominant narratives.'' 

\subsubsection{Exploration of uncertainty} 
Whenever futures are discussed, they bring uncertainties with them, with our study being no exception. Many papers in our corpus display an explorative approach to uncertainties.

Exploration of uncertainty displays close links to two categories in the SPIN framework -- epistemic stance and contingency perceptions.
For example, papers adopting an explorative epistemic stance \cite[][ etc.]{alves-oliveira_collection_2021, herdel_above_2022, homewood_tracing_2021} and normative ones \cite[such as ][]{smith_designing_2017, sondergaard_fabulation_2023, prost_awareness_2015} demonstrate heightened sensitivity to the uncertainties bundled with interactive technologies and with their relations to human and other environments. Those papers also manifest open approaches to contingencies by discussing such unpredictable aspects of technology as the ``disconnect'' between urban informatics and nature \cite{smith_designing_2017} or the host of roles that robots could take in society \cite{alves-oliveira_collection_2021}.

In an effort to explore uncertainties via other means than forecasting of technological trends, explorative researchers have engaged in speculation and fabulation to envision alternative futures. Their papers highlight potential risks \cite{wong_eliciting_2017,tseng_dark_2022, eghtebas_co-speculating_2023}, illuminate unexpected outcomes \cite{pater_no_2022, park_social_2022}, and inquire into desirable and utopian futures \cite{chopra_negotiating_2022, sondergaard_fabulation_2023, bardzell_utopias_2018}:

\begin{quote}
  \textit{Presented as a U.S. university-based fictional memo describing a post-hoc IRB} [institutional review board] \textit{review of a research study about social media and public health, this design fiction draws inspiration from current debates and uncertainties in the HCI and social computing communities around issues such as the use of public data, privacy, open science, and unintended consequences, in order to highlight the limitations of regulatory bodies as arbiters of ethics and the importance of forward-thinking ethical considerations from researchers and research communities. Though the following illustrative example is fictional, it is inspired by real examples as well as current ethical debates within the HCI community.} \cite[p. 2]{pater_no_2022} 
\end{quote}

\begin{quote}
    \textit{We also agreed that a key alignment was a desire for our images to be utopian: we intended for them to be interpreted as visual stories for positive change and to counter dystopian narratives that often aim to critique societal values by portraying bleak future scenarios. Striving for utopianism was a direct response to the challenges we had faced in doing critical work with bodily fluids, including stigma, ethics, emotion work, and institutional resistance; from which we needed a space of joyful allyship and optimism.} \cite[p. 1698]{sondergaard_fabulation_2023}
\end{quote}

A more speculative, explorative approach to uncertainty reflects deeper engagement with the complexities of future technologies, acknowledging both their transformative potential and their unpredictability.
Often we could identify an overarching narrative of existential unease. This narrative raises concerns about the implications of emerging technologies, bringing to the fore such issues as the ethics-related tensions of responsible AI \cite{rakova2021responsible}, the increasingly indistinct boundaries between humans and machines \cite{seering_beyond_2019, brand_design_2021}, and societal disruptions (such as mass unemployment) wrought by automation \cite{oh_us_2017}.

Reflections on the ambiguity surrounding human--machine interactions spark existential unease, as illustrated in this extract:

\begin{quote}
    \textit{What is striking, however, is that even quite positive participants seem uncertain about the relationship that will emerge from interacting with an sVA.} [social voice assistant] \textit{Is Kiro the end of "awkward silence" or will it just create experiences similar to the human-human conversation? Will Kiro use technology the same way people do, although it is a technology itself? While the notion to become social with a machine appeared interesting and stimulating, participants seemed slightly puzzled about whether the fact that the social partner remains a technology will negatively or positively impact their experiences.} \cite[p. 13]{ringfort-felner_kiro_2022}
\end{quote}

Similarly, depictions of tension between perceptions of AI as a potential danger and as a helpful tool underscore the uncertainty surrounding its role in societies and a need for control over its use:
\begin{quote}
    \textit{Through the interviews, we identified that the participants had preconceptions and fixed ideas about AI: (a) AI could be a source of potential danger, and (b) AI agents should help humans. Although these two stereotypes seem to be contradictory, one seeing AI as a potential danger and the other seeing it as a tool, they are connected in terms of control over the technology. The idea that AI could be dangerous to humans can be extended to the idea that it should be controlled so that it can play a beneficial and helpful role for us.} \cite[p. 2530]{oh_us_2017}
\end{quote}

\subsubsection{Prevailing short time horizons}    
As introduced above, the pattern of taking technology as grounding for envisioning futures is visible in the comprehensive-futuring sample too. A techno-centric orientation directly shaped the time frames of the envisioned futures, binding them to the technological possibilities of the present.

 We found that techno-centrism cut short attention to long-term futures, such as those spanning 50 or 100 years, yet precisely these extended time frames prove critical for understanding how interactions evolve over time \cite{nathan_envisioning_2008}, how technology adoption unfolds \cite{lindley_implications_2017}, and ways to design for a dynamic and unpredictable future \cite{kozubaev_expanding_2020}. 

Even in studies employing design fiction and speculative design that probes alternative futures \cite{sondergaard_intimate_2018,khan_speculative_2021, ballard2019judgment}, there was often an implicit assumption of linear progression from existing conditions, with speculative scenarios plainly extended from current technologies and societal behavior. Only a few papers \cite{nathan_envisioning_2008, tseng_dark_2022, sanders_designing_2014, speicher_what_2019, chopra_negotiating_2022, carroll_wild_2013} detail any long-term considerations; far less rare are vague references to ``the long term,'' and concrete time frames, such as five years or a decade, seldom feature.

This limitation is especially notable in contexts requiring longer time horizons to address systemic challenges.
For example, \citeauthor{chopra_negotiating_2022} highlighted the importance of designing for extended time\-scales to address pressing urban challenges:

\begin{quote}
    \textit{While struggling to balance between boundless speculation and the uncompromising realities of the situated everyday. Speculative design and associated approaches for instance, have been drawn on to highlight damaging anthropocentric consequences in the near future [\ldots]. Here, designing for longer timescales is becoming particularly prescient for many urban communities due to the scale of the challenges and ever increasing threats presented by governance, degrading environments, and growing urban populations [\ldots].} \cite[p. 2]{chopra_negotiating_2022}
\end{quote}
Similarly, \citeauthor{nathan_envisioning_2008} underscored how systemic interactions and the impacts of technologies often take years to become evident, highlighting longer timeframes:
\begin{quote}
    \textit{Yet most successfully deployed technologies remain in use in society far longer, on the order of 3, 5, or 10[-]plus years. Moreover, systemic interactions emerge over time. Thus, we are more likely to notice these interactions 5 years rather than 5 months out.} \cite[p. 3]{nathan_envisioning_2008}
\end{quote}

\noindent Temporal delineations of the future need not focus exclusively on moving forward in time, though \cite{kozubaev_expanding_2020}. Counterfactuals \cite{forlano_speculative_2023}, fabulation \cite{sondergaard_fabulation_2023}, and relationships with more-than-human entities \cite{liu_design_2018}  merge temporalities, present alternative futures, and bring out uncertainties through such mechanisms as alternative pasts, parallel timelines, and data obtained from living organisms. 

We appeared to detect the beginnings of a more sensitive approach to uncertainties, nuanced beyond risk factors \cite{kozubaev_expanding_2020, forlano_speculative_2023, benjamin_machine_2021, bodker_participatory_2018}. 
This sensitivity was exhibited by approaches that frame temporality as flexible and contestable, as \citeauthor{kozubaev_expanding_2020} have suggested:

\begin{quote}
    \textit{For HCI design to be reflective about temporality, we suggest framing temporality as malleable and contestable, thereby opening new possibilities and ways of speculating about the future. First, HCI designers can explore alternative and novel notions of temporality and make them more visible and interactive. For example, Odom et al.’s work on slow design illustrates how HCI design can support reflections and subjective experiences of time such as anticipation, memory and re-visiting the past. Soro et al. propose an alternative take on the futures cone by flipping its orientation, much like the Aymara, to designing for the past.} \cite[pp. 5-6]{kozubaev_expanding_2020}
\end{quote}
\citeauthor{benjamin_machine_2021} extended this perspective by examining how addressing uncertainty through speculative design can challenge normative assumptions about human--technology relations:
\begin{quote}
\textit{Designing for thingly uncertainty with futures creep and pattern leakage can shed light on how human subjectivities become entangled with ML} [machine-learning]\textit{-driven artefacts. This is not only a symbolic or aesthetic exercise, but rather a potentially powerful way of investigating how standard thinking on human--ML relations rely on normative assumptions (e.g., anthropocentric, capitalist, hetero-normative) about the technological as much as the human side of those relations.} \cite[p. 11]{benjamin_machine_2021} 
\end{quote}

\subsubsection{Focus on human experience}
We also identified a characteristic whereby HCI seems to fill a unique role that sets it apart especially from the traditions of futures studies: expression of an interest in individuals and their experiences. This focus seems absent from STEEPLE and the other futuring typologies reviewed in the course of developing the SPIN framework.
Hence, upon repeatedly encountering individual- and experience-oriented papers that resisted ready mapping to existing typologies, we chose to introduce individual-level experience as a key dimension within SPIN's systemic-integration category. 

The introduction of individual experience into SPIN proved immediately useful. In the comprehensive futuring papers, we observed a related implicit precautionary narrative framing emerging technologies and their unexpected consequences as threats to humanity. This narrative often reflected an anthropocentric bias in how future scenarios are envisioned.
These papers often explored the ethical, psychological, and social impacts of technologies in human life, digging into how they might impinge on human actions by driving behavior change \cite{brand_design_2021, sauppe_social_2015}, disrupting social norms \cite{blackwell_harassment_2019}, perpetuating inequality \cite{eghtebas_co-speculating_2023}, or even challenging our conceptions of privacy and autonomy \cite{jakobi2019privacy, pierce_smart_2019}. 

The focus on human-experience future scenarios is illustrated well by \citeauthor{brand_design_2021}'s seven design proposals for introspective AI. 
\begin{quote}
    \textit{Deep Talk Report preserves and enhances records of deep social exchanges as interactive resources. This approach suggests an opportunity for designers to generate more interpersonally-oriented Introspective AI applications that engage directly with the social relations that shape a person’s current and future ideal self. Nevertheless, there is a need for future work to explore the extent to which divorcing these exchanges from their original context and reducing them to interconnected bits might alter their perceived value and lead to added social expectations.} \cite[p. 1614]{brand_design_2021}
\end{quote}
\citeauthor{eghtebas_co-speculating_2023} spotlighted human experience similarly, by underscoring the risks of socio-economic imbalances (re)produced in the rollout of ubiquitous technologies, and called for interdisciplinary approaches so as to mitigate the attendant challenges:
\begin{quote}
    \textit{Technology cannot fix all of society’s problems, but it can inherit them. The discussion motivated from our analysis highlights how multi-disciplinary perspectives, including technological views, but also political and social sciences will be required to be able to fully mitigate what may become of our envisioned dark scenarios. As alluded to in several of our scenarios, UAR} [ubiquitous(ly) Augmented Reality] \textit{is inherently asymmetric: unequal access to costly hardware, or the lack of accommodating of peoples various roles and abilities as UAR continues to develop, can cause societal imbalances.} \cite[p. 2404]{eghtebas_co-speculating_2023}
\end{quote}

\noindent On the other hand, HCI's disciplinary strength in examining the everyday interactions with technology also presents opportunities to investigate futures at the micro level, which is often lacking in more systemic analyses from futures studies \cite{candy_futures_2010}.
A micro-level focus can help elucidate how individuals and their communities experience, adapt to, and resist technological changes in day-to-day life.

By granting prioritizing micro-scale interactions between people and technology, individual-experience-oriented papers with comprehensive futuring have propagated nuanced insight connected with broader impacts of technologies' integration and adoption \cite{carroll_wild_2013, pierce_smart_2019, lindley_implications_2017, elsden_speculative_2017}. 
A speculative approach to design has afforded deeper probing of the potential futures of marginal communities, highlighting matters regularly overlooked in mainstream technology design and addressing inequality of several sorts \cite{harrington2019deconstructing, keyes_reimagining_2020, harrington_eliciting_2021, khan_speculative_2021}, culture-specific elements \cite{bray_speculative_2021}, and unique needs of underrepresented groups \cite{morrissey_value_2017, bennett2019biographical, bardzell2010feminist, haimson_designing_2020, dillahunt_eliciting_2023}. 

For instance, \citeauthor{harrington_eliciting_2021} illustrate how participants struggled to envision technology-based futures free from systemic social issues, such as racism:
\begin{quote}
    \textit{Our results uncovered key constructs that are difficult or perhaps impossible to separate from design. Common among our findings was that students have a difficult time imagining a future without the existing social issues they face today. Among many of the utopian concepts were still elements of identified dystopian challenges that seemingly could not be detached from concept generation. In all cases, students’ technology-based futures encapsulated some form of racism and it was difficult for them to imagine technology that exists in a world without it. Previous literature suggests that human imagination is bounded and people might have difficulties imagining the future as situations become more distant in likelihood, perspective, time, and place. However, our analysis provides insight into the unique difficulty of envisioning the future that is confounded by race and social class, which was present among our participants} \cite[p. 10]{harrington_eliciting_2021}
\end{quote}

\subsubsection{Contestation of dominant narratives}
When we examined the comprehensive-futuring papers' narratives, several publications stood out for their critical examination of how design decisions privilege certain individuals or groups over others, highlighting issues of power dynamics embedded in the design process \cite[e.g., ][]{pendse_treatment_2022, keyes_reimagining_2020, winchester_realizing_2010, breuer_how_2023, suresh_beyond_2021, adamu_no_2023, avle_designing_2016, dell_ins_2016}.
These works reveal varying degrees of bias -- whether stemming from the designer or the focus on expert knowledge --
which can culminate in missed opportunities for overlooked communities. For example, \citeauthor{breuer_how_2023} adopts a science \& technology studies lens to analyse the ethical, social, and cultural assumptions that underlie the design of healthcare robotics \cite{breuer_how_2023}. Similarly, \citeauthor{suresh_beyond_2021} illustrate how machine learning designers' recurrent focus on expert knowledge, which leads to assigning all other potential users the one-dimensional label ``non-experts'' and virtually guarantees that opportunities for inventing more equitable, adaptable, and context-sensitive machine learning systems remain overlooked \cite{suresh_beyond_2021}.
Other work stressing designers' accountability for their position of power has been done under the ``human--computer interaction for development,'' or HCI4D, umbrella \cite{avle_designing_2016, dell_ins_2016, saha_towards_2022, adamu_no_2023}.
Despite these works and the expansion of interactive systems' design beyond ``the West'', Western standards and narratives continue to dominate. 

\begin{quote}
\textit{The concept of interest convergence, which stems from critical race theory, holds that those in power tend to support goals that serve their own interests. In other words, without actively involving stakeholders whose interests are in opposition to existing power structures, and considering their input crucial, resultant interpretability systems will fit the standards and needs of those in power -- for example, executives with a vested interest in maintaining the status quo, or engineers and researchers who might communicate about model decisions in a way that is not understandable to people without formal ML} [machine learning] \textit{knowledge. Involving stakeholders with different interests first requires reflexivity, or explicitly acknowledging what our own backgrounds and interests are.} \cite[p. 12]{suresh_beyond_2021}
\end{quote}

\begin{quote}
    \textit{Several participants called for a stronger focus on designing for non-traditional computing environments. For example, P5 said, “Why would you have an office, QWERTY keyboard, desktop metaphor, textual interface for people who don’t think about things in that way? The traditional appliances and systems embed middle-aged white guys from the Pacific North-west. They are the ones in the corner office whose language is premised in QWERTY. Not only their spoken language, but they’re also print literate. The appliance is really focused on that context and no wonder it can be alienating to different contexts. HCI4D is about breaking out of these rich, white, male, US systems into all kinds of other systems. What would a tropical computing environment look like?”} \cite[p. 2228]{dell_ins_2016}
\end{quote}

\noindent In addition to addressing the social and ethics factors obviously raised by HCI, works in the sub-fields of sustainable HCI and more-than-human design \cite{homewood_tracing_2021} underline the webs of relations that link technological developments with ecological systems \cite{nathan_sustainably_2009}. Some researchers dealing with these areas explore connections between cultures and perspectives \cite{lindtner2016reconstituting,bennett_promise_2019, brynjarsdottir_sustainably_2012, chopra_negotiating_2022}, alongside how these intersect with environmental ecosystems \cite{hansson2021decade, smith_designing_2017}, advocating accordingly for sustainable transformation and inclusive design practices that reflect more-than-human concerns \cite{liu_design_2018, frauenberger_entanglement_2020, tsaknaki_fabulating_2022, sondergaard_feminist_2023}. 

One case in point comes from conceptualizing tools that support diverse human--fungus interactions: \citeauthor{liu_design_2018} reimagines design as a means to enable collaborative survival, with the nature of the constituent relations left open-ended and adaptable.
\begin{quote}
    \textit{What the variety of tools makes evident about collaborative survival is that it is well suited to attend}[ing] \textit{to the multiplicity of human--fungi relationships. The metaphor allowed us to honor fungi in its multiple forms and expressions--a ubiquitous underground network, a barometer for ecosystem health, a delicacy to eat, or a specimen to identify (to name a few). There is not one true or correct life for a human to attend to and thus, there are multiple ways of becoming entangled with fungi in both its physical and digital manifestations. For instance, Liu did not design the tools to enact particular narratives of protecting or conserving fungi. Instead, the concept suggested that we design only to make a particular relationship with fungi possible, and it is up to each human as to what that relationship can entail and what arrangements will be mutually beneficial.} \cite[p. 9]{liu_design_2018}
\end{quote}
Taking a distinct stance, \citeauthor{tsaknaki_fabulating_2022} explored what more-than-human design can bring to the table to reframe collab-relationships across human and non-human entities: 
\begin{quote}
    \textit{Taken together, these concepts explore more-than-human agencies living, knowing, and collaborating with humans. Acknowledging these strands of thought do not all knit together perfectly, this theme invites expansive ideation on how relations and collaborations between human and non-human bodies. Instead of foregrounding the authority of exclusively human bodies and biodata, this theme seeks to also account for other companion species [\ldots] in or outside our bodies, ranging from microorganisms to animals, to materials as vibrant bodies [\ldots], to technologies.} \cite[p. 1181]{tsaknaki_fabulating_2022}
\end{quote}

While the foregoing examples demonstrate reaction against entrenched phenomena such as techno-centrism, they also point to our field's rich legacy and its focus on human factors in computing. 
They illustrate a shift toward breaking the hold of Western narratives \cite{alvaradogarcia2021decolonial, adamu_no_2023} and anthropocentric discourse \cite{yoo2023more, sondergaard_feminist_2023}. 
By incorporating diverse perspectives and acknowledging the limitations of human-centric design, HCI researchers of today are embracing sustainable, equitable futures that account for multi-system knots' challenges by factoring in communities, cultures, and non-human organisms.

\subsection{Summary of the Findings}

Our findings present a heterogeneous landscape of futuring in HCI. 
While there is a growing rate of future-oriented papers, Figure~\ref{fig:heatmap} presented a dominance of fleeting futures. 
This dominace likely stems from the aforementioned prevalence of a technology-use focus as the primary driver for envisioning HCI futures, a trait visible even within the comprehensive-futuring category. However, there are encouraging signs from the latter sample, which exhibits greater depth and variation. It expresses robust engagement along the dimensions of all four SPIN categories. 
Circumscribing those categories has helped us illuminate precisely how HCI papers conceptualize, articulate, and impart futures -- and how they fall short. It aids especially in crystallizing what is distinct in the fleeting vs. comprehensive approaches.

In terms of epistemic stance (S), comprehensive futures demonstrate a more diverse engagement, incorporating explorative and normative stances that seek to imagine alternative trajectories or advocate for desirable futures aligned with societal or community values, contrasting with the narrower, predictive focus of fleeting futures. 

Concerning contingency perception (P), comprehensive futures embrace an open approach to uncertainty although more work is needed in leveraging it as a resource to explore non-linear and long-term temporalities. Clearly, forward-looking HCI research could delve far more deeply into the complex relations and unpredictabilities of future scenarios.

Comprehensive futures recognize the importance of systemic integration (I). Their foci acknowledge community experiences, cultural nuances, and more-than-human positionings. However, scholars' treatment of interrelation in this regard could advance further, in that only a single contingency element receives focus at a time. There are significant opportunities for exploring several contingencies in combination.

Finally, narratives (N) play a significant role in shaping HCI futures. However, we identified a weak spot in the way in which narratives often prioritize Western, anthropocentric perspectives, perpetuating systemic inequities and marginalizing more-than-human and Global South viewpoints. Addressing this gap requires an intentional shift in HCI research to challenge normative assumptions, incorporate underrepresented perspectives, and foster decolonial and inclusive approaches to futuring.

\section{Discussion} \label{sec:discussion}

Both our initial survey of related research (see Section~\ref{sec:related-research}) and our findings offer evidence of how HCI has evolved in its futuring's orientation over the years. Overall, the field has progressed from narrow visions focused on technological development toward more holistic and exploratory efforts to account for voices less often heard 
(e.g., of minorities and non-human entities), 
yet the landscape remains uneven. Some of the remaining gaps we found were highlighted already by \citeauthor{bell_yesterdays_2007} in \citeyear{bell_yesterdays_2007}. 
They attributed the field's techno-centric visions to the enduring influence of \citeauthor{weiser1991computer}'s vision of ubiquitous computing \cite{weiser1991computer}, which not only shoehorned HCI work toward developer time horizons narrowed to daily living in near-term futures but also expressed ideals specific to a North American upper-middle-class future \cite{bell_yesterdays_2007}. While attention from authors such as \citeauthor{bell_yesterdays_2007} may have helped spur bridge-building in HCI's future-orientation, some relatively recent work discussed above \cite[e.g., ][]{avle_designing_2016, adamu_no_2023, dillahunt_eliciting_2023} has continued to spotlight the associated gap. Indeed, it persists in even the most current HCI research.

In this section, we zoom in on two overarching gaps -- expanding the futuring's foundations beyond technology, and scaling HCI's insights through multilateral engagements -- from which we can gain a deeper understanding of the reasons why the gaps highlighted by \citeauthor{bell_yesterdays_2007} still persist today despite HCI's expanded future-orientation. We also propose five actionable opportunities to respond to the challenges. They adopt a normative yet adaptable approach that broadens HCI’s future-orientation. Each opportunity discussed below aims to enhance HCI’s contribution to futuring by promoting more inclusive, interdisciplinary practices.

\subsection{Expanding the Basis of Futuring Beyond Technology}
The undercurrent of techno-centrism, visible in HCI's fleeting and comprehensive futuring alike, has been pointed out again and again over the years. \citeauthor{bell_yesterdays_2007} were not alone. Several publications featured in our literature review point out this latent tendency, all the way from 2008 to 2023 (e.g., \citeauthor{reeves_envisioning_2012}'s \cite{reeves_envisioning_2012} from \citeyear{reeves_envisioning_2012}, \citeauthor{pargmanSustainabilityImaginedFuture2017}'s \cite{pargmanSustainabilityImaginedFuture2017} from \citeyear{pargmanSustainabilityImaginedFuture2017}, and publications by \citeauthor{soden_time_2021} \cite{soden_time_2021} and \citeauthor{forlano_speculative_2023} \cite{forlano_speculative_2023} in the 2021 and 2023 respectively), 
yet it still prevails. In response, some researchers have proposed drawing from disciplines experienced with systemic challenges, such as futures studies \cite{nathan_envisioning_2008, mankoffLookingYesterdayTomorrow2013a, salovaaraEvaluationPrototypesProblem2017, pargmanSustainabilityImaginedFuture2017, lightCollaborativeSpeculationAnticipation2021a, epp2022reinventing, moesgenDesigningUncertainFutures2023}; however, their contributions do not eliminate the risk of endorsing technology-oriented solutions that \emph{appear} transformative yet only provide imagined convenience. If we fail to address the underlying pressing needs by looking past technology \cite{nathan_envisioning_2008, tomlinson_collapse_2013, brynjarsdottir_sustainably_2012, harding_hci_2015}, ongoing reliance on technology-driven futures could even create new forms of dependency, beyond present disparities. We might exacerbate the very problems we set out to solve \cite{bell_yesterdays_2007, kinsleyFuturesMakingPractices2012}. 

\begin{description}
\item[Opportunity 1:] Turn to non-technological domains for drivers of HCI futuring. 
\end{description}

Two papers in our comprehensive-futuring sample 
identify possible deep roots to the challenge of technology-driven futuring. 
Firstly, \citeauthor{reeves_envisioning_2012} argued that a certain techno-determinism distinctive of HCI envisioning promotes a sense that, since technology advances linearly, we can predict societal implications with ease. 
The crutch of regarding technological progress as linear and independent of other catalysts of social change is attractive in that it frees HCI researchers from addressing the complexity of wider implications \cite{reeves_envisioning_2012}.

The other paper, by \citeauthor{do_thats_2023}, points out that computer scientists' research process does not naturally consider unintended consequences, let alone the web of factors created by them \cite{do_thats_2023}. 
\citeauthor{do_thats_2023} cited two main reasons for this blind spot: 
Computer scientists are seldom trained in means and guidance for envisioning such consequences. 
While our survey of prior research (see Section~\ref{sec:related-research}) found that HCI is not lacking in tools for envisioning possible futures, popular ones such as envisioning cards \cite{friedman_envisioning_2012} and design fiction \cite{bleecker_design_2009, sterling_design_2009} tend to get regarded as the province of designers and practitioners. This preconceptions has held back their adoption in research settings \cite{do_thats_2023, forlano_speculative_2023}.

Moreover, feedback loops between research priorities and institutional funding, publication timelines, etc. sustain techno-centric norms in HCI -- 
pressure for swift publishing leaves researchers little or no time to think about unintended consequences \cite{do_thats_2023}; 
meanwhile, incentives for financeable innovation and feasible implementation push the field toward incremental advancements, thereby suppressing its capacity to explore many, far futures \cite{reeves_envisioning_2012, do_thats_2023}.

The dominance of normative narratives centered on techno-driven futures highlights a critical tension in the field. 
One one side there are active groups of scholars forwarding systemic \cite{nathan_envisioning_2008, reeves_envisioning_2012, mankoffLookingYesterdayTomorrow2013a, salovaaraEvaluationPrototypesProblem2017, kozubaev_expanding_2020}, speculative \cite{blythe_research_2014, lindley_pushing_2016, wong2018speculative, elsden_speculative_2017, fox_vivewell_2019, chopra_negotiating_2022, bray_speculative_2021, forlano_speculative_2023}, pluralistic \cite{dell_ins_2016, avle_designing_2016, harrington_eliciting_2021, adamu_no_2023, kozubaev_expanding_2020, bardzell_utopias_2018, saha_towards_2022} and more-than-human \cite{liu_design_2018, key_feminist_2022, sondergaard_feminist_2023, sondergaard_fabulation_2023} approaches just to list some examples.
These groups are pulling against inertia from ``publish or perish'' pressure, discomfort with unfamiliar methods \cite{do_thats_2023}, and other impediments to applied envisioning in computer-science research. 

While the comprehensive-futuring work sheds considerable light on HCI's path, fleeting futuring too plays a critical role in shaping our field's future-orientation discourse. The narrowly scoped futures often operate as low-stakes tools for experimentation, whereby researchers explore emerging ideas or perform quick testing along speculative lines. Additionally, busy scholars can reap practical benefits via a focus on rapid prototyping and immediate utility, especially in industry-driven projects. 
For broader consideration of social, cultural, and ethics dimensions within this ``fleeting'' work, we refer to the suggestions of \citeauthor{lindley_implications_2017} and \citeauthor{kozubaev_expanding_2020}. 
Researchers who give futuring a fleeting glimpse can take advantage of opportunities to evaluate technological innovation, prototypes, etc. amid the everyday and mundane. Interventions that entail evaluating a particular future in day-to-day settings support profound learning about how people would interact with particular technologies in real life \cite{kozubaev_expanding_2020}. Researchers applying such a lens also have a tool for examining these technologies' long-term adoption in connection with the standard evaluation and for reflecting on the implications of that adoption \cite{lindley_implications_2017}. 

For comprehensive futuring, in turn, we argue that future-orien\-tation could be further enhanced by incorporating generative uses of uncertainty. Generative application readily permits explorations that attend to future challenges yet still form new possibilities in the here and now \cite{akamaDesignEthnographyFutures2015, epp_uncertainties_2024, sodenModesUncertaintyHCI2022}. 
For example, uncertainty can invite alternative futures that are not bound by current technological trends or assumptions \cite{tomlinson_collapse_2013}, but also support HCI researchers reflect on what they know or do not know, and reveal factors that have been ignored \cite{sardarThreeTomorrowsPostnormal2016}. The increased attention to design fiction and speculation has paved the path for HCI to consider the future as dynamic and open-ended. Adopting this ontology of the future \cite{poli_anticipation_2019} presents HCI with the opportunity to address complexity beyond technology while also the responsibility to consider and care for non-predominant stakeholders and ways of knowing so that futures being envisioned are meaningful to the stakeholders they are representing \cite{adam_futures_2011, kozubaev_expanding_2020}.

\begin{description}
\item[Opportunity 2:] Engage with generative modes of uncertainty to tap into unknown possibilities.
\end{description}

\subsection{Scaling Up HCI's Strengths Beyond Its Field}
In our pursuit of a literature review that captures how the HCI field applies its future-orientation at \emph{multiple} levels, 
we have considered the wider expanse via what futures studies scholarship could bring to the table. While we have identified papers that call for integration of futures studies concepts and methods into HCI, 
an opportunity exists also for boundary-crossing work in the other direction: reaching out to contribute to the outputs from futures studies.
We see three traits in particular through which HCI could add value to that neighboring field.

Firstly, HCI research can contribute to wider future-oriented discourses by bringing the angle of individuals and small groups to such arenas as futures studies and policymaking. Development of the SPIN framework (see Subsection~\ref{sec:qual-analysis}) delineated this unique feature, which is demonstrated well by studies of individuals' experiences with technology \cite{pierce_beyond_2012, brynjarsdottir_sustainably_2012,su_dolls_2019}. Futures studies has only recently cultivated an interest in the participatory and experiential so could learn much from this standpoint \cite{vorosIntegralFuturesApproach2008, candy_futures_2010, garcia2021designing}. Scholars of HCI could respond to the impetus for expansion of futures studies at such borders.

Secondly, HCI exercises the rare capability of studying futures ``in action'' by staging prototype-based user studies \cite{salovaaraEvaluationPrototypesProblem2017,elsden_speculative_2017,simeone2022immersive}. This methodological asset holds promise likewise for studies anchored not in HCI research agendas but in goals of other fields.

HCI's future-oriented work can build on a designerly approach that experiments with possibilities and ponders the outcomes. This third opportunity, through which less typically intervention-focused disciplines gain flexibility, is most clearly evident in ``research through design'' \cite[e.g., ][]{zimmerman_analysis_2010} but seized most directly via \citeauthor{kozubaev_expanding_2020}'s positing of ``five modes of reflection'' for HCI, where ``designerly formgiving,'' ``real-world engagement,'' and ``knowledge production through design'' articulate the fundaments poignantly \cite{kozubaev_expanding_2020}. 
Fruitful understandings produced thereby could generate excitement far beyond HCI. 

Our field, as an inherently interdisciplinary and participatory one, occupies a unique position, from which it can facilitate multilateral discussions. Merging perspectives from design, psychology, engineering, ecology, and the social sciences, it is adept at coming to integrative understanding of a myriad viewpoints. 
A more reflective \cite{kozubaev_expanding_2020}, responsible \cite{adam_futures_2011} stance to futuring would enable the HCI field to recast its role in dialogues about where societies, Earth, and other environments are headed. Ultimately, the stances we take decisively influence how futures get envisioned, developed, and implemented, across contextual boundaries.

\begin{description}
\item[Opportunity 3:] Embrace methods and tools for experiencing futures, and demonstrate the strengths they offer other fields.
\end{description}

\noindent Embracing more-than-human approaches, of which we see glimmers in the HCI field, helps overcome the anthropocentric bias that characterizes futuring research generally, whether in academic or policy domains \cite{adam_futures_2011, sardar_postnormal_2015}. Though knowing what technologies do in human life is undeniably important, attending exclusively to this can obscure the interdependencies linking humans and other elements of our planetary ecosystem \cite{yoo2023more, forlanoPosthumanismDesign2017, wakkary2021things}. 
To understand the vital part these dependency relations play in our world’s complex systems, especially as technological interventions casts an ever larger shadow, HCI futuring need look no further than our own field to witness the power of collaboration involving more-than-human organisms and of the relations' power in general \cite{liu_design_2018, homewood_tracing_2021, sondergaard_feminist_2023, sondergaard_fabulation_2023}.

\begin{description}
\item[Opportunity 4:] Cultivate reflexivity, and incorporate perspectives of \textit{otherness}.
\end{description}

\noindent Finally, to amplify the impact of its futures work, HCI can reorient itself by subverting dominant narratives \cite{light_ecologies_2022}. To reach this goal, the field can connect reflexively with its political position, thus demonstrating a willingness to usher in collective change rather than solely support adoption by individuals \cite{dourish_hci_2010, ashby_fourth-wave_2019}. By transforming into a proactive force, HCI can remain open to other perspectives, grapple with the mutability of the future, and maintain capacity to act on it in the present \cite{kozubaev_expanding_2020, millerSensingMakingsenseFutures2018}. To disrupt entrenched power dynamics, HCI researchers can cultivate sufficient humility to listen to the needs of \textit{others}, commit to long-term collaboration able to unveil various desirable futures, and steer actions toward pursuing those futures that emerge as valuable \cite{carroll_wild_2013,chopra_negotiating_2022, clarke_situated_2016, dell_ins_2016, avle_designing_2016, adamu_no_2023}.

\begin{description}
\item[Opportunity 5:] Engage in activism and political action.
\end{description}

\section{Conclusion}
As we had hoped, our literature-based review shed light on the state of HCI research's future-orientation. We proved able both to examine the extent of futuring in HCI and to identify telling characteristics of that HCI research directed toward more comprehensive futuring activity. Futures studies informed our research design -- from sensitizing the retrieval stage's search terms, through aiding with inclusion/exclusion, to supporting the qualitative analysis (the final stage) by helping populate our four-category SPIN framework for examining how futuring is approached in HCI -- yet the work sharply illuminated our own discipline specifically. Together, the most future-oriented papers in the field, and the sub-sample of $N=205$ highly cited comprehensive-futures publications retrieved for our literature review gave us the final SPIN framework and valuable findings specific to our field's most influential comprehensively future-oriented papers.

Our findings highlight that futuring is a growing trend in HCI, with comprehensive futuring having shown considerable growth since 2017. 
However, fleeting futuring has remained the dominant category, and HCI futuring can be characterized as mostly techno-centric. While our review identified a very real risk of reproducing tunnel vision and of scenarios that echo notions of linear societal progress, 
analysis of the ``comprehensive futuring'' category uncovered encouraging signs: a significant proportion of the work features active exploration of uncertainty, focus on human experience, and contestation of dominant narratives -- all of which help mitigate such issues. 
Both the gaps and such strides toward overcoming them fueled our reflections as to why techno-centrism and short-term visions of the future still prevail in HCI, and on five opportunities to rectify these problems.


\printbibliography[keyword={references}]

\end{refsection}
\newpage
\appendix

\begin{refsection}

\section{Inclusion/Exclusion Criteria}\label{sec:inclusion-exclusion}
\subsection{Exclusion Criteria: 
}
\textbf{1. Missing future-oriented terminology}
Papers that do not include any future-related search terms in the relevant sections (all sections for Interactions magazine, and the introduction or discussion/conclusion for non-Interactions magazine's articles) are excluded.

\textbf{2. Future terms not used in a future-oriented context}
If future-related terms appear but refer solely to future work/research without exploring future scenarios, the paper is excluded.

Example:“In the future work, more details will be perfected, and more levels will be developed for further enhancing children’s skills, which will also be corroborated in more rigorous experiments.” [p. 14]\cite{10.1145/3586183.3606755}

\textbf{3. Call for future research without elaboration.}
Papers invoking future terms in a call for new research practices or approaches but failing to elaborate on them are excluded.

Examples:
“Where the sanguine rhetoric of democratizing technology and open data envision transformation through the availability of new tools, it does so with little regard for the human and community costs involved in that transformation. However, a mode of intervention that is based in community practice shifts the power to the community.” \cite[p. 792]{10.1145/2598510.2598563}

“How best to train ML researchers and practitioners to engage in creative speculation or to otherwise anticipate potential consequences of their work is an area where more research is needed.” \cite[p. 23]{10.1145/3555760}

\textbf{4. Existing technology use scenarios without futuring}
Articles proposing use-scenarios for technologies already available at the time of writing and not explicitly referring to the future are excluded.

Example:
“There are also several applications for low-cost audience polling outside of a classroom context. We envision that the technology developed in this paper would apply equally well in these scenarios.” \cite[p. 54]{10.1145/2380116.2380124}

\subsection{Inclusion Criteria:}
\textbf{1. Using future-oriented methods}
Papers proposing or utilizing methods for anticipating or envisioning future scenarios are included, including research through design, design fictions, speculative design, etc.

\textbf{2. Future visions and new artifacts}
Papers presenting new artifacts that convey future visions are included, with future visions encompassing: a) future implications for specific communities or society, b) technological implications, c) considerations on how the envisioned future could be achieved, d) discussions within the STEEPLE framework (Social, Technological, Economic, Environmental, Political, Legal, and Ethical considerations).

\textbf{3. Conceptual future discussions}
Papers discussing or critiquing future scenarios, imaginaries, or visions are included, especially if they problematize common future predictions.

\textbf{3. References to future thinking outside of the scope} 
Future search terms appear outside the analysis scope (introduction, discussion, and conclusion sections) but more detailed futuring is discussed the referenced section.

Example:
“The design work we have presented here follows the traditions of participatory design with some affinity toward speculative design.” \cite[p. 2304]{asad_creating_2017}

\subsection{Fleeting Futuring}
Papers that present superficial descriptions of future scenarios or common imaginaries with little elaboration are considered "fleeting futuring." Such mentions typically only include one sentence or a brief reference to a future vision.

\printbibliography[keyword={references}, title={References for Inclusion/Exclusion Criteria}]
\end{refsection}

\begin{refsection}
\section{Comprehensive Futuring Articles}
The following list enumerates the articles that manifest comprehensive futuring. 

\nocite{adamu2023, adibSmartHomesThat2015, alanTariffAgentInteracting2016, alkhatibExaminingCrowdWork2017, alvarezdelavegaUnderstandingPlatformMediated2023, alves-oliveiraCollectionMetaphorsHumanRobot2021, asadCreatingSociotechnicalAPI2017, asadPrefigurativeDesignMethod2019, avleDesigningHereThere2016, ballardJudgmentCallGame2019a, bardzellCriticalDesignCritical2012a, bardzellReadingCriticalDesigns2014, bardzellUtopiasParticipationFeminism2018, bardzellWhatCriticalCritical2013, baumerEvaluatingDesignFiction2020, benfordPerformanceLedResearchWild2013, benjaminMachineLearningUncertainty2021a, bennettAccessibilityCrowdedSidewalk2021, bennettPromiseEmpathyDesign2019, blackwellHarassmentSocialVirtual2019, blytheResearchDesignFiction2014a, blytheResearchFictionStorytelling2017a, blytheSolutionistStrategiesSeriously2016, bodkerParticipatoryDesignThat2018, boehnerDataDesignCivics2016, bonnailMemoryManipulationsExtended2023, boydAutomatedEmotionRecognition2023, brandDesignInquiryIntrospective2021, braySpeculativeBlacknessConsidering2021a, breuerHowEngineersImaginaries2023, brushHomeAutomationWild2011, brynjarsdottirSustainablyUnpersuadedHow2012, cambreOneVoiceFits2019, carrollWildHomeNeighborhood2013, chenLearningHomeMixedMethods2021, cheokEmpatheticLivingMedia2008, chignellEvolutionHCIHuman2023a, chopraNegotiatingSustainableFutures2022, choTopophiliaPlacemakingBoundary2022, chungIntersectionUsersRoles2021, churchillPsQsDesigningDigital2008, cilaProductsAgentsMetaphors2017, clarkeSituatedEncountersSocially2016, costanzaDesignableVisualMarkers2009, costanzaDoingLaundryAgents2014, crivellaroContestingCityEnacting2015, dellInsOutsHCI2016, desjardinsBespokeBookletsMethod2019, desjardinsLivingPrototypeReconfigured2016, devendorfDontWantWear2016, dhanorkarWhoNeedsKnow2021, dillahuntElicitingAlternativeEconomic2023, disalvoNourishingGroundSustainable2009, dorkInformationFlaneurFresh2011, doThatsImportantHow2023, drugaFamilyThirdSpace2022, eghtebasCoSpeculatingDarkScenarios2023, ehsanExpandingExplainabilitySocial2021c, ellisonFEATURESocialNetworkSites2009, elsdenSpeculativeEnactments2017a, eslamiFirstItThen2016, faasLongitudinalVideoStudy2020, forlanoSpeculativeHistoriesJust2023a, foxVivewellSpeculatingFuture2019, frauenbergerEntanglementHCINext2020a, freemanRediscoveringPhysicalBody2022a, friedmanEnvisioningCardsToolkit2012, gargSocialContextsAgency2022, gaverAnnotatedPortfolios2012, grammenosFEATURETheAmbientMirror2009, greenfieldFEATUREAtEndWorld2009, grinterInsOutsHome2009, grudinChatbotsHumbotsQuest2019, gugenheimerShareVREnablingCoLocated2017, haimsonDesigningTransTechnology2020a, hardingHCICivicEngagement2015, harringtonDeconstructingCommunityBasedCollaborative2019a, harringtonElicitingTechFutures2021b, herdelScopingReviewDomains2022, holzImplantedUserInterfaces2012, homewoodTracingConceptionsBody2021, iivariCriticalDesignResearch2017, impioGiveManFish2010, iraniStoriesWeTell2016, iraniTurkopticonInterruptingWorker2013, ishiiRadicalAtomsTangible2012a, ismailImaginingCaringFutures2022, jakobiItWhatThey2019, jhaverDesigningWordFilter2022, kaneSharedDecisionMaking2013, kawakamiSensingWellbeingWorkplace2023, kawakamiWhyCareWhats2022, kayeMoneyTalksTracking2014, keyesReimaginingWomensHealth2020, keyFeministCareAnthropocene2022, khanSpeculativeDesignEducation2021, kingsleyGiveEverybodyLittle2022, kirshEmbodiedCognitionMagical2013, klopfensteinRiseBotsSurvey2017, kocielnikReflectionCompanionConversational2018,  kozubaevExpandingModesReflection2020, lazarovasquezIntroducingSustainablePrototyping2020, leeWeBuildAIParticipatoryFramework2019, liArtifactPowerLens2023, lightDesignExistentialCrisis2017, lightEcologiesSubversionTroubling2022b, lindleyImplicationsAdoption2017, lindleyPlacingAgeTransitioning2015, lindleyPushingLimitsDesign2016a, lindtnerEmergingSitesHCI2014a, lindtnerReconstitutingUtopianVision2016a, liuDesignCollaborativeSurvival2018a, loglerMetaphorCardsHowGuide2018, luCodingBiasUse2021, luFeelItMy2019, luoEseedShapeChangingInterfaces2020, luriaReEmbodimentCoEmbodimentExploration2019, manciniContravisionExploringUsers2010a, maoHowDataScientistsWork2019, matviienkoBabyYouCan2022, maUncoveringGigWorkerCentered2023, meyersImpoverishedVisionsSustainability2016, michaelisCollaborativeSimplyUncaged2020, morrisseyValueExperienceCentredDesign2017, muellerJoggingQuadcopter2015, muellerUnderstandingDesignIntertwined2023, murray-rustBlockchainUnderstandingBlockchains2022, murrayFEATUREResearchStrategiesFuture2009, nathanEnvisioningSystemicEffects2008a, nathanSUSTAINABLYOURSInformation2009, neustaedterSharingDomesticLife2015, odomFieldworkFutureUser2012, odomPhotoboxDesignSlow2012, odomPlacelessnessSpacelessnessFormlessness2014, odomTechnologyHeirloomsConsiderations2012, ohaganPrivacyEnhancingTechnologyEveryday2022, ohUsVsThem2017, oogjesDesigningOtherHome2018, parkGenerativeAgentsInteractive2023, parkSocialSimulacraCreating2022, paterNoHumansHere2022, pendseTreatmentHealingEnvisioning2022, pierceDesignProposalWireless2016, pierceEnergyMonitorsInteraction2012, pierceSmartHomeSecurity2019a, plodererProcessEngagementEngaging2012, poupyrevProjectJacquardInteractive2016, prostAwarenessEmpowermentUsing2015, raeFrameworkUnderstandingDesigning2015, rakovaWhereResponsibleAI2021, raziSlidingMyDMs2023, reevesEnvisioningUbiquitousComputing2012, ribesRepresentingCommunityKnowing2008, ringfort-felnerKiroDesignFiction2022, rogersNeverTooOld2014, sabieDecadeInternationalMigration2022, saffoRemoteCollaborativeVirtual2021, sahaSustainableICTDBangladesh2022, salehiWeAreDynamo2015, sanchez-cortesMoodVlogMultimodal2015, sandersDesigningCodesigningCollective2014, saraijiMetaArmsBodyRemapping2018, sauppeSocialImpactRobot2015, schulenbergLeveragingAIbasedModeration2023, seeringDyadicInteractionsConsidering2019, seeringReconsideringSelfModerationRole2020, simeoneSubstitutionalRealityUsing2015, smithDesigningCohabitationNaturecultures2017, sodenTimeHistoricismCSCW2021, sondergaardFabulationApproachDesign2023, sondergaardFeministPosthumanistDesign2023, sondergaardIntimateFuturesStaying2018, speicherWhatMixedReality2019, steinhardtAnticipationWorkCultivating2015a, sterlingCOVERSTORYDesignFiction2009, strengersSmartEnergyEveryday2014, suDollsMenAnticipating2019, sureshExpertiseRolesFramework2021, sykownikSomethingPersonalMetaverse2022, tanenbaumDesignFictionalInteractions2014, tanenbaumSteampunkDesignFiction2012, thiemeDesigningHumancenteredAI2023, tomlinsonCollapseInformaticsAugmenting2012a, tomlinsonCollapseInformaticsPractice2013a, tomlinsonWhatIfSustainability2012, tsaknakiFabulatingBiodataFutures2022, tsengDarkSidePerceptual2022, tuliRethinkingMenstrualTrackers2022, vanderbeekenFEATURETakingBroaderView2009, varanasiItCurrentlyHodgepodge2023, vermaRethinkingRoleAI2023, vinesAgeOldProblemExamining2015a, vinesJoyChequesTrust2012, volkelElicitingAnalysingUsers2021, wakkaryMorseThingsDesign2017, wakkarySustainableDesignFiction2013, williamsPerpetualWorkLife2019, winchesterREALizingOurMessy2010, wongElicitingValuesReflections2017a, wongRealFictionalEntanglementsUsing2017b, wongSeeingToolkitHow2023, yildirimHowExperiencedDesigners2022, yooValueSensitiveActionreflection2013, zamfirescu-pereiraHerdingAICats2023, zimmermanAnalysisCritiqueResearch2010a}
\printbibliography[keyword={includes}, heading=none]
\end{refsection}

\begin{refsection}
\section{Fleeting Futuring Articles}
The following list enumerates the articles that manifest fleeting futuring. 

\nocite{abokhodairDissectingSocialBotnet2015,ackermansEffectsExplicitIntention2020, ahujaEduSensePracticalClassroom2019, akashClassificationModelSensing2018, alexanderGrandChallengesShapeChanging2018, altInteractionTechniquesCreating2013a, altUsingEyetrackingSupport2014, amoresEssenceOlfactoryInterfaces2017a, andersonInteractionsLookingBroadly2009, anjaniWhyPeopleWatch2020, anThermorphDemocratizing4D2018, baharinSonicAIRSupportingIndependent2015a, bakerSchoolsBackScaffolding2021, bakerUnderstandingTrustRepair2018, bedriEarBitUsingWearable2017, bedriFitByteAutomaticDiet2020, bellUMeExploringHuman2023, benkoSphereMultitouchInteractions2008, bermanHowDIYMetaDesignTools2021, bernsteinCrowdsTwoSeconds2011, besmerMovingUntaggingPhoto2010a, bhatWeAreHalfdoctors2023, boringTouchProjectorMobile2010, breidebandHomeLifeWorkRhythm2022a, brownTroubleAutopilotsAssisted2017, brubakerFocusingSharedExperiences2012, brudySurfaceFleetExploringDistributed2020, brushDigitalNeighborhoodWatch2013, burgessHealthcareAITreatment2023, buschelMIRIAMixedReality2021, caoLargeScaleAnalysis2021a, chancellorWhoHumanHumanCentered2019, chandrasekharanBagCommunitiesIdentifying2017, chenCaringCaregiversDesigning2013, chenDuetExploringJoint2014, chenFinexusTrackingPrecise2016, chiouDesigningAIExploration2023, churchillTeachingLearningHumancomputer2013, conversyVizirDomainSpecificGraphical2018, correllEthicalDimensionsVisualization2019a, dangTextGenerationSupporting2022, danzicoDesignSerendipityNot2010, dasswainAlgorithmicPowerPunishment2023, dekaZIPTZeroIntegrationPerformance2017, dementyevDualBlinkWearableDevice2017, dementyevRovablesMiniatureBody2016, dementyevSensorTapeModularProgrammable2015, dementyevWristFlexLowpowerGesture2014, devitoHowTransfeminineTikTok2022, disalvoFruitAreHeavy2017a, disalvoNavigatingTerrainSustainable2010, dodgeExplainingModelsEmpirical2019, doveUXDesignInnovation2017, duDepthLabRealtime3D2020, ehsanAutomatedRationaleGeneration2019a, ekhtiarGoalsGoalSetting2023, elkinAlignedRankTransform2021, elsdenMakingSenseBlockchain2018, engelbutzederSurplusScarcityAbundance2023, epsteinSuppressingSearchEngine2017, ernalaLinguisticMarkersIndicating2017, estevesOrbitsGazeInteraction2015, fastMetaEnablingProgramming2016, feitEverydayGazeInput2017, fenderCausalitypreservingAsynchronousReality2022a, fengHowUXPractitioners2023, feuchtnerExtendingBodyInteraction2017, fischerRecommendingEnergyTariffs2013, follmerInFORMDynamicPhysical2013, forlizziWhereShouldTurn2010, foxPatchworkHiddenHuman2023, freedStalkersParadiseHow2018, freemanWorkingTogetherApart2022, freyBreezeSharingBiofeedback2018, fridmanCognitiveLoadEstimation2018, furloRethinkingDatingApps2021, geroSparksInspirationScience2022, gheranGesturesSmartRings2018, goedickeVROOMVirtualReality2018, gomesMorePhoneStudyActuated2013, grillAttitudesFolkTheories2022, groegerObjectSkinAugmentingEveryday2018, gugenheimerFaceTouchEnablingTouch2016, guoFacadeAutogeneratingTactile2017, haeslerConnectedSelfOrganizedCitizens2021, hanHydroRingSupportingMixed2018, hartikainenSafeSextingAdvice2021, hassanFootStrikerEMSbasedFoot2017, hassibEmotionActuatorEmbodied2017, headAugmentingScientificPapers2021, held3DPuppetryKinectbased2012, hewittAssessingPublicPerception2019, hirschNothingCompilationHow2024, hollanderTaxonomyVulnerableRoad2021, hongSmartphonebasedSensingPlatform2014a, honnetPolySenseAugmentingTextiles2020, huangOrecchioExtendingBodyLanguage2018, huangWovenProbeProbingPossibilities2021, huhHealthVlogsSocial2014, hurstAutomaticallyDetectingPointing2008, hwangSocialbotsVoicesFronts2012, iqbalOasisFrameworkLinking2010, ishiiOpticalMarionetteGraphical2016, jainHeadMountedDisplayVisualizations2015, jiangHandAvatarEmbodyingNonHumanoid2023, jokelaDiaryStudyCombining2015, kaimotoSketchedRealitySketching2022, kaneAccessOverlaysImproving2011, kapurAlterEgoPersonalizedWearable2018a, karranFrameworkPsychophysiologicalClassification2015, khotUnderstandingPhysicalActivity2014, kimCellsGeneratorsLenses2023, kimDatadrivenInteractionTechniques2014, kimExploringChartQuestion2023a, kimInflatableMouseVolumeadjustable2008a, kimOmniTrackFlexibleSelfTracking2017, kirkHomeVideoCommunication2010, kirkHumanRemainsValues2010, kirkOpeningFamilyArchive2010, kleinbergerSupportingElderConnectedness2019, klokmoseWebstratesShareableDynamic2015, kongFramesSlantsTitles2018, kraemerBatteryfreeMakeCodeAccessible2022, krumAllRoadsLead2008a, kundingerDriverDrowsinessAutomated2020a, lafreniereCrowdsourcedFabrication2016, laputSensingFineGrainedHand2019, lawsonProblematisingUpstreamTechnology2015a, leeSleepGuruPersonalizedSleep2022a, leeZeroNMidairTangible2011, leithingerPhysicalTelepresenceShape2014, leithingerSublimateStatechangingVirtual2013, liaoAllWorkNo2018, lightDigitalInterdependenceHow2011, liMeasuringUnderstandingPhoto2019, linAdasaConversationalVehicle2018, linSYNCCrowdsourcingPlatform2021a, liTangibleGridTangibleWeb2022, liu3DALLEIntegratingTextImage2023, liuLivingHeritageHistoric2012, luceroMobileCollocatedInteractions2013, lugerHavingReallyBad2016b, luParticipatoryNoticingPhotovoice2023, maloneyTalkingVoiceUnderstanding2020, marcusTwitinfoAggregatingVisualizing2011, markResilienceCollaborationTechnology2008, marlowActivityTracesSignals2013, matthiesEarFieldSensingNovelEar2017a, mccarthyContextContentCommunity2008, mcgillCreatingAugmentingKeyboards2022, mcnallyCodesigningMobileOnline2018, mentisInvisibleEmotionInformation2010a, messerschmidtANISMAPrototypingToolkit2022, miltonSeeMeHere2023, mizrahiDigitalGastronomyMethods2016, muellerJoggingDistanceEurope2010, muellerNextStepsHumanComputer2020, mullenbachExploringAffectiveCommunication2014, mullerLookingGlassField2012, mulliganThisThingCalled2019, nakagakiChainFORMLinearIntegrated2016, nakajimaReflectingHumanBehavior2008, niiyamaPoimoPortableInflatable2020a, nunesSharingDigitalPhotographs2008, obristOpportunitiesOdorExperiences2014, oduorFrustrationsBenefitsMobile2016, oelenCrowdsourcingScholarlyDiscourse2021, oharaBlendedInteractionSpaces2011, olesonTeachingInclusiveDesign2023, ouCilllia3DPrinted2016, parkManifestationDepressionLoneliness2015, parkMetaverseWorkspaceOpportunities2023, peiHandInterfacesUsing2022, perraultWatchitSimpleGestures2013, pfeifferCruiseControlPedestrians2015, pierceInterfaceUserExploratory2018, porfirioAuthoringVerifyingHumanRobot2018, prakashAutoDescFacilitatingConvenient2023, qianInferringMotionDirection2017a, qianScalARAuthoringSemantically2022, rajanTaskLoadEstimation2016, rappExploringLivedExperience2023, reichertsItsGoodTalk2022, retelnyExpertCrowdsourcingFlash2014, richardsmaldonadoReaderQuizzerAugmentingResearch2023a, roffarelloAchievingDigitalWellbeing2023, ruanComparingSpeechKeyboard2018, sahaLanguageLGBTQMinority2019, sanchesMindBodyDesigning2010a, satriadiMapsMe3D2020, scheuermanHowComputersSee2019, scheuermanHowWeveTaught2020, scheuermanSafeSpacesSafe2018a, schlesingerLetsTalkRace2018, schmidtCrossdeviceInteractionStyle2012a, schoenebeckGivingTwitterLent2014, schrillsHowUsersExperience2023, schroederPocketSkillsConversational2018, schweikardtSUSTAINABLYOURSUserCentered2009, seboRobotsGroupsTeams2020, shiMarkitTalkitLowBarrier2017, silvaVehicleDriverRecognition2012a, sodenInformatingCrisisExpanding2018, sohnDiaryStudyMobile2008, steichenUseradaptiveInformationVisualization2013, steimleFlexpadHighlyFlexible2013, stewartCharacteristicsPressurebasedInput2010a, streliHOOVHandOutView2023, strohmayerTechnologiesSocialJustice2019, suDesigningNomadicWork2008a, suhAISocialGlue2021, suhSensecapeEnablingMultilevel2023, sunInvestigatingExplainabilityGenerative2022, suraleTabletInVRExploringDesign2019a, sutherlandGigEconomyInformation2017, suzukiAugmentedRealityRobotics2022, suzukiShapeBotsShapechangingSwarm2019, teibrichPatchingPhysicalObjects2015, tolmieThisHasBe2016, toTheyJustDont2020, trajkovaAlexaToyExploring2020, trujilloMakeRedditGreat2022, tsaknakiExpandingWabiSabiDesign2016a, tuddenhamGraspablesRevisitedMultitouch2010, uhlTangibleImmersiveTrauma2023a, vaishTwitchCrowdsourcingCrowd2014, valentineFlashOrganizationsCrowdsourcing2017, vazquez3DPrintingPneumatic2015a, vinesQuestionableConceptsCritique2012, vonsawitzkyHazardNotificationsCyclists2022, wangAutoDSHumanCenteredAutomation2021, wangCarNoteReducingMisunderstanding2017, wangCASSBuildingSocialSupport2021, wangHumanAICollaborationData2019, wangSeismoBloodPressure2018, weibelPaperSketchPaperdigitalCollaborative2011, westQuantifiedPatientDoctors2016, whitmireHapticRevolverTouch2018, whitneyCOVERSTORYTheCounterfeit2009, wibergMaterialityMattersexperienceMaterials2013, willisPrintedOptics3D2012, willisSideBySideAdhocMultiuser2011, wuAIChainsTransparent2022, wuUnfabricateDesigningSmart2020, xiaCrossTalkIntelligentSubstrates2023, xuEnablingHandGesture2022, yanEnhancingAudienceEngagement2016a, yangMakingSustainabilitySustainable2014, yangSimuLearnFastAccurate2020, yenVisibleHeartsVisible2018, yeoStickEarMakingEveryday2013a, zagalskyEmergenceGitHubCollaborative2015, zhangAlgorithmicManagementReimagined2022, zhangHowDataScience2020, zhangSkinTrackUsingBody2016, zhangSpeeChinSmartNecklace2021, zhangTomoWearableLowCost2015, zhengTellingStoriesComputational2022, zhouCyanochromicInterfaceAligning2023}
\printbibliography[keyword={fleeting}, heading=none]

@inproceedings{abokhodairDissectingSocialBotnet2015,
	title = {Dissecting a {{Social Botnet}}: {{Growth}}, {{Content}} and {{Influence}} in {{Twitter}}},
	shorttitle = {Dissecting a {{Social Botnet}}},
	booktitle = {Proceedings of the 18th {{ACM Conference}} on {{Computer Supported Cooperative Work}} \& {{Social Computing}}},
	author = {Abokhodair, Norah and Yoo, Daisy and McDonald, David W.},
	year = 2015,
	month = feb,
	pages = {839--851},
	publisher = {ACM},
	address = {Vancouver BC Canada},
	doi = {10.1145/2675133.2675208},
	urldate = {2024-09-13},
	isbn = {978-1-4503-2922-4},
	langid = {english},
}

@inproceedings{ackermansEffectsExplicitIntention2020,
	title = {The {{Effects}} of {{Explicit Intention Communication}}, {{Conspicuous Sensors}}, and {{Pedestrian Attitude}} in {{Interactions}} with {{Automated Vehicles}}},
	booktitle = {Proceedings of the 2020 {{CHI Conference}} on {{Human Factors}} in {{Computing Systems}}},
	author = {Ackermans, Sander and Dey, Debargha and Ruijten, Peter and Cuijpers, Raymond H. and Pfleging, Bastian},
	year = 2020,
	month = apr,
	pages = {1--14},
	publisher = {ACM},
	address = {Honolulu HI USA},
	doi = {10.1145/3313831.3376197},
	urldate = {2024-09-13},
	isbn = {978-1-4503-6708-0},
	langid = {english}
}

@article{ahujaEduSensePracticalClassroom2019,
	title = {{{EduSense}}: {{Practical Classroom Sensing}} at {{Scale}}},
	shorttitle = {{EduSense}},
	author = {Ahuja, Karan and Kim, Dohyun and Xhakaj, Franceska and Varga, Virag and Xie, Anne and Zhang, Stanley and Townsend, Jay Eric and Harrison, Chris and Ogan, Amy and Agarwal, Yuvraj},
	year = 2019,
	month = sep,
	journal = {Proceedings of the ACM on Interactive, Mobile, Wearable and Ubiquitous Technologies},
	volume = 3,
	number = 3,
	pages = {1--26},
	issn = {2474-9567},
	doi = {10.1145/3351229},
	urldate = {2024-09-13},
	langid = {english},
}

@article{akashClassificationModelSensing2018,
	title = {A {{Classification Model}} for {{Sensing Human Trust}} in {{Machines Using EEG}} and {{GSR}}},
	author = {Akash, Kumar and Hu, Wan-Lin and Jain, Neera and Reid, Tahira},
	year = 2018,
	month = dec,
	journal = {ACM Transactions on Interactive Intelligent Systems},
	volume = 8,
	number = 4,
	pages = {1--20},
	issn = {2160-6455, 2160-6463},
	doi = {10.1145/3132743},
	urldate = {2024-09-13},
	langid = {english},
}

@inproceedings{alexanderGrandChallengesShapeChanging2018,
	title = {Grand {{Challenges}} in {{Shape-Changing Interface Research}}},
	booktitle = {Proceedings of the 2018 {{CHI Conference}} on {{Human Factors}} in {{Computing Systems}}},
	author = {Alexander, Jason and Roudaut, Anne and Steimle, J{\"u}rgen and Hornb{\ae}k, Kasper and Bruns Alonso, Miguel and Follmer, Sean and Merritt, Timothy},
	year = 2018,
	month = apr,
	pages = {1--14},
	publisher = {ACM},
	address = {Montreal QC Canada},
	doi = {10.1145/3173574.3173873},
	urldate = {2024-09-13},
	isbn = {978-1-4503-5620-6},
	langid = {english},
}

@inproceedings{altInteractionTechniquesCreating2013a,
	title = {Interaction Techniques for Creating and Exchanging Content with Public Displays},
	booktitle = {Proceedings of the {{SIGCHI Conference}} on {{Human Factors}} in {{Computing Systems}}},
	author = {Alt, Florian and Shirazi, Alireza Sahami and Kubitza, Thomas and Schmidt, Albrecht},
	year = 2013,
	month = apr,
	pages = {1709--1718},
	publisher = {ACM},
	address = {Paris France},
	doi = {10.1145/2470654.2466226},
	urldate = {2024-09-13},
	isbn = {978-1-4503-1899-0},
	langid = {english},
}

@inproceedings{altUsingEyetrackingSupport2014,
	title = {Using Eye-Tracking to Support Interaction with Layered {{3D}} Interfaces on Stereoscopic Displays},
	booktitle = {Proceedings of the 19th International Conference on {{Intelligent User Interfaces}}},
	author = {Alt, Florian and Schneegass, Stefan and Auda, Jonas and Rzayev, Rufat and Broy, Nora},
	year = 2014,
	month = feb,
	pages = {267--272},
	publisher = {ACM},
	address = {Haifa Israel},
	doi = {10.1145/2557500.2557518},
	urldate = {2024-09-13},
	isbn = {978-1-4503-2184-6},
	langid = {english},
}

@inproceedings{amoresEssenceOlfactoryInterfaces2017a,
	title = {Essence: {{Olfactory Interfaces}} for {{Unconscious Influence}} of {{Mood}} and {{Cognitive Performance}}},
	shorttitle = {Essence},
	booktitle = {Proceedings of the 2017 {{CHI Conference}} on {{Human Factors}} in {{Computing Systems}}},
	author = {Amores, Judith and Maes, Pattie},
	year = 2017,
	month = may,
	pages = {28--34},
	publisher = {ACM},
	address = {Denver Colorado USA},
	doi = {10.1145/3025453.3026004},
	urldate = {2024-09-13},
	isbn = {978-1-4503-4655-9},
	langid = {english}
}

@article{andersonInteractionsLookingBroadly2009,
	title = {{\emph{Interactions}}: Looking Broadly to the Future},
	shorttitle = {{\emph{Interactions}}},
	author = {Anderson, Richard and Kolko, Jon},
	year = 2009,
	month = sep,
	journal = {Interactions},
	volume = 16,
	number = 5,
	pages = {5--5},
	issn = {1072-5520, 1558-3449},
	doi = {10.1145/1572626.1572627},
	urldate = {2024-09-13},
	langid = {english}
}

@inproceedings{anjaniWhyPeopleWatch2020,
	title = {Why Do People Watch Others Eat Food? {{An Empirical Study}} on the {{Motivations}} and {{Practices}} of {{Mukbang Viewers}}},
	shorttitle = {Why Do People Watch Others Eat Food?},
	booktitle = {Proceedings of the 2020 {{CHI Conference}} on {{Human Factors}} in {{Computing Systems}}},
	author = {Anjani, Laurensia and Mok, Terrance and Tang, Anthony and Oehlberg, Lora and Goh, Wooi Boon},
	year = 2020,
	month = apr,
	pages = {1--13},
	publisher = {ACM},
	address = {Honolulu HI USA},
	doi = {10.1145/3313831.3376567},
	urldate = {2024-09-13},
	isbn = {978-1-4503-6708-0},
	langid = {english}
}

@inproceedings{anThermorphDemocratizing4D2018,
	title = {Thermorph: {{Democratizing 4D Printing}} of {{Self-Folding Materials}} and {{Interfaces}}},
	shorttitle = {Thermorph},
	booktitle = {Proceedings of the 2018 {{CHI Conference}} on {{Human Factors}} in {{Computing Systems}}},
	author = {An, Byoungkwon and Tao, Ye and Gu, Jianzhe and Cheng, Tingyu and Chen, Xiang 'Anthony' and Zhang, Xiaoxiao and Zhao, Wei and Do, Youngwook and Takahashi, Shigeo and Wu, Hsiang-Yun and Zhang, Teng and Yao, Lining},
	year = 2018,
	month = apr,
	pages = {1--12},
	publisher = {ACM},
	address = {Montreal QC Canada},
	doi = {10.1145/3173574.3173834},
	urldate = {2024-09-13},
	isbn = {978-1-4503-5620-6},
	langid = {english}
}

@article{baharinSonicAIRSupportingIndependent2015a,
	title = {{{SonicAIR}}: {{Supporting Independent Living}} with {{Reciprocal Ambient Audio Awareness}}},
	shorttitle = {{SonicAIR}},
	author = {Baharin, Hanif and Viller, Stephen and Rintel, Sean},
	year = 2015,
	month = jul,
	journal = {ACM Transactions on Computer-Human Interaction},
	volume = 22,
	number = 4,
	pages = {1--23},
	issn = {1073-0516, 1557-7325},
	doi = {10.1145/2754165},
	urldate = {2024-09-13},
	langid = {english},
}

@article{bakerSchoolsBackScaffolding2021,
	title = {School's {{Back}}: {{Scaffolding Reminiscence}} in {{Social Virtual Reality}} with {{Older Adults}}},
	shorttitle = {School's {{Back}}},
	author = {Baker, Steven and Kelly, Ryan M. and Waycott, Jenny and Carrasco, Romina and Bell, Roger and Joukhadar, Zaher and Hoang, Thuong and Ozanne, Elizabeth and Vetere, Frank},
	year = 2021,
	month = jan,
	journal = {Proceedings of the ACM on Human-Computer Interaction},
	volume = 4,
	number = {CSCW3},
	pages = {1--25},
	issn = {2573-0142},
	doi = {10.1145/3434176},
	urldate = {2024-09-13},
	langid = {english},
}

@article{bakerUnderstandingTrustRepair2018,
	title = {Toward an {{Understanding}} of {{Trust Repair}} in {{Human-Robot Interaction}}: {{Current Research}} and {{Future Directions}}},
	shorttitle = {Toward an {{Understanding}} of {{Trust Repair}} in {{Human-Robot Interaction}}},
	author = {Baker, Anthony L. and Phillips, Elizabeth K. and Ullman, Daniel and Keebler, Joseph R.},
	year = 2018,
	month = dec,
	journal = {ACM Transactions on Interactive Intelligent Systems},
	volume = 8,
	number = 4,
	pages = {1--30},
	issn = {2160-6455, 2160-6463},
	doi = {10.1145/3181671},
	urldate = {2024-09-13},
	langid = {english}
}

@article{bedriEarBitUsingWearable2017,
	title = {{{EarBit}}: {{Using Wearable Sensors}} to {{Detect Eating Episodes}} in {{Unconstrained Environments}}},
	shorttitle = {{EarBit}},
	author = {Bedri, Abdelkareem and Li, Richard and Haynes, Malcolm and Kosaraju, Raj Prateek and Grover, Ishaan and Prioleau, Temiloluwa and Beh, Min Yan and Goel, Mayank and Starner, Thad and Abowd, Gregory},
	year = 2017,
	month = sep,
	journal = {Proceedings of the ACM on Interactive, Mobile, Wearable and Ubiquitous Technologies},
	volume = 1,
	number = 3,
	pages = {1--20},
	issn = {2474-9567},
	doi = {10.1145/3130902},
	urldate = {2024-09-13},
	langid = {english},
}

@inproceedings{bedriFitByteAutomaticDiet2020,
	title = {{{FitByte}}: {{Automatic Diet Monitoring}} in {{Unconstrained Situations Using Multimodal Sensing}} on {{Eyeglasses}}},
	shorttitle = {{FitByte}},
	booktitle = {Proceedings of the 2020 {{CHI Conference}} on {{Human Factors}} in {{Computing Systems}}},
	author = {Bedri, Abdelkareem and Li, Diana and Khurana, Rushil and Bhuwalka, Kunal and Goel, Mayank},
	year = 2020,
	month = apr,
	pages = {1--12},
	publisher = {ACM},
	address = {Honolulu HI USA},
	doi = {10.1145/3313831.3376869},
	urldate = {2024-09-13},
	isbn = {978-1-4503-6708-0},
	langid = {english},
}

@inproceedings{bellUMeExploringHuman2023,
	title = {{{{\textmu}Me}}: {{Exploring}} the {{Human Microbiome}} as an {{Intimate Material}} for {{Living Interfaces}}},
	shorttitle = {{{\textmu}Me}},
	booktitle = {Proceedings of the 2023 {{ACM Designing Interactive Systems Conference}}},
	author = {Bell, Fiona and Ramsahoye, Michelle and Coffie, Joshua and Tung, Julia and Alistar, Mirela},
	year = 2023,
	month = jul,
	pages = {2019--2033},
	publisher = {ACM},
	address = {Pittsburgh PA USA},
	doi = {10.1145/3563657.3596133},
	urldate = {2024-09-13},
	isbn = {978-1-4503-9893-0},
	langid = {english},
}

@inproceedings{benkoSphereMultitouchInteractions2008,
	title = {Sphere: Multi-Touch Interactions on a Spherical Display},
	shorttitle = {Sphere},
	booktitle = {Proceedings of the 21st Annual {{ACM}} Symposium on {{User}} Interface Software and Technology},
	author = {Benko, Hrvoje and Wilson, Andrew D. and Balakrishnan, Ravin},
	year = 2008,
	month = oct,
	pages = {77--86},
	publisher = {ACM},
	address = {Monterey CA USA},
	doi = {10.1145/1449715.1449729},
	urldate = {2024-09-13},
	isbn = {978-1-59593-975-3},
	langid = {english}
}

@inproceedings{bermanHowDIYMetaDesignTools2021,
	title = {{{HowDIY}}: {{Towards Meta-Design Tools}} to {{Support Anyone}} to {{3D Print Anywhere}}},
	shorttitle = {{HowDIY}},
	booktitle = {26th {{International Conference}} on {{Intelligent User Interfaces}}},
	author = {Berman, Alexander and Thakare, Ketan and Howell, Joshua and Quek, Francis and Kim, Jeeeun},
	year = 2021,
	month = apr,
	pages = {491--503},
	publisher = {ACM},
	address = {College Station TX USA},
	doi = {10.1145/3397481.3450638},
	urldate = {2024-09-13},
	isbn = {978-1-4503-8017-1},
	langid = {english}
}

@inproceedings{bernsteinCrowdsTwoSeconds2011,
	title = {Crowds in Two Seconds: Enabling Realtime Crowd-Powered Interfaces},
	shorttitle = {Crowds in Two Seconds},
	booktitle = {Proceedings of the 24th Annual {{ACM}} Symposium on {{User}} Interface Software and Technology},
	author = {Bernstein, Michael S. and Brandt, Joel and Miller, Robert C. and Karger, David R.},
	year = 2011,
	month = oct,
	pages = {33--42},
	publisher = {ACM},
	address = {Santa Barbara California USA},
	doi = {10.1145/2047196.2047201},
	urldate = {2024-09-13},
	isbn = {978-1-4503-0716-1},
	langid = {english},
}

@inproceedings{besmerMovingUntaggingPhoto2010a,
	title = {Moving beyond Untagging: Photo Privacy in a Tagged World},
	shorttitle = {Moving beyond Untagging},
	booktitle = {Proceedings of the {{SIGCHI Conference}} on {{Human Factors}} in {{Computing Systems}}},
	author = {Besmer, Andrew and Richter Lipford, Heather},
	year = 2010,
	month = apr,
	pages = {1563--1572},
	publisher = {ACM},
	address = {Atlanta Georgia USA},
	doi = {10.1145/1753326.1753560},
	urldate = {2024-09-13},
	isbn = {978-1-60558-929-9},
	langid = {english}
}

@article{bhatWeAreHalfdoctors2023,
	title = {"{{We}} Are Half-Doctors": {{Family Caregivers}} as {{Boundary Actors}} in {{Chronic Disease Management}}},
	shorttitle = {"{{We}} Are Half-Doctors"},
	author = {Bhat, Karthik S. and Hall, Amanda K. and Kuo, Tiffany and Kumar, Neha},
	year = 2023,
	month = apr,
	journal = {Proceedings of the ACM on Human-Computer Interaction},
	volume = 7,
	number = {CSCW1},
	pages = {1--29},
	issn = {2573-0142},
	doi = {10.1145/3579545},
	urldate = {2024-09-13},
	langid = {english},
}

@inproceedings{boringTouchProjectorMobile2010,
	title = {Touch Projector: Mobile Interaction through Video},
	shorttitle = {Touch Projector},
	booktitle = {Proceedings of the {{SIGCHI Conference}} on {{Human Factors}} in {{Computing Systems}}},
	author = {Boring, Sebastian and Baur, Dominikus and Butz, Andreas and Gustafson, Sean and Baudisch, Patrick},
	year = 2010,
	month = apr,
	pages = {2287--2296},
	publisher = {ACM},
	address = {Atlanta Georgia USA},
	doi = {10.1145/1753326.1753671},
	urldate = {2024-09-13},
	isbn = {978-1-60558-929-9},
	langid = {english}
}

@article{breidebandHomeLifeWorkRhythm2022a,
	title = {Home-{{Life}} and {{Work Rhythm Diversity}} in {{Distributed Teamwork}}: {{A Study}} with {{Information Workers}} during the {{COVID-19 Pandemic}}},
	shorttitle = {Home-{{Life}} and {{Work Rhythm Diversity}} in {{Distributed Teamwork}}},
	author = {Breideband, Thomas and Talkad Sukumar, Poorna and Mark, Gloria and Caruso, Megan and D'Mello, Sidney and Striegel, Aaron D.},
	year = 2022,
	month = mar,
	journal = {Proceedings of the ACM on Human-Computer Interaction},
	volume = 6,
	number = {CSCW1},
	pages = {1--23},
	issn = {2573-0142},
	doi = {10.1145/3512942},
	urldate = {2024-09-13},
	langid = {english},
}

@inproceedings{brownTroubleAutopilotsAssisted2017,
	title = {The {{Trouble}} with {{Autopilots}}: {{Assisted}} and {{Autonomous Driving}} on the {{Social Road}}},
	shorttitle = {The {{Trouble}} with {{Autopilots}}},
	booktitle = {Proceedings of the 2017 {{CHI Conference}} on {{Human Factors}} in {{Computing Systems}}},
	author = {Brown, Barry and Laurier, Eric},
	year = 2017,
	month = may,
	pages = {416--429},
	publisher = {ACM},
	address = {Denver Colorado USA},
	doi = {10.1145/3025453.3025462},
	urldate = {2024-09-13},
	isbn = {978-1-4503-4655-9},
	langid = {english}
}

@inproceedings{brubakerFocusingSharedExperiences2012,
	title = {Focusing on Shared Experiences: Moving beyond the Camera in Video Communication},
	shorttitle = {Focusing on Shared Experiences},
	booktitle = {Proceedings of the {{Designing Interactive Systems Conference}}},
	author = {Brubaker, Jed R. and Venolia, Gina and Tang, John C.},
	year = 2012,
	month = jun,
	pages = {96--105},
	publisher = {ACM},
	address = {Newcastle Upon Tyne United Kingdom},
	doi = {10.1145/2317956.2317973},
	urldate = {2024-09-13},
	isbn = {978-1-4503-1210-3},
	langid = {english}
}

@inproceedings{brudySurfaceFleetExploringDistributed2020,
	title = {{{SurfaceFleet}}: {{Exploring Distributed Interactions Unbounded}} from {{Device}}, {{Application}}, {{User}}, and {{Time}}},
	shorttitle = {{SurfaceFleet}},
	booktitle = {Proceedings of the 33rd {{Annual ACM Symposium}} on {{User Interface Software}} and {{Technology}}},
	author = {Brudy, Frederik and Ledo, David and Pahud, Michel and Henry Riche, Nathalie and Holz, Christian and Waghmare, Anand and Surale, Hemant Bhaskar and Peinado, Marcus and Zhang, Xiaokuan and Joyner, Shannon and Chandramouli, Badrish and Minhas, Umar Farooq and Goldstein, Jonathan and Buxton, William and Hinckley, Ken},
	year = 2020,
	month = oct,
	pages = {7--21},
	publisher = {ACM},
	address = {Virtual Event USA},
	doi = {10.1145/3379337.3415874},
	urldate = {2024-09-13},
	isbn = {978-1-4503-7514-6},
	langid = {english}
}

@inproceedings{brushDigitalNeighborhoodWatch2013,
	title = {Digital Neighborhood Watch: Investigating the Sharing of Camera Data amongst Neighbors},
	shorttitle = {Digital Neighborhood Watch},
	booktitle = {Proceedings of the 2013 Conference on {{Computer}} Supported Cooperative Work},
	author = {Brush, A.J. Bernheim and Jung, Jaeyeon and Mahajan, Ratul and Martinez, Frank},
	year = 2013,
	month = feb,
	pages = {693--700},
	publisher = {ACM},
	address = {San Antonio Texas USA},
	doi = {10.1145/2441776.2441853},
	urldate = {2024-09-13},
	isbn = {978-1-4503-1331-5},
	langid = {english}
}

@inproceedings{burgessHealthcareAITreatment2023,
	title = {Healthcare {{AI Treatment Decision Support}}: {{Design Principles}} to {{Enhance Clinician Adoption}} and {{Trust}}},
	shorttitle = {Healthcare {{AI Treatment Decision Support}}},
	booktitle = {Proceedings of the 2023 {{CHI Conference}} on {{Human Factors}} in {{Computing Systems}}},
	author = {Burgess, Eleanor R. and Jankovic, Ivana and Austin, Melissa and Cai, Nancy and Kapu{\'s}ci{\'n}ska, Adela and Currie, Suzanne and Overhage, J. Marc and Poole, Erika S and Kaye, Jofish},
	year = 2023,
	month = apr,
	pages = {1--19},
	publisher = {ACM},
	address = {Hamburg Germany},
	doi = {10.1145/3544548.3581251},
	urldate = {2024-09-13},
	isbn = {978-1-4503-9421-5},
	langid = {english},
}

@inproceedings{buschelMIRIAMixedReality2021,
	title = {{{MIRIA}}: {{A Mixed Reality Toolkit}} for the {{In-Situ Visualization}} and {{Analysis}} of {{Spatio-Temporal Interaction Data}}},
	shorttitle = {{MIRIA}},
	booktitle = {Proceedings of the 2021 {{CHI Conference}} on {{Human Factors}} in {{Computing Systems}}},
	author = {B{\"u}schel, Wolfgang and Lehmann, Anke and Dachselt, Raimund},
	year = 2021,
	month = may,
	pages = {1--15},
	publisher = {ACM},
	address = {Yokohama Japan},
	doi = {10.1145/3411764.3445651},
	urldate = {2024-09-13},
	isbn = {978-1-4503-8096-6},
	langid = {english},
}

@inproceedings{caoLargeScaleAnalysis2021a,
	title = {Large {{Scale Analysis}} of {{Multitasking Behavior During Remote Meetings}}},
	booktitle = {Proceedings of the 2021 {{CHI Conference}} on {{Human Factors}} in {{Computing Systems}}},
	author = {Cao, Hancheng and Lee, Chia-Jung and Iqbal, Shamsi and Czerwinski, Mary and Wong, Priscilla N Y and Rintel, Sean and Hecht, Brent and Teevan, Jaime and Yang, Longqi},
	year = 2021,
	month = may,
	pages = {1--13},
	publisher = {ACM},
	address = {Yokohama Japan},
	doi = {10.1145/3411764.3445243},
	urldate = {2024-09-13},
	isbn = {978-1-4503-8096-6},
	langid = {english},
}

@article{chancellorWhoHumanHumanCentered2019,
	title = {Who Is the "{{Human}}" in {{Human-Centered Machine Learning}}: {{The Case}} of {{Predicting Mental Health}} from {{Social Media}}},
	shorttitle = {Who Is the "{{Human}}" in {{Human-Centered Machine Learning}}},
	author = {Chancellor, Stevie and Baumer, Eric P. S. and De Choudhury, Munmun},
	year = 2019,
	month = nov,
	journal = {Proceedings of the ACM on Human-Computer Interaction},
	volume = 3,
	number = {CSCW},
	pages = {1--32},
	issn = {2573-0142},
	doi = {10.1145/3359249},
	urldate = {2024-09-13},
	langid = {english},
}

@inproceedings{chandrasekharanBagCommunitiesIdentifying2017,
	title = {The {{Bag}} of {{Communities}}: {{Identifying Abusive Behavior Online}} with {{Preexisting Internet Data}}},
	shorttitle = {The {{Bag}} of {{Communities}}},
	booktitle = {Proceedings of the 2017 {{CHI Conference}} on {{Human Factors}} in {{Computing Systems}}},
	author = {Chandrasekharan, Eshwar and Samory, Mattia and Srinivasan, Anirudh and Gilbert, Eric},
	year = 2017,
	month = may,
	pages = {3175--3187},
	publisher = {ACM},
	address = {Denver Colorado USA},
	doi = {10.1145/3025453.3026018},
	urldate = {2024-09-13},
	isbn = {978-1-4503-4655-9},
	langid = {english}
}

@inproceedings{chenCaringCaregiversDesigning2013,
	title = {Caring for Caregivers: Designing for Integrality},
	shorttitle = {Caring for Caregivers},
	booktitle = {Proceedings of the 2013 Conference on {{Computer}} Supported Cooperative Work},
	author = {Chen, Yunan and Ngo, Victor and Park, Sun Young},
	year = 2013,
	month = feb,
	pages = {91--102},
	publisher = {ACM},
	address = {San Antonio Texas USA},
	doi = {10.1145/2441776.2441789},
	urldate = {2024-09-13},
	isbn = {978-1-4503-1331-5},
	langid = {english}
}

@inproceedings{chenDuetExploringJoint2014,
	title = {Duet: Exploring Joint Interactions on a Smart Phone and a Smart Watch},
	shorttitle = {Duet},
	booktitle = {Proceedings of the {{SIGCHI Conference}} on {{Human Factors}} in {{Computing Systems}}},
	author = {Chen, Xiang 'Anthony' and Grossman, Tovi and Wigdor, Daniel J. and Fitzmaurice, George},
	year = 2014,
	month = apr,
	pages = {159--168},
	publisher = {ACM},
	address = {Toronto Ontario Canada},
	doi = {10.1145/2556288.2556955},
	urldate = {2024-09-13},
	isbn = {978-1-4503-2473-1},
	langid = {english}
}

@inproceedings{chenFinexusTrackingPrecise2016,
	title = {Finexus: {{Tracking Precise Motions}} of {{Multiple Fingertips Using Magnetic Sensing}}},
	shorttitle = {Finexus},
	booktitle = {Proceedings of the 2016 {{CHI Conference}} on {{Human Factors}} in {{Computing Systems}}},
	author = {Chen, Ke-Yu and Patel, Shwetak N. and Keller, Sean},
	year = 2016,
	month = may,
	pages = {1504--1514},
	publisher = {ACM},
	address = {San Jose California USA},
	doi = {10.1145/2858036.2858125},
	urldate = {2024-09-13},
	isbn = {978-1-4503-3362-7},
	langid = {english}
}

@inproceedings{chiouDesigningAIExploration2023,
	title = {Designing with {{AI}}: {{An Exploration}} of {{Co-Ideation}} with {{Image Generators}}},
	shorttitle = {Designing with {{AI}}},
	booktitle = {Proceedings of the 2023 {{ACM Designing Interactive Systems Conference}}},
	author = {Chiou, Li-Yuan and Hung, Peng-Kai and Liang, Rung-Huei and Wang, Chun-Teng},
	year = 2023,
	month = jul,
	pages = {1941--1954},
	publisher = {ACM},
	address = {Pittsburgh PA USA},
	doi = {10.1145/3563657.3596001},
	urldate = {2024-09-13},
	isbn = {978-1-4503-9893-0},
	langid = {english}
}

@article{churchillTeachingLearningHumancomputer2013,
	title = {Teaching and Learning Human-Computer Interaction: Past, Present, and Future},
	shorttitle = {Teaching and Learning Human-Computer Interaction},
	author = {Churchill, Elizabeth F. and Bowser, Anne and Preece, Jennifer},
	year = 2013,
	month = mar,
	journal = {Interactions},
	volume = 20,
	number = 2,
	pages = {44--53},
	issn = {1072-5520, 1558-3449},
	doi = {10.1145/2427076.2427086},
	urldate = {2024-09-13},
	langid = {english}
}

@inproceedings{conversyVizirDomainSpecificGraphical2018,
	title = {Vizir: {{A Domain-Specific Graphical Language}} for {{Authoring}} and {{Operating Airport Automations}}},
	shorttitle = {Vizir},
	booktitle = {Proceedings of the 31st {{Annual ACM Symposium}} on {{User Interface Software}} and {{Technology}}},
	author = {Conversy, St{\'e}phane and Garcia, Jeremie and Buisan, Guilhem and Cousy, Mathieu and Poirier, Mathieu and Saporito, Nicolas and Taurino, Damiano and Frau, Giuseppe and Debattista, Johan},
	year = 2018,
	month = oct,
	pages = {261--273},
	publisher = {ACM},
	address = {Berlin Germany},
	doi = {10.1145/3242587.3242623},
	urldate = {2024-09-13},
	isbn = {978-1-4503-5948-1},
	langid = {english},
}

@inproceedings{correllEthicalDimensionsVisualization2019a,
	title = {Ethical {{Dimensions}} of {{Visualization Research}}},
	booktitle = {Proceedings of the 2019 {{CHI Conference}} on {{Human Factors}} in {{Computing Systems}}},
	author = {Correll, Michael},
	year = 2019,
	month = may,
	pages = {1--13},
	publisher = {ACM},
	address = {Glasgow Scotland Uk},
	doi = {10.1145/3290605.3300418},
	urldate = {2024-09-13},
	isbn = {978-1-4503-5970-2},
	langid = {english},
}

@inproceedings{dangTextGenerationSupporting2022,
	title = {Beyond {{Text Generation}}: {{Supporting Writers}} with {{Continuous Automatic Text Summaries}}},
	shorttitle = {Beyond {{Text Generation}}},
	booktitle = {Proceedings of the 35th {{Annual ACM Symposium}} on {{User Interface Software}} and {{Technology}}},
	author = {Dang, Hai and Benharrak, Karim and Lehmann, Florian and Buschek, Daniel},
	year = 2022,
	month = oct,
	pages = {1--13},
	publisher = {ACM},
	address = {Bend OR USA},
	doi = {10.1145/3526113.3545672},
	urldate = {2024-09-13},
	isbn = {978-1-4503-9320-1},
	langid = {english},
}

@article{danzicoDesignSerendipityNot2010,
	title = {The Design of Serendipity Is Not by Chance},
	author = {Danzico, Liz},
	year = 2010,
	month = sep,
	journal = {Interactions},
	volume = 17,
	number = 5,
	pages = {16--18},
	issn = {1072-5520, 1558-3449},
	doi = {10.1145/1836216.1836220},
	urldate = {2024-09-13},
	langid = {english}
}

@inproceedings{dasswainAlgorithmicPowerPunishment2023,
	title = {Algorithmic {{Power}} or {{Punishment}}: {{Information Worker Perspectives}} on {{Passive Sensing Enabled AI Phenotyping}} of {{Performance}} and {{Wellbeing}}},
	shorttitle = {Algorithmic {{Power}} or {{Punishment}}},
	booktitle = {Proceedings of the 2023 {{CHI Conference}} on {{Human Factors}} in {{Computing Systems}}},
	author = {Das Swain, Vedant and Gao, Lan and Wood, William A and Matli, Srikruthi C and Abowd, Gregory D. and De Choudhury, Munmun},
	year = 2023,
	month = apr,
	pages = {1--17},
	publisher = {ACM},
	address = {Hamburg Germany},
	doi = {10.1145/3544548.3581376},
	urldate = {2024-09-13},
	isbn = {978-1-4503-9421-5},
	langid = {english},
}

@inproceedings{dekaZIPTZeroIntegrationPerformance2017,
	title = {{{ZIPT}}: {{Zero-Integration Performance Testing}} of {{Mobile App Designs}}},
	shorttitle = {{ZIPT}},
	booktitle = {Proceedings of the 30th {{Annual ACM Symposium}} on {{User Interface Software}} and {{Technology}}},
	author = {Deka, Biplab and Huang, Zifeng and Franzen, Chad and Nichols, Jeffrey and Li, Yang and Kumar, Ranjitha},
	year = 2017,
	month = oct,
	pages = {727--736},
	publisher = {ACM},
	address = {Qu{\'e}bec City QC Canada},
	doi = {10.1145/3126594.3126647},
	urldate = {2024-09-13},
	isbn = {978-1-4503-4981-9},
	langid = {english},
}

@article{dementyevDualBlinkWearableDevice2017,
	title = {{{DualBlink}}: {{A Wearable Device}} to {{Continuously Detect}}, {{Track}}, and {{Actuate Blinking For Alleviating Dry Eyes}} and {{Computer Vision Syndrome}}},
	shorttitle = {{DualBlink}},
	author = {Dementyev, Artem and Holz, Christian},
	year = 2017,
	month = mar,
	journal = {Proceedings of the ACM on Interactive, Mobile, Wearable and Ubiquitous Technologies},
	volume = 1,
	number = 1,
	pages = {1--19},
	issn = {2474-9567},
	doi = {10.1145/3053330},
	urldate = {2024-09-13},
	langid = {english}
}

@inproceedings{dementyevRovablesMiniatureBody2016,
	title = {Rovables: {{Miniature On-Body Robots}} as {{Mobile Wearables}}},
	shorttitle = {Rovables},
	booktitle = {Proceedings of the 29th {{Annual Symposium}} on {{User Interface Software}} and {{Technology}}},
	author = {Dementyev, Artem and Kao, Hsin-Liu (Cindy) and Choi, Inrak and Ajilo, Deborah and Xu, Maggie and Paradiso, Joseph A. and Schmandt, Chris and Follmer, Sean},
	year = 2016,
	month = oct,
	pages = {111--120},
	publisher = {ACM},
	address = {Tokyo Japan},
	doi = {10.1145/2984511.2984531},
	urldate = {2024-09-13},
	isbn = {978-1-4503-4189-9},
	langid = {english},
}

@inproceedings{dementyevSensorTapeModularProgrammable2015,
	title = {{{SensorTape}}: {{Modular}} and {{Programmable 3D-Aware Dense Sensor Network}} on a {{Tape}}},
	shorttitle = {{SensorTape}},
	booktitle = {Proceedings of the 28th {{Annual ACM Symposium}} on {{User Interface Software}} \& {{Technology}}},
	author = {Dementyev, Artem and Kao, Hsin-Liu (Cindy) and Paradiso, Joseph A.},
	year = 2015,
	month = nov,
	pages = {649--658},
	publisher = {ACM},
	address = {Charlotte NC USA},
	doi = {10.1145/2807442.2807507},
	urldate = {2024-09-13},
	isbn = {978-1-4503-3779-3},
	langid = {english},
}

@inproceedings{dementyevWristFlexLowpowerGesture2014,
	title = {{{WristFlex}}: Low-Power Gesture Input with Wrist-Worn Pressure Sensors},
	shorttitle = {{WristFlex}},
	booktitle = {Proceedings of the 27th Annual {{ACM}} Symposium on {{User}} Interface Software and Technology},
	author = {Dementyev, Artem and Paradiso, Joseph A.},
	year = 2014,
	month = oct,
	pages = {161--166},
	publisher = {ACM},
	address = {Honolulu Hawaii USA},
	doi = {10.1145/2642918.2647396},
	urldate = {2024-09-13},
	isbn = {978-1-4503-3069-5},
	langid = {english},
}

@article{devitoHowTransfeminineTikTok2022,
	title = {How {{Transfeminine TikTok Creators Navigate}} the {{Algorithmic Trap}} of {{Visibility Via Folk Theorization}}},
	author = {DeVito, Michael Ann},
	year = 2022,
	month = nov,
	journal = {Proceedings of the ACM on Human-Computer Interaction},
	volume = 6,
	number = {CSCW2},
	pages = {1--31},
	issn = {2573-0142},
	doi = {10.1145/3555105},
	urldate = {2024-09-13},
	langid = {english},
}

@inproceedings{disalvoFruitAreHeavy2017a,
	title = {Fruit {{Are Heavy}}: {{A Prototype Public IoT System}} to {{Support Urban Foraging}}},
	shorttitle = {Fruit {{Are Heavy}}},
	booktitle = {Proceedings of the 2017 {{Conference}} on {{Designing Interactive Systems}}},
	author = {DiSalvo, Carl and Jenkins, Tom},
	year = 2017,
	month = jun,
	pages = {541--553},
	publisher = {ACM},
	address = {Edinburgh United Kingdom},
	doi = {10.1145/3064663.3064748},
	urldate = {2024-09-13},
	isbn = {978-1-4503-4922-2},
	langid = {english},
}

@article{disalvoNavigatingTerrainSustainable2010,
	title = {Navigating the Terrain of Sustainable {{HCI}}},
	author = {DiSalvo, Carl and Sengers, Phoebe and Brynjarsd{\'o}ttir, Hr{\"o}nn},
	year = 2010,
	month = jul,
	journal = {Interactions},
	volume = 17,
	number = 4,
	pages = {22--25},
	issn = {1072-5520, 1558-3449},
	doi = {10.1145/1806491.1806497},
	urldate = {2024-09-13},
	langid = {english}
}

@inproceedings{dodgeExplainingModelsEmpirical2019,
	title = {Explaining Models: An Empirical Study of How Explanations Impact Fairness Judgment},
	shorttitle = {Explaining Models},
	booktitle = {Proceedings of the 24th {{International Conference}} on {{Intelligent User Interfaces}}},
	author = {Dodge, Jonathan and Liao, Q. Vera and Zhang, Yunfeng and Bellamy, Rachel K. E. and Dugan, Casey},
	year = 2019,
	month = mar,
	pages = {275--285},
	publisher = {ACM},
	address = {Marina del Ray California},
	doi = {10.1145/3301275.3302310},
	urldate = {2024-09-13},
	isbn = {978-1-4503-6272-6},
	langid = {english},
}

@inproceedings{doveUXDesignInnovation2017,
	title = {{{UX Design Innovation}}: {{Challenges}} for {{Working}} with {{Machine Learning}} as a {{Design Material}}},
	shorttitle = {{UX Design Innovation}},
	booktitle = {Proceedings of the 2017 {{CHI Conference}} on {{Human Factors}} in {{Computing Systems}}},
	author = {Dove, Graham and Halskov, Kim and Forlizzi, Jodi and Zimmerman, John},
	year = 2017,
	month = may,
	pages = {278--288},
	publisher = {ACM},
	address = {Denver Colorado USA},
	doi = {10.1145/3025453.3025739},
	urldate = {2024-09-13},
	isbn = {978-1-4503-4655-9},
	langid = {english}
}

@inproceedings{duDepthLabRealtime3D2020,
	title = {{{DepthLab}}: {{Real-time 3D Interaction}} with {{Depth Maps}} for {{Mobile Augmented Reality}}},
	shorttitle = {{DepthLab}},
	booktitle = {Proceedings of the 33rd {{Annual ACM Symposium}} on {{User Interface Software}} and {{Technology}}},
	author = {Du, Ruofei and Turner, Eric and Dzitsiuk, Maksym and Prasso, Luca and Duarte, Ivo and Dourgarian, Jason and Afonso, Joao and Pascoal, Jose and Gladstone, Josh and Cruces, Nuno and Izadi, Shahram and Kowdle, Adarsh and Tsotsos, Konstantine and Kim, David},
	year = 2020,
	month = oct,
	pages = {829--843},
	publisher = {ACM},
	address = {Virtual Event USA},
	doi = {10.1145/3379337.3415881},
	urldate = {2024-09-13},
	isbn = {978-1-4503-7514-6},
	langid = {english},
}

@inproceedings{ehsanAutomatedRationaleGeneration2019a,
	title = {Automated Rationale Generation: A Technique for Explainable {{AI}} and Its Effects on Human Perceptions},
	shorttitle = {Automated Rationale Generation},
	booktitle = {Proceedings of the 24th {{International Conference}} on {{Intelligent User Interfaces}}},
	author = {Ehsan, Upol and Tambwekar, Pradyumna and Chan, Larry and Harrison, Brent and Riedl, Mark O.},
	year = 2019,
	month = mar,
	pages = {263--274},
	publisher = {ACM},
	address = {Marina del Ray California},
	doi = {10.1145/3301275.3302316},
	urldate = {2024-09-13},
	isbn = {978-1-4503-6272-6},
	langid = {english},
}

@inproceedings{ekhtiarGoalsGoalSetting2023,
	title = {Goals for {{Goal Setting}}: {{A Scoping Review}} on {{Personal Informatics}}},
	shorttitle = {Goals for {{Goal Setting}}},
	booktitle = {Proceedings of the 2023 {{ACM Designing Interactive Systems Conference}}},
	author = {Ekhtiar, Tina and Karahano{\u g}lu, Arma{\u g}an and Gouveia, R{\'u}ben and Ludden, Geke},
	year = 2023,
	month = jul,
	pages = {2625--2641},
	publisher = {ACM},
	address = {Pittsburgh PA USA},
	doi = {10.1145/3563657.3596087},
	urldate = {2024-09-13},
	isbn = {978-1-4503-9893-0},
	langid = {english},
}

@inproceedings{elkinAlignedRankTransform2021,
	title = {An {{Aligned Rank Transform Procedure}} for {{Multifactor Contrast Tests}}},
	booktitle = {The 34th {{Annual ACM Symposium}} on {{User Interface Software}} and {{Technology}}},
	author = {Elkin, Lisa A. and Kay, Matthew and Higgins, James J. and Wobbrock, Jacob O.},
	year = 2021,
	month = oct,
	pages = {754--768},
	publisher = {ACM},
	address = {Virtual Event USA},
	doi = {10.1145/3472749.3474784},
	urldate = {2024-09-13},
	isbn = {978-1-4503-8635-7},
	langid = {english},
}

@inproceedings{elsdenMakingSenseBlockchain2018,
	title = {Making {{Sense}} of {{Blockchain Applications}}: {{A Typology}} for {{HCI}}},
	shorttitle = {Making {{Sense}} of {{Blockchain Applications}}},
	booktitle = {Proceedings of the 2018 {{CHI Conference}} on {{Human Factors}} in {{Computing Systems}}},
	author = {Elsden, Chris and Manohar, Arthi and Briggs, Jo and Harding, Mike and Speed, Chris and Vines, John},
	year = 2018,
	month = apr,
	pages = {1--14},
	publisher = {ACM},
	address = {Montreal QC Canada},
	doi = {10.1145/3173574.3174032},
	urldate = {2024-09-13},
	isbn = {978-1-4503-5620-6},
	langid = {english},
}

@article{engelbutzederSurplusScarcityAbundance2023,
	title = {From {{Surplus}} and {{Scarcity}} toward {{Abundance}}: {{Understanding}} the {{Use}} of {{ICT}} in {{Food Resource Sharing Practices}}},
	shorttitle = {From {{Surplus}} and {{Scarcity}} toward {{Abundance}}},
	author = {Engelbutzeder, Philip and Randell, Dave and Landwehr, Marvin and Aal, Konstantin and Stevens, Gunnar and Wulf, Volker},
	year = 2023,
	month = oct,
	journal = {ACM Transactions on Computer-Human Interaction},
	volume = 30,
	number = 5,
	pages = {1--31},
	issn = {1073-0516, 1557-7325},
	doi = {10.1145/3589957},
	urldate = {2024-09-13},
	langid = {english},
}

@article{epsteinSuppressingSearchEngine2017,
	title = {Suppressing the {{Search Engine Manipulation Effect}} ({{SEME}})},
	author = {Epstein, Robert and Robertson, Ronald E. and Lazer, David and Wilson, Christo},
	year = 2017,
	month = dec,
	journal = {Proceedings of the ACM on Human-Computer Interaction},
	volume = 1,
	number = {CSCW},
	pages = {1--22},
	issn = {2573-0142},
	doi = {10.1145/3134677},
	urldate = {2024-09-13},
	langid = {english}
}

@article{ernalaLinguisticMarkersIndicating2017,
	title = {Linguistic {{Markers Indicating Therapeutic Outcomes}} of {{Social Media Disclosures}} of {{Schizophrenia}}},
	author = {Ernala, Sindhu Kiranmai and Rizvi, Asra F. and Birnbaum, Michael L. and Kane, John M. and De Choudhury, Munmun},
	year = 2017,
	month = dec,
	journal = {Proceedings of the ACM on Human-Computer Interaction},
	volume = 1,
	number = {CSCW},
	pages = {1--27},
	issn = {2573-0142},
	doi = {10.1145/3134678},
	urldate = {2024-09-13},
	langid = {english}
}

@inproceedings{estevesOrbitsGazeInteraction2015,
	title = {Orbits: {{Gaze Interaction}} for {{Smart Watches}} Using {{Smooth Pursuit Eye Movements}}},
	shorttitle = {Orbits},
	booktitle = {Proceedings of the 28th {{Annual ACM Symposium}} on {{User Interface Software}} \& {{Technology}}},
	author = {Esteves, Augusto and Velloso, Eduardo and Bulling, Andreas and Gellersen, Hans},
	year = 2015,
	month = nov,
	pages = {457--466},
	publisher = {ACM},
	address = {Charlotte NC USA},
	doi = {10.1145/2807442.2807499},
	urldate = {2024-09-13},
	isbn = {978-1-4503-3779-3},
	langid = {english}
}

@inproceedings{fastMetaEnablingProgramming2016,
	title = {Meta: {{Enabling Programming Languages}} to {{Learn}} from the {{Crowd}}},
	shorttitle = {Meta},
	booktitle = {Proceedings of the 29th {{Annual Symposium}} on {{User Interface Software}} and {{Technology}}},
	author = {Fast, Ethan and Bernstein, Michael S.},
	year = 2016,
	month = oct,
	pages = {259--270},
	publisher = {ACM},
	address = {Tokyo Japan},
	doi = {10.1145/2984511.2984532},
	urldate = {2024-09-13},
	isbn = {978-1-4503-4189-9},
	langid = {english}
}

@inproceedings{feitEverydayGazeInput2017,
	title = {Toward {{Everyday Gaze Input}}: {{Accuracy}} and {{Precision}} of {{Eye Tracking}} and {{Implications}} for {{Design}}},
	shorttitle = {Toward {{Everyday Gaze Input}}},
	booktitle = {Proceedings of the 2017 {{CHI Conference}} on {{Human Factors}} in {{Computing Systems}}},
	author = {Feit, Anna Maria and Williams, Shane and Toledo, Arturo and Paradiso, Ann and Kulkarni, Harish and Kane, Shaun and Morris, Meredith Ringel},
	year = 2017,
	month = may,
	pages = {1118--1130},
	publisher = {ACM},
	address = {Denver Colorado USA},
	doi = {10.1145/3025453.3025599},
	urldate = {2024-09-13},
	isbn = {978-1-4503-4655-9},
	langid = {english},
}

@inproceedings{fenderCausalitypreservingAsynchronousReality2022a,
	title = {Causality-Preserving {{Asynchronous Reality}}},
	booktitle = {{{CHI Conference}} on {{Human Factors}} in {{Computing Systems}}},
	author = {Fender, Andreas Rene and Holz, Christian},
	year = 2022,
	month = apr,
	pages = {1--15},
	publisher = {ACM},
	address = {New Orleans LA USA},
	doi = {10.1145/3491102.3501836},
	urldate = {2024-09-13},
	isbn = {978-1-4503-9157-3},
	langid = {english},
}

@inproceedings{fengHowUXPractitioners2023,
	title = {How {{Do UX Practitioners Communicate AI}} as a {{Design Material}}? {{Artifacts}}, {{Conceptions}}, and {{Propositions}}},
	shorttitle = {How {{Do UX Practitioners Communicate AI}} as a {{Design Material}}?},
	booktitle = {Proceedings of the 2023 {{ACM Designing Interactive Systems Conference}}},
	author = {Feng, K. J. Kevin and Coppock, Maxwell James and McDonald, David W.},
	year = 2023,
	month = jul,
	pages = {2263--2280},
	publisher = {ACM},
	address = {Pittsburgh PA USA},
	doi = {10.1145/3563657.3596101},
	urldate = {2024-09-13},
	isbn = {978-1-4503-9893-0},
	langid = {english},
}

@inproceedings{feuchtnerExtendingBodyInteraction2017,
	title = {Extending the {{Body}} for {{Interaction}} with {{Reality}}},
	booktitle = {Proceedings of the 2017 {{CHI Conference}} on {{Human Factors}} in {{Computing Systems}}},
	author = {Feuchtner, Tiare and M{\"u}ller, J{\"o}rg},
	year = 2017,
	month = may,
	pages = {5145--5157},
	publisher = {ACM},
	address = {Denver Colorado USA},
	doi = {10.1145/3025453.3025689},
	urldate = {2024-09-13},
	isbn = {978-1-4503-4655-9},
	langid = {english},
}

@inproceedings{fischerRecommendingEnergyTariffs2013,
	title = {Recommending Energy Tariffs and Load Shifting Based on Smart Household Usage Profiling},
	booktitle = {Proceedings of the 2013 International Conference on {{Intelligent}} User Interfaces},
	author = {Fischer, Joel E. and Ramchurn, Sarvapali D. and Osborne, Michael and Parson, Oliver and Huynh, Trung Dong and Alam, Muddasser and Pantidi, Nadia and Moran, Stuart and Bachour, Khaled and Reece, Steve and Costanza, Enrico and Rodden, Tom and Jennings, Nicholas R.},
	year = 2013,
	month = mar,
	pages = {383--394},
	publisher = {ACM},
	address = {Santa Monica California USA},
	doi = {10.1145/2449396.2449446},
	urldate = {2024-09-13},
	isbn = {978-1-4503-1965-2},
	langid = {english},
}

@inproceedings{follmerInFORMDynamicPhysical2013,
	title = {{{inFORM}}: Dynamic Physical Affordances and Constraints through Shape and Object Actuation},
	shorttitle = {{inFORM}},
	booktitle = {Proceedings of the 26th Annual {{ACM}} Symposium on {{User}} Interface Software and Technology},
	author = {Follmer, Sean and Leithinger, Daniel and Olwal, Alex and Hogge, Akimitsu and Ishii, Hiroshi},
	year = 2013,
	month = oct,
	pages = {417--426},
	publisher = {ACM},
	address = {St. Andrews Scotland, United Kingdom},
	doi = {10.1145/2501988.2502032},
	urldate = {2024-09-13},
	isbn = {978-1-4503-2268-3},
	langid = {english},
}

@inproceedings{forlizziWhereShouldTurn2010,
	title = {Where Should i Turn: Moving from Individual to Collaborative Navigation Strategies to Inform the Interaction Design of Future Navigation Systems},
	shorttitle = {Where Should i Turn},
	booktitle = {Proceedings of the {{SIGCHI Conference}} on {{Human Factors}} in {{Computing Systems}}},
	author = {Forlizzi, Jodi and Barley, William C. and Seder, Thomas},
	year = 2010,
	month = apr,
	pages = {1261--1270},
	publisher = {ACM},
	address = {Atlanta Georgia USA},
	doi = {10.1145/1753326.1753516},
	urldate = {2024-09-13},
	isbn = {978-1-60558-929-9},
	langid = {english}
}

@article{foxPatchworkHiddenHuman2023,
	title = {Patchwork: {{The Hidden}}, {{Human Labor}} of {{AI Integration}} within {{Essential Work}}},
	shorttitle = {Patchwork},
	author = {Fox, Sarah E. and Shorey, Samantha and Kang, Esther Y. and Montiel Valle, Dominique and Rodriguez, Estefania},
	year = 2023,
	month = apr,
	journal = {Proceedings of the ACM on Human-Computer Interaction},
	volume = 7,
	number = {CSCW1},
	pages = {1--20},
	issn = {2573-0142},
	doi = {10.1145/3579514},
	urldate = {2024-09-13},
	langid = {english},
}

@inproceedings{freedStalkersParadiseHow2018,
	title = {``{{A Stalker}}'s {{Paradise}}'': {{How Intimate Partner Abusers Exploit Technology}}},
	shorttitle = {``{{A Stalker}}'s {{Paradise}}''},
	booktitle = {Proceedings of the 2018 {{CHI Conference}} on {{Human Factors}} in {{Computing Systems}}},
	author = {Freed, Diana and Palmer, Jackeline and Minchala, Diana and Levy, Karen and Ristenpart, Thomas and Dell, Nicola},
	year = 2018,
	month = apr,
	pages = {1--13},
	publisher = {ACM},
	address = {Montreal QC Canada},
	doi = {10.1145/3173574.3174241},
	urldate = {2024-09-13},
	isbn = {978-1-4503-5620-6},
	langid = {english}
}

@article{freemanWorkingTogetherApart2022,
	title = {Working {{Together Apart}} through {{Embodiment}}: {{Engaging}} in {{Everyday Collaborative Activities}} in {{Social Virtual Reality}}},
	shorttitle = {Working {{Together Apart}} through {{Embodiment}}},
	author = {Freeman, Guo and Acena, Dane and McNeese, Nathan J. and Schulenberg, Kelsea},
	year = 2022,
	month = jan,
	journal = {Proceedings of the ACM on Human-Computer Interaction},
	volume = 6,
	number = {GROUP},
	pages = {1--25},
	issn = {2573-0142},
	doi = {10.1145/3492836},
	urldate = {2024-09-13},
	langid = {english},
}

@inproceedings{freyBreezeSharingBiofeedback2018,
	title = {Breeze: {{Sharing Biofeedback}} through {{Wearable Technologies}}},
	shorttitle = {Breeze},
	booktitle = {Proceedings of the 2018 {{CHI Conference}} on {{Human Factors}} in {{Computing Systems}}},
	author = {Frey, J{\'e}r{\'e}my and Grabli, May and Slyper, Ronit and Cauchard, Jessica R.},
	year = 2018,
	month = apr,
	pages = {1--12},
	publisher = {ACM},
	address = {Montreal QC Canada},
	doi = {10.1145/3173574.3174219},
	urldate = {2024-09-13},
	isbn = {978-1-4503-5620-6},
	langid = {english},
}

@inproceedings{fridmanCognitiveLoadEstimation2018,
	title = {Cognitive {{Load Estimation}} in the {{Wild}}},
	booktitle = {Proceedings of the 2018 {{CHI Conference}} on {{Human Factors}} in {{Computing Systems}}},
	author = {Fridman, Lex and Reimer, Bryan and Mehler, Bruce and Freeman, William T.},
	year = 2018,
	month = apr,
	pages = {1--9},
	publisher = {ACM},
	address = {Montreal QC Canada},
	doi = {10.1145/3173574.3174226},
	urldate = {2024-09-13},
	isbn = {978-1-4503-5620-6},
	langid = {english},
}

@inproceedings{furloRethinkingDatingApps2021,
	title = {Rethinking {{Dating Apps}} as {{Sexual Consent Apps}}: {{A New Use Case}} for {{AI-Mediated Communication}}},
	shorttitle = {Rethinking {{Dating Apps}} as {{Sexual Consent Apps}}},
	booktitle = {Companion {{Publication}} of the 2021 {{Conference}} on {{Computer Supported Cooperative Work}} and {{Social Computing}}},
	author = {Furlo, Nicholas and Gleason, Jacob and Feun, Karen and Zytko, Douglas},
	year = 2021,
	month = oct,
	pages = {53--56},
	publisher = {ACM},
	address = {Virtual Event USA},
	doi = {10.1145/3462204.3481770},
	urldate = {2024-09-13},
	isbn = {978-1-4503-8479-7},
	langid = {english}
}

@inproceedings{geroSparksInspirationScience2022,
	title = {Sparks: {{Inspiration}} for {{Science Writing}} Using {{Language Models}}},
	shorttitle = {Sparks},
	booktitle = {Designing {{Interactive Systems Conference}}},
	author = {Gero, Katy Ilonka and Liu, Vivian and Chilton, Lydia},
	year = 2022,
	month = jun,
	pages = {1002--1019},
	publisher = {ACM},
	address = {Virtual Event Australia},
	doi = {10.1145/3532106.3533533},
	urldate = {2024-09-13},
	isbn = {978-1-4503-9358-4},
	langid = {english},
}

@inproceedings{gheranGesturesSmartRings2018,
	title = {Gestures for {{Smart Rings}}: {{Empirical Results}}, {{Insights}}, and {{Design Implications}}},
	shorttitle = {Gestures for {{Smart Rings}}},
	booktitle = {Proceedings of the 2018 {{Designing Interactive Systems Conference}}},
	author = {Gheran, Bogdan-Florin and Vanderdonckt, Jean and Vatavu, Radu-Daniel},
	year = 2018,
	month = jun,
	pages = {623--635},
	publisher = {ACM},
	address = {Hong Kong China},
	doi = {10.1145/3196709.3196741},
	urldate = {2024-09-13},
	isbn = {978-1-4503-5198-0},
	langid = {english}
}

@inproceedings{goedickeVROOMVirtualReality2018,
	title = {{{VR-OOM}}: {{Virtual Reality On-rOad}} Driving {{siMulation}}},
	shorttitle = {{VR-OOM}},
	booktitle = {Proceedings of the 2018 {{CHI Conference}} on {{Human Factors}} in {{Computing Systems}}},
	author = {Goedicke, David and Li, Jamy and Evers, Vanessa and Ju, Wendy},
	year = 2018,
	month = apr,
	pages = {1--11},
	publisher = {ACM},
	address = {Montreal QC Canada},
	doi = {10.1145/3173574.3173739},
	urldate = {2024-09-13},
	isbn = {978-1-4503-5620-6},
	langid = {english}
}

@inproceedings{gomesMorePhoneStudyActuated2013,
	title = {{{MorePhone}}: A Study of Actuated Shape Deformations for Flexible Thin-Film Smartphone Notifications},
	shorttitle = {{MorePhone}},
	booktitle = {Proceedings of the {{SIGCHI Conference}} on {{Human Factors}} in {{Computing Systems}}},
	author = {Gomes, Antonio and Nesbitt, Andrea and Vertegaal, Roel},
	year = 2013,
	month = apr,
	pages = {583--592},
	publisher = {ACM},
	address = {Paris France},
	doi = {10.1145/2470654.2470737},
	urldate = {2024-09-13},
	isbn = {978-1-4503-1899-0},
	langid = {english}
}

@article{grillAttitudesFolkTheories2022,
	title = {Attitudes and {{Folk Theories}} of {{Data Subjects}} on {{Transparency}} and {{Accuracy}} in {{Emotion Recognition}}},
	author = {Grill, Gabriel and Andalibi, Nazanin},
	year = 2022,
	month = mar,
	journal = {Proceedings of the ACM on Human-Computer Interaction},
	volume = 6,
	number = {CSCW1},
	pages = {1--35},
	issn = {2573-0142},
	doi = {10.1145/3512925},
	urldate = {2024-09-13},
	langid = {english}
}

@article{groegerObjectSkinAugmentingEveryday2018,
	title = {{{ObjectSkin}}: {{Augmenting Everyday Objects}} with {{Hydroprinted Touch Sensors}} and {{Displays}}},
	shorttitle = {{ObjectSkin}},
	author = {Groeger, Daniel and Steimle, J{\"u}rgen},
	year = 2018,
	month = jan,
	journal = {Proceedings of the ACM on Interactive, Mobile, Wearable and Ubiquitous Technologies},
	volume = 1,
	number = 4,
	pages = {1--23},
	issn = {2474-9567},
	doi = {10.1145/3161165},
	urldate = {2024-09-13},
	langid = {english}
}

@inproceedings{gugenheimerFaceTouchEnablingTouch2016,
	title = {{{FaceTouch}}: {{Enabling Touch Interaction}} in {{Display Fixed UIs}} for {{Mobile Virtual Reality}}},
	shorttitle = {{FaceTouch}},
	booktitle = {Proceedings of the 29th {{Annual Symposium}} on {{User Interface Software}} and {{Technology}}},
	author = {Gugenheimer, Jan and Dobbelstein, David and Winkler, Christian and Haas, Gabriel and Rukzio, Enrico},
	year = 2016,
	month = oct,
	pages = {49--60},
	publisher = {ACM},
	address = {Tokyo Japan},
	doi = {10.1145/2984511.2984576},
	urldate = {2024-09-13},
	isbn = {978-1-4503-4189-9},
	langid = {english}
}

@inproceedings{guoFacadeAutogeneratingTactile2017,
	title = {Facade: {{Auto-generating Tactile Interfaces}} to {{Appliances}}},
	shorttitle = {Facade},
	booktitle = {Proceedings of the 2017 {{CHI Conference}} on {{Human Factors}} in {{Computing Systems}}},
	author = {Guo, Anhong and Kim, Jeeeun and Chen, Xiang 'Anthony' and Yeh, Tom and Hudson, Scott E. and Mankoff, Jennifer and Bigham, Jeffrey P.},
	year = 2017,
	month = may,
	pages = {5826--5838},
	publisher = {ACM},
	address = {Denver Colorado USA},
	doi = {10.1145/3025453.3025845},
	urldate = {2024-09-13},
	isbn = {978-1-4503-4655-9},
	langid = {english}
}

@inproceedings{haeslerConnectedSelfOrganizedCitizens2021,
	title = {Connected {{Self-Organized Citizens}} in {{Crises}}: {{An Interdisciplinary Resilience Concept}} for {{Neighborhoods}}},
	shorttitle = {Connected {{Self-Organized Citizens}} in {{Crises}}},
	booktitle = {Companion {{Publication}} of the 2021 {{Conference}} on {{Computer Supported Cooperative Work}} and {{Social Computing}}},
	author = {Haesler, Steffen and Mogk, Ragnar and Putz, Florentin and Logan, Kevin T. and Thiessen, Nadja and Kleinschnitger, Katharina and Baumg{\"a}rtner, Lars and Stroscher, Jan-Philipp and Reuter, Christian and Knodt, Michele and Hollick, Matthias},
	year = 2021,
	month = oct,
	pages = {62--66},
	publisher = {ACM},
	address = {Virtual Event USA},
	doi = {10.1145/3462204.3481749},
	urldate = {2024-09-13},
	isbn = {978-1-4503-8479-7},
	langid = {english}
}

@inproceedings{hanHydroRingSupportingMixed2018,
	title = {{{HydroRing}}: {{Supporting Mixed Reality Haptics Using Liquid Flow}}},
	shorttitle = {{HydroRing}},
	booktitle = {Proceedings of the 31st {{Annual ACM Symposium}} on {{User Interface Software}} and {{Technology}}},
	author = {Han, Teng and Anderson, Fraser and Irani, Pourang and Grossman, Tovi},
	year = 2018,
	month = oct,
	pages = {913--925},
	publisher = {ACM},
	address = {Berlin Germany},
	doi = {10.1145/3242587.3242667},
	urldate = {2024-09-13},
	isbn = {978-1-4503-5948-1},
	langid = {english}
}

@article{hartikainenSafeSextingAdvice2021,
	title = {Safe {{Sexting}}: {{The Advice}} and {{Support Adolescents Receive}} from {{Peers}} Regarding {{Online Sexual Risks}}},
	shorttitle = {Safe {{Sexting}}},
	author = {Hartikainen, Heidi and Razi, Afsaneh and Wisniewski, Pamela},
	year = 2021,
	month = apr,
	journal = {Proceedings of the ACM on Human-Computer Interaction},
	volume = 5,
	number = {CSCW1},
	pages = {1--31},
	issn = {2573-0142},
	doi = {10.1145/3449116},
	urldate = {2024-09-13},
	langid = {english}
}

@article{hassanFootStrikerEMSbasedFoot2017,
	title = {{{FootStriker}}: {{An EMS-based Foot Strike Assistant}} for {{Running}}},
	shorttitle = {{FootStriker}},
	author = {Hassan, Mahmoud and Daiber, Florian and Wiehr, Frederik and Kosmalla, Felix and Kr{\"u}ger, Antonio},
	year = 2017,
	month = mar,
	journal = {Proceedings of the ACM on Interactive, Mobile, Wearable and Ubiquitous Technologies},
	volume = 1,
	number = 1,
	pages = {1--18},
	issn = {2474-9567},
	doi = {10.1145/3053332},
	urldate = {2024-09-13},
	langid = {english}
}

@inproceedings{hassibEmotionActuatorEmbodied2017,
	title = {Emotion {{Actuator}}: {{Embodied Emotional Feedback}} through {{Electroencephalography}} and {{Electrical Muscle Stimulation}}},
	shorttitle = {Emotion {{Actuator}}},
	booktitle = {Proceedings of the 2017 {{CHI Conference}} on {{Human Factors}} in {{Computing Systems}}},
	author = {Hassib, Mariam and Pfeiffer, Max and Schneegass, Stefan and Rohs, Michael and Alt, Florian},
	year = 2017,
	month = may,
	pages = {6133--6146},
	publisher = {ACM},
	address = {Denver Colorado USA},
	doi = {10.1145/3025453.3025953},
	urldate = {2024-09-13},
	isbn = {978-1-4503-4655-9},
	langid = {english}
}

@inproceedings{headAugmentingScientificPapers2021,
	title = {Augmenting {{Scientific Papers}} with {{Just-in-Time}}, {{Position-Sensitive Definitions}} of {{Terms}} and {{Symbols}}},
	booktitle = {Proceedings of the 2021 {{CHI Conference}} on {{Human Factors}} in {{Computing Systems}}},
	author = {Head, Andrew and Lo, Kyle and Kang, Dongyeop and Fok, Raymond and Skjonsberg, Sam and Weld, Daniel S. and Hearst, Marti A.},
	year = 2021,
	month = may,
	pages = {1--18},
	publisher = {ACM},
	address = {Yokohama Japan},
	doi = {10.1145/3411764.3445648},
	urldate = {2024-09-13},
	isbn = {978-1-4503-8096-6},
	langid = {english},
}

@inproceedings{held3DPuppetryKinectbased2012,
	title = {{{3D}} Puppetry: A Kinect-Based Interface for {{3D}} Animation},
	shorttitle = {{{3D}} Puppetry},
	booktitle = {Proceedings of the 25th Annual {{ACM}} Symposium on {{User}} Interface Software and Technology},
	author = {Held, Robert and Gupta, Ankit and Curless, Brian and Agrawala, Maneesh},
	year = 2012,
	month = oct,
	pages = {423--434},
	publisher = {ACM},
	address = {Cambridge Massachusetts USA},
	doi = {10.1145/2380116.2380170},
	urldate = {2024-09-13},
	isbn = {978-1-4503-1580-7},
	langid = {english}
}

@inproceedings{hewittAssessingPublicPerception2019,
	title = {Assessing Public Perception of Self-Driving Cars: The Autonomous Vehicle Acceptance Model},
	shorttitle = {Assessing Public Perception of Self-Driving Cars},
	booktitle = {Proceedings of the 24th {{International Conference}} on {{Intelligent User Interfaces}}},
	author = {Hewitt, Charlie and Politis, Ioannis and Amanatidis, Theocharis and Sarkar, Advait},
	year = 2019,
	month = mar,
	pages = {518--527},
	publisher = {ACM},
	address = {Marina del Ray California},
	doi = {10.1145/3301275.3302268},
	urldate = {2024-09-13},
	isbn = {978-1-4503-6272-6},
	langid = {english}
}

@article{hirschNothingCompilationHow2024,
	title = {Nothing {{Like Compilation}}: {{How Professional Digital Fabrication Workflows Go Beyond Extruding}}, {{Milling}}, and {{Machines}}},
	shorttitle = {Nothing {{Like Compilation}}},
	author = {Hirsch, Mare and Benabdallah, Gabrielle and Jacobs, Jennifer and Peek, Nadya},
	year = 2024,
	month = feb,
	journal = {ACM Transactions on Computer-Human Interaction},
	volume = 31,
	number = 1,
	pages = {1--45},
	issn = {1073-0516, 1557-7325},
	doi = {10.1145/3609328},
	urldate = {2024-09-13},
	langid = {english},
}

@inproceedings{hollanderTaxonomyVulnerableRoad2021,
	title = {A {{Taxonomy}} of {{Vulnerable Road Users}} for {{HCI Based On A Systematic Literature Review}}},
	booktitle = {Proceedings of the 2021 {{CHI Conference}} on {{Human Factors}} in {{Computing Systems}}},
	author = {Holl{\"a}nder, Kai and Colley, Mark and Rukzio, Enrico and Butz, Andreas},
	year = 2021,
	month = may,
	pages = {1--13},
	publisher = {ACM},
	address = {Yokohama Japan},
	doi = {10.1145/3411764.3445480},
	urldate = {2024-09-13},
	isbn = {978-1-4503-8096-6},
	langid = {english}
}

@inproceedings{hongSmartphonebasedSensingPlatform2014a,
	title = {A Smartphone-Based Sensing Platform to Model Aggressive Driving Behaviors},
	booktitle = {Proceedings of the {{SIGCHI Conference}} on {{Human Factors}} in {{Computing Systems}}},
	author = {Hong, Jin-Hyuk and Margines, Ben and Dey, Anind K.},
	year = 2014,
	month = apr,
	pages = {4047--4056},
	publisher = {ACM},
	address = {Toronto Ontario Canada},
	doi = {10.1145/2556288.2557321},
	urldate = {2024-09-13},
	isbn = {978-1-4503-2473-1},
	langid = {english},
}

@inproceedings{honnetPolySenseAugmentingTextiles2020,
	title = {{{PolySense}}: {{Augmenting Textiles}} with {{Electrical Functionality}} Using {{In-Situ Polymerization}}},
	shorttitle = {{PolySense}},
	booktitle = {Proceedings of the 2020 {{CHI Conference}} on {{Human Factors}} in {{Computing Systems}}},
	author = {Honnet, Cedric and {Perner-Wilson}, Hannah and Teyssier, Marc and Fruchard, Bruno and Steimle, J{\"u}rgen and Baptista, Ana C. and Strohmeier, Paul},
	year = 2020,
	month = apr,
	pages = {1--13},
	publisher = {ACM},
	address = {Honolulu HI USA},
	doi = {10.1145/3313831.3376841},
	urldate = {2024-09-13},
	isbn = {978-1-4503-6708-0},
	langid = {english},
}

@inproceedings{huangOrecchioExtendingBodyLanguage2018,
	title = {Orecchio: {{Extending Body-Language}} through {{Actuated Static}} and {{Dynamic Auricular Postures}}},
	shorttitle = {Orecchio},
	booktitle = {Proceedings of the 31st {{Annual ACM Symposium}} on {{User Interface Software}} and {{Technology}}},
	author = {Huang, Da-Yuan and Seyed, Teddy and Li, Linjun and Gong, Jun and Yao, Zhihao and Jiao, Yuchen and Chen, Xiang 'Anthony' and Yang, Xing-Dong},
	year = 2018,
	month = oct,
	pages = {697--710},
	publisher = {ACM},
	address = {Berlin Germany},
	doi = {10.1145/3242587.3242629},
	urldate = {2024-09-13},
	isbn = {978-1-4503-5948-1},
	langid = {english}
}

@inproceedings{huangWovenProbeProbingPossibilities2021,
	title = {{{WovenProbe}}: {{Probing Possibilities}} for {{Weaving Fully-Integrated On-Skin Systems Deployable}} in the {{Field}}},
	shorttitle = {{WovenProbe}},
	booktitle = {Designing {{Interactive Systems Conference}} 2021},
	author = {Huang, Kunpeng and Sun, Ruojia and Zhang, Ximeng and Islam Molla, Md. Tahmidul and Dunne, Margaret and Guimbretiere, Francois and Kao, Cindy Hsin-Liu},
	year = 2021,
	month = jun,
	pages = {1143--1158},
	publisher = {ACM},
	address = {Virtual Event USA},
	doi = {10.1145/3461778.3462105},
	urldate = {2024-09-13},
	isbn = {978-1-4503-8476-6},
	langid = {english}
}

@article{huhHealthVlogsSocial2014,
	title = {Health {{Vlogs}} as {{Social Support}} for {{Chronic Illness Management}}},
	author = {Huh, Jina and Liu, Leslie S. and Neogi, Tina and Inkpen, Kori and Pratt, Wanda},
	year = 2014,
	month = aug,
	journal = {ACM Transactions on Computer-Human Interaction},
	volume = 21,
	number = 4,
	pages = {1--31},
	issn = {1073-0516, 1557-7325},
	doi = {10.1145/2630067},
	urldate = {2024-09-13},
	langid = {english},
}

@inproceedings{hurstAutomaticallyDetectingPointing2008,
	title = {Automatically Detecting Pointing Performance},
	booktitle = {Proceedings of the 13th International Conference on {{Intelligent}} User Interfaces},
	author = {Hurst, Amy and Hudson, Scott E. and Mankoff, Jennifer and Trewin, Shari},
	year = 2008,
	month = jan,
	pages = {11--19},
	publisher = {ACM},
	address = {Gran Canaria Spain},
	doi = {10.1145/1378773.1378776},
	urldate = {2024-09-13},
	isbn = {978-1-59593-987-6},
	langid = {english},
}

@article{hwangSocialbotsVoicesFronts2012,
	title = {Socialbots: Voices from the Fronts},
	shorttitle = {Socialbots},
	author = {Hwang, Tim and Pearce, Ian and Nanis, Max},
	year = 2012,
	month = mar,
	journal = {Interactions},
	volume = 19,
	number = 2,
	pages = {38--45},
	issn = {1072-5520, 1558-3449},
	doi = {10.1145/2090150.2090161},
	urldate = {2024-09-13},
	langid = {english}
}

@article{iqbalOasisFrameworkLinking2010,
	title = {Oasis: {{A}} Framework for Linking Notification Delivery to the Perceptual Structure of Goal-Directed Tasks},
	shorttitle = {Oasis},
	author = {Iqbal, Shamsi T. and Bailey, Brian P.},
	year = 2010,
	month = dec,
	journal = {ACM Transactions on Computer-Human Interaction},
	volume = 17,
	number = 4,
	pages = {1--28},
	issn = {1073-0516, 1557-7325},
	doi = {10.1145/1879831.1879833},
	urldate = {2024-09-13},
	langid = {english}
}

@inproceedings{ishiiOpticalMarionetteGraphical2016,
	title = {Optical {{Marionette}}: {{Graphical Manipulation}} of {{Human}}'s {{Walking Direction}}},
	shorttitle = {Optical {{Marionette}}},
	booktitle = {Proceedings of the 29th {{Annual Symposium}} on {{User Interface Software}} and {{Technology}}},
	author = {Ishii, Akira and Suzuki, Ippei and Sakamoto, Shinji and Kanai, Keita and Takazawa, Kazuki and Doi, Hiraku and Ochiai, Yoichi},
	year = 2016,
	month = oct,
	pages = {705--716},
	publisher = {ACM},
	address = {Tokyo Japan},
	doi = {10.1145/2984511.2984545},
	urldate = {2024-09-13},
	isbn = {978-1-4503-4189-9},
	langid = {english}
}

@inproceedings{jainHeadMountedDisplayVisualizations2015,
	title = {Head-{{Mounted Display Visualizations}} to {{Support Sound Awareness}} for the {{Deaf}} and {{Hard}} of {{Hearing}}},
	booktitle = {Proceedings of the 33rd {{Annual ACM Conference}} on {{Human Factors}} in {{Computing Systems}}},
	author = {Jain, Dhruv and Findlater, Leah and Gilkeson, Jamie and Holland, Benjamin and Duraiswami, Ramani and Zotkin, Dmitry and Vogler, Christian and Froehlich, Jon E.},
	year = 2015,
	month = apr,
	pages = {241--250},
	publisher = {ACM},
	address = {Seoul Republic of Korea},
	doi = {10.1145/2702123.2702393},
	urldate = {2024-09-13},
	isbn = {978-1-4503-3145-6},
	langid = {english}
}

@inproceedings{jiangHandAvatarEmbodyingNonHumanoid2023,
	title = {{{HandAvatar}}: {{Embodying Non-Humanoid Virtual Avatars}} through {{Hands}}},
	shorttitle = {{HandAvatar}},
	booktitle = {Proceedings of the 2023 {{CHI Conference}} on {{Human Factors}} in {{Computing Systems}}},
	author = {Jiang, Yu and Li, Zhipeng and He, Mufei and Lindlbauer, David and Yan, Yukang},
	year = 2023,
	month = apr,
	pages = {1--17},
	publisher = {ACM},
	address = {Hamburg Germany},
	doi = {10.1145/3544548.3581027},
	urldate = {2024-09-13},
	isbn = {978-1-4503-9421-5},
	langid = {english}
}

@inproceedings{jokelaDiaryStudyCombining2015,
	title = {A {{Diary Study}} on {{Combining Multiple Information Devices}} in {{Everyday Activities}} and {{Tasks}}},
	booktitle = {Proceedings of the 33rd {{Annual ACM Conference}} on {{Human Factors}} in {{Computing Systems}}},
	author = {Jokela, Tero and Ojala, Jarno and Olsson, Thomas},
	year = 2015,
	month = apr,
	pages = {3903--3912},
	publisher = {ACM},
	address = {Seoul Republic of Korea},
	doi = {10.1145/2702123.2702211},
	urldate = {2024-09-13},
	isbn = {978-1-4503-3145-6},
	langid = {english}
}

@inproceedings{kaimotoSketchedRealitySketching2022,
	title = {Sketched {{Reality}}: {{Sketching Bi-Directional Interactions Between Virtual}} and {{Physical Worlds}} with {{AR}} and {{Actuated Tangible UI}}},
	shorttitle = {Sketched {{Reality}}},
	booktitle = {Proceedings of the 35th {{Annual ACM Symposium}} on {{User Interface Software}} and {{Technology}}},
	author = {Kaimoto, Hiroki and Monteiro, Kyzyl and Faridan, Mehrad and Li, Jiatong and Farajian, Samin and Kakehi, Yasuaki and Nakagaki, Ken and Suzuki, Ryo},
	year = 2022,
	month = oct,
	pages = {1--12},
	publisher = {ACM},
	address = {Bend OR USA},
	doi = {10.1145/3526113.3545626},
	urldate = {2024-09-13},
	isbn = {978-1-4503-9320-1},
	langid = {english},
}

@inproceedings{kaneAccessOverlaysImproving2011,
	title = {Access Overlays: Improving Non-Visual Access to Large Touch Screens for Blind Users},
	shorttitle = {Access Overlays},
	booktitle = {Proceedings of the 24th Annual {{ACM}} Symposium on {{User}} Interface Software and Technology},
	author = {Kane, Shaun K. and Morris, Meredith Ringel and Perkins, Annuska Z. and Wigdor, Daniel and Ladner, Richard E. and Wobbrock, Jacob O.},
	year = 2011,
	month = oct,
	pages = {273--282},
	publisher = {ACM},
	address = {Santa Barbara California USA},
	doi = {10.1145/2047196.2047232},
	urldate = {2024-09-13},
	isbn = {978-1-4503-0716-1},
	langid = {english}
}

@inproceedings{kapurAlterEgoPersonalizedWearable2018a,
	title = {{{AlterEgo}}: {{A Personalized Wearable Silent Speech Interface}}},
	shorttitle = {{AlterEgo}},
	booktitle = {23rd {{International Conference}} on {{Intelligent User Interfaces}}},
	author = {Kapur, Arnav and Kapur, Shreyas and Maes, Pattie},
	year = 2018,
	month = mar,
	pages = {43--53},
	publisher = {ACM},
	address = {Tokyo Japan},
	doi = {10.1145/3172944.3172977},
	urldate = {2024-09-13},
	isbn = {978-1-4503-4945-1},
	langid = {english}
}

@article{karranFrameworkPsychophysiologicalClassification2015,
	title = {A {{Framework}} for {{Psychophysiological Classification}} within a {{Cultural Heritage Context Using Interest}}},
	author = {Karran, Alexander J. and Fairclough, Stephen H. and Gilleade, Kiel},
	year = 2015,
	month = jan,
	journal = {ACM Transactions on Computer-Human Interaction},
	volume = 21,
	number = 6,
	pages = {1--19},
	issn = {1073-0516, 1557-7325},
	doi = {10.1145/2687925},
	urldate = {2024-09-13},
	langid = {english},
}

@inproceedings{khotUnderstandingPhysicalActivity2014,
	title = {Understanding Physical Activity through {{3D}} Printed Material Artifacts},
	booktitle = {Proceedings of the {{SIGCHI Conference}} on {{Human Factors}} in {{Computing Systems}}},
	author = {Khot, Rohit Ashok and Hjorth, Larissa and Mueller, Florian 'Floyd'},
	year = 2014,
	month = apr,
	pages = {3835--3844},
	publisher = {ACM},
	address = {Toronto Ontario Canada},
	doi = {10.1145/2556288.2557144},
	urldate = {2024-09-13},
	isbn = {978-1-4503-2473-1},
	langid = {english}
}

@inproceedings{kimCellsGeneratorsLenses2023,
	title = {Cells, {{Generators}}, and {{Lenses}}: {{Design Framework}} for {{Object-Oriented Interaction}} with {{Large Language Models}}},
	shorttitle = {Cells, {{Generators}}, and {{Lenses}}},
	booktitle = {Proceedings of the 36th {{Annual ACM Symposium}} on {{User Interface Software}} and {{Technology}}},
	author = {Kim, Tae Soo and Lee, Yoonjoo and Chang, Minsuk and Kim, Juho},
	year = 2023,
	month = oct,
	pages = {1--18},
	publisher = {ACM},
	address = {San Francisco CA USA},
	doi = {10.1145/3586183.3606833},
	urldate = {2024-09-13},
	isbn = 9798400701320,
	langid = {english},
}

@inproceedings{kimDatadrivenInteractionTechniques2014,
	title = {Data-Driven Interaction Techniques for Improving Navigation of Educational Videos},
	booktitle = {Proceedings of the 27th Annual {{ACM}} Symposium on {{User}} Interface Software and Technology},
	author = {Kim, Juho and Guo, Philip J. and Cai, Carrie J. and Li, Shang-Wen (Daniel) and Gajos, Krzysztof Z. and Miller, Robert C.},
	year = 2014,
	month = oct,
	pages = {563--572},
	publisher = {ACM},
	address = {Honolulu Hawaii USA},
	doi = {10.1145/2642918.2647389},
	urldate = {2024-09-13},
	isbn = {978-1-4503-3069-5},
	langid = {english},
}

@inproceedings{kimExploringChartQuestion2023a,
	title = {Exploring {{Chart Question Answering}} for {{Blind}} and {{Low Vision Users}}},
	booktitle = {Proceedings of the 2023 {{CHI Conference}} on {{Human Factors}} in {{Computing Systems}}},
	author = {Kim, Jiho and Srinivasan, Arjun and Kim, Nam Wook and Kim, Yea-Seul},
	year = 2023,
	month = apr,
	pages = {1--15},
	publisher = {ACM},
	address = {Hamburg Germany},
	doi = {10.1145/3544548.3581532},
	urldate = {2024-09-13},
	isbn = {978-1-4503-9421-5},
	langid = {english}
}

@inproceedings{kimInflatableMouseVolumeadjustable2008a,
	title = {Inflatable Mouse: Volume-Adjustable Mouse with Air-Pressure-Sensitive Input and Haptic Feedback},
	shorttitle = {Inflatable Mouse},
	booktitle = {Proceedings of the {{SIGCHI Conference}} on {{Human Factors}} in {{Computing Systems}}},
	author = {Kim, Seoktae and Kim, Hyunjung and Lee, Boram and Nam, Tek-Jin and Lee, Woohun},
	year = 2008,
	month = apr,
	pages = {211--224},
	publisher = {ACM},
	address = {Florence Italy},
	doi = {10.1145/1357054.1357090},
	urldate = {2024-09-13},
	isbn = {978-1-60558-011-1},
	langid = {english}
}

@article{kimOmniTrackFlexibleSelfTracking2017,
	title = {{{OmniTrack}}: {{A Flexible Self-Tracking Approach Leveraging Semi-Automated Tracking}}},
	shorttitle = {{OmniTrack}},
	author = {Kim, Young-Ho and Jeon, Jae Ho and Lee, Bongshin and Choe, Eun Kyoung and Seo, Jinwook},
	year = 2017,
	month = sep,
	journal = {Proceedings of the ACM on Interactive, Mobile, Wearable and Ubiquitous Technologies},
	volume = 1,
	number = 3,
	pages = {1--28},
	issn = {2474-9567},
	doi = {10.1145/3130930},
	urldate = {2024-09-13},
	langid = {english}
}

@inproceedings{kirkHomeVideoCommunication2010,
	title = {Home Video Communication: Mediating 'Closeness'},
	shorttitle = {Home Video Communication},
	booktitle = {Proceedings of the 2010 {{ACM}} Conference on {{Computer}} Supported Cooperative Work},
	author = {Kirk, David S. and Sellen, Abigail and Cao, Xiang},
	year = 2010,
	month = feb,
	pages = {135--144},
	publisher = {ACM},
	address = {Savannah Georgia USA},
	doi = {10.1145/1718918.1718945},
	urldate = {2024-09-13},
	isbn = {978-1-60558-795-0},
	langid = {english}
}

@article{kirkHumanRemainsValues2010,
	title = {On Human Remains: {{Values}} and Practice in the Home Archiving of Cherished Objects},
	shorttitle = {On Human Remains},
	author = {Kirk, David S. and Sellen, Abigail},
	year = 2010,
	month = jul,
	journal = {ACM Transactions on Computer-Human Interaction},
	volume = 17,
	number = 3,
	pages = {1--43},
	issn = {1073-0516, 1557-7325},
	doi = {10.1145/1806923.1806924},
	urldate = {2024-09-13},
	langid = {english}
}

@inproceedings{kirkOpeningFamilyArchive2010,
	title = {Opening up the Family Archive},
	booktitle = {Proceedings of the 2010 {{ACM}} Conference on {{Computer}} Supported Cooperative Work},
	author = {Kirk, David S. and Izadi, Shahram and Sellen, Abigail and Taylor, Stuart and Banks, Richard and Hilliges, Otmar},
	year = 2010,
	month = feb,
	pages = {261--270},
	publisher = {ACM},
	address = {Savannah Georgia USA},
	doi = {10.1145/1718918.1718968},
	urldate = {2024-09-13},
	isbn = {978-1-60558-795-0},
	langid = {english}
}

@inproceedings{kleinbergerSupportingElderConnectedness2019,
	title = {Supporting {{Elder Connectedness}} through {{Cognitively Sustainable Design Interactions}} with the {{Memory Music Box}}},
	booktitle = {Proceedings of the 32nd {{Annual ACM Symposium}} on {{User Interface Software}} and {{Technology}}},
	author = {Kleinberger, Rebecca and Rieger, Alexandra and Sands, Janelle and Baker, Janet},
	year = 2019,
	month = oct,
	pages = {355--369},
	publisher = {ACM},
	address = {New Orleans LA USA},
	doi = {10.1145/3332165.3347877},
	urldate = {2024-09-13},
	isbn = {978-1-4503-6816-2},
	langid = {english},
}

@inproceedings{klokmoseWebstratesShareableDynamic2015,
	title = {{\emph{Webstrates}}: {{Shareable Dynamic Media}}},
	shorttitle = {{\emph{Webstrates}}},
	booktitle = {Proceedings of the 28th {{Annual ACM Symposium}} on {{User Interface Software}} \& {{Technology}}},
	author = {Klokmose, Clemens N. and Eagan, James R. and Baader, Siemen and Mackay, Wendy and {Beaudouin-Lafon}, Michel},
	year = 2015,
	month = nov,
	pages = {280--290},
	publisher = {ACM},
	address = {Charlotte NC USA},
	doi = {10.1145/2807442.2807446},
	urldate = {2024-09-13},
	isbn = {978-1-4503-3779-3},
	langid = {english},
}

@article{kraemerBatteryfreeMakeCodeAccessible2022,
	title = {Battery-Free {{MakeCode}}: {{Accessible Programming}} for {{Intermittent Computing}}},
	shorttitle = {Battery-Free {{MakeCode}}},
	author = {Kraemer, Christopher and Guo, Amy and Ahmed, Saad and Hester, Josiah},
	year = 2022,
	month = mar,
	journal = {Proceedings of the ACM on Interactive, Mobile, Wearable and Ubiquitous Technologies},
	volume = 6,
	number = 1,
	pages = {1--35},
	issn = {2474-9567},
	doi = {10.1145/3517236},
	urldate = {2024-09-13},
	langid = {english}
}

@inproceedings{krumAllRoadsLead2008a,
	title = {All Roads Lead to {{CHI}}: Interaction in the Automobile},
	shorttitle = {All Roads Lead to {{CHI}}},
	booktitle = {{{CHI}} '08 {{Extended Abstracts}} on {{Human Factors}} in {{Computing Systems}}},
	author = {Krum, David M. and Faenger, Jens and Lathrop, Brian and Sison, Jo Ann and Lien, Annie},
	year = 2008,
	month = apr,
	pages = {2387--2390},
	publisher = {ACM},
	address = {Florence Italy},
	doi = {10.1145/1358628.1358691},
	urldate = {2024-09-13},
	isbn = {978-1-60558-012-8},
	langid = {english}
}

@inproceedings{kundingerDriverDrowsinessAutomated2020a,
	title = {Driver Drowsiness in Automated and Manual Driving: Insights from a Test Track Study},
	shorttitle = {Driver Drowsiness in Automated and Manual Driving},
	booktitle = {Proceedings of the 25th {{International Conference}} on {{Intelligent User Interfaces}}},
	author = {Kundinger, Thomas and Riener, Andreas and Sofra, Nikoletta and Weigl, Klemens},
	year = 2020,
	month = mar,
	pages = {369--379},
	publisher = {ACM},
	address = {Cagliari Italy},
	doi = {10.1145/3377325.3377506},
	urldate = {2024-09-13},
	isbn = {978-1-4503-7118-6},
	langid = {english}
}

@inproceedings{lafreniereCrowdsourcedFabrication2016,
	title = {Crowdsourced {{Fabrication}}},
	booktitle = {Proceedings of the 29th {{Annual Symposium}} on {{User Interface Software}} and {{Technology}}},
	author = {Lafreniere, Benjamin and Grossman, Tovi and Anderson, Fraser and Matejka, Justin and Kerrick, Heather and Nagy, Danil and Vasey, Lauren and Atherton, Evan and Beirne, Nicholas and Coelho, Marcelo H. and Cote, Nicholas and Li, Steven and Nogueira, Andy and Nguyen, Long and Schwinn, Tobias and Stoddart, James and Thomasson, David and Wang, Ray and White, Thomas and Benjamin, David and Conti, Maurice and Menges, Achim and Fitzmaurice, George},
	year = 2016,
	month = oct,
	pages = {15--28},
	publisher = {ACM},
	address = {Tokyo Japan},
	doi = {10.1145/2984511.2984553},
	urldate = {2024-09-13},
	isbn = {978-1-4503-4189-9},
	langid = {english}
}

@inproceedings{laputSensingFineGrainedHand2019,
	title = {Sensing {{Fine-Grained Hand Activity}} with {{Smartwatches}}},
	booktitle = {Proceedings of the 2019 {{CHI Conference}} on {{Human Factors}} in {{Computing Systems}}},
	author = {Laput, Gierad and Harrison, Chris},
	year = 2019,
	month = may,
	pages = {1--13},
	publisher = {ACM},
	address = {Glasgow Scotland Uk},
	doi = {10.1145/3290605.3300568},
	urldate = {2024-09-13},
	isbn = {978-1-4503-5970-2},
	langid = {english}
}

@inproceedings{lawsonProblematisingUpstreamTechnology2015a,
	title = {Problematising {{Upstream Technology}} through {{Speculative Design}}: {{The Case}} of {{Quantified Cats}} and {{Dogs}}},
	shorttitle = {Problematising {{Upstream Technology}} through {{Speculative Design}}},
	booktitle = {Proceedings of the 33rd {{Annual ACM Conference}} on {{Human Factors}} in {{Computing Systems}}},
	author = {Lawson, Shaun and Kirman, Ben and Linehan, Conor and Feltwell, Tom and Hopkins, Lisa},
	year = 2015,
	month = apr,
	pages = {2663--2672},
	publisher = {ACM},
	address = {Seoul Republic of Korea},
	doi = {10.1145/2702123.2702260},
	urldate = {2024-09-13},
	isbn = {978-1-4503-3145-6},
	langid = {english},
}

@inproceedings{leeSleepGuruPersonalizedSleep2022a,
	title = {{{SleepGuru}}: {{Personalized Sleep Planning System}} for {{Real-life Actionability}} and {{Negotiability}}},
	shorttitle = {{SleepGuru}},
	booktitle = {Proceedings of the 35th {{Annual ACM Symposium}} on {{User Interface Software}} and {{Technology}}},
	author = {Lee, Jungeun and Kim, Sungnam and Cheon, Minki and Ju, Hyojin and Lee, JaeEun and Hwang, Inseok},
	year = 2022,
	month = oct,
	pages = {1--16},
	publisher = {ACM},
	address = {Bend OR USA},
	doi = {10.1145/3526113.3545709},
	urldate = {2024-09-13},
	isbn = {978-1-4503-9320-1},
	langid = {english}
}

@inproceedings{leeZeroNMidairTangible2011,
	title = {{{ZeroN}}: Mid-Air Tangible Interaction Enabled by Computer Controlled Magnetic Levitation},
	shorttitle = {{ZeroN}},
	booktitle = {Proceedings of the 24th Annual {{ACM}} Symposium on {{User}} Interface Software and Technology},
	author = {Lee, Jinha and Post, Rehmi and Ishii, Hiroshi},
	year = 2011,
	month = oct,
	pages = {327--336},
	publisher = {ACM},
	address = {Santa Barbara California USA},
	doi = {10.1145/2047196.2047239},
	urldate = {2024-09-13},
	isbn = {978-1-4503-0716-1},
	langid = {english}
}

@inproceedings{leithingerPhysicalTelepresenceShape2014,
	title = {Physical Telepresence: Shape Capture and Display for Embodied, Computer-Mediated Remote Collaboration},
	shorttitle = {Physical Telepresence},
	booktitle = {Proceedings of the 27th Annual {{ACM}} Symposium on {{User}} Interface Software and Technology},
	author = {Leithinger, Daniel and Follmer, Sean and Olwal, Alex and Ishii, Hiroshi},
	year = 2014,
	month = oct,
	pages = {461--470},
	publisher = {ACM},
	address = {Honolulu Hawaii USA},
	doi = {10.1145/2642918.2647377},
	urldate = {2024-09-13},
	isbn = {978-1-4503-3069-5},
	langid = {english}
}

@inproceedings{leithingerSublimateStatechangingVirtual2013,
	title = {Sublimate: State-Changing Virtual and Physical Rendering to Augment Interaction with Shape Displays},
	shorttitle = {Sublimate},
	booktitle = {Proceedings of the {{SIGCHI Conference}} on {{Human Factors}} in {{Computing Systems}}},
	author = {Leithinger, Daniel and Follmer, Sean and Olwal, Alex and Luescher, Samuel and Hogge, Akimitsu and Lee, Jinha and Ishii, Hiroshi},
	year = 2013,
	month = apr,
	pages = {1441--1450},
	publisher = {ACM},
	address = {Paris France},
	doi = {10.1145/2470654.2466191},
	urldate = {2024-09-13},
	isbn = {978-1-4503-1899-0},
	langid = {english},
}

@inproceedings{liaoAllWorkNo2018,
	title = {All {{Work}} and {{No Play}}?},
	booktitle = {Proceedings of the 2018 {{CHI Conference}} on {{Human Factors}} in {{Computing Systems}}},
	author = {Liao, Q. Vera and {Mas-ud Hussain}, Muhammed and Chandar, Praveen and Davis, Matthew and Khazaeni, Yasaman and Crasso, Marco Patricio and Wang, Dakuo and Muller, Michael and Shami, N. Sadat and Geyer, Werner},
	year = 2018,
	month = apr,
	pages = {1--13},
	publisher = {ACM},
	address = {Montreal QC Canada},
	doi = {10.1145/3173574.3173577},
	urldate = {2024-09-13},
	isbn = {978-1-4503-5620-6},
	langid = {english}
}

@article{lightDigitalInterdependenceHow2011,
	title = {Digital Interdependence and How to Design for It},
	author = {Light, Ann},
	year = 2011,
	month = mar,
	journal = {Interactions},
	volume = 18,
	number = 2,
	pages = {34--39},
	issn = {1072-5520, 1558-3449},
	doi = {10.1145/1925820.1925829},
	urldate = {2024-09-13},
	langid = {english}
}

@inproceedings{liMeasuringUnderstandingPhoto2019,
	title = {Measuring and {{Understanding Photo Sharing Experiences}} in {{Social Virtual Reality}}},
	booktitle = {Proceedings of the 2019 {{CHI Conference}} on {{Human Factors}} in {{Computing Systems}}},
	author = {Li, Jie and Kong, Yiping and R{\"o}ggla, Thomas and De Simone, Francesca and Ananthanarayan, Swamy and De Ridder, Huib and El Ali, Abdallah and Cesar, Pablo},
	year = 2019,
	month = may,
	pages = {1--14},
	publisher = {ACM},
	address = {Glasgow Scotland Uk},
	doi = {10.1145/3290605.3300897},
	urldate = {2024-09-13},
	isbn = {978-1-4503-5970-2},
	langid = {english},
}

@inproceedings{linAdasaConversationalVehicle2018,
	title = {Adasa: {{A Conversational In-Vehicle Digital Assistant}} for {{Advanced Driver Assistance Features}}},
	shorttitle = {Adasa},
	booktitle = {Proceedings of the 31st {{Annual ACM Symposium}} on {{User Interface Software}} and {{Technology}}},
	author = {Lin, Shih-Chieh and Hsu, Chang-Hong and Talamonti, Walter and Zhang, Yunqi and Oney, Steve and Mars, Jason and Tang, Lingjia},
	year = 2018,
	month = oct,
	pages = {531--542},
	publisher = {ACM},
	address = {Berlin Germany},
	doi = {10.1145/3242587.3242593},
	urldate = {2024-09-13},
	isbn = {978-1-4503-5948-1},
	langid = {english}
}

@inproceedings{linSYNCCrowdsourcingPlatform2021a,
	title = {{{SYNC}}: {{A Crowdsourcing Platform}} for {{News Co-editing}}},
	shorttitle = {{SYNC}},
	booktitle = {Companion {{Publication}} of the 2021 {{Conference}} on {{Computer Supported Cooperative Work}} and {{Social Computing}}},
	author = {Lin, Jin-An and Hsu, Feng-Yi and Yao, Hsin-Yu and Lu, Shang-Hsun and Kuo, Tsai-Yu and Lin, Chieh-Kai and Chang, Yung-Ju},
	year = 2021,
	month = oct,
	pages = {126--129},
	publisher = {ACM},
	address = {Virtual Event USA},
	doi = {10.1145/3462204.3481775},
	urldate = {2024-09-13},
	isbn = {978-1-4503-8479-7},
	langid = {english}
}

@inproceedings{liTangibleGridTangibleWeb2022,
	title = {{{TangibleGrid}}: {{Tangible Web Layout Design}} for {{Blind Users}}},
	shorttitle = {{TangibleGrid}},
	booktitle = {Proceedings of the 35th {{Annual ACM Symposium}} on {{User Interface Software}} and {{Technology}}},
	author = {Li, Jiasheng and Yan, Zeyu and Jarjue, Ebrima Haddy and Shetty, Ashrith and Peng, Huaishu},
	year = 2022,
	month = oct,
	pages = {1--12},
	publisher = {ACM},
	address = {Bend OR USA},
	doi = {10.1145/3526113.3545627},
	urldate = {2024-09-13},
	isbn = {978-1-4503-9320-1},
	langid = {english},
}

@inproceedings{liu3DALLEIntegratingTextImage2023,
	title = {{{3DALL-E}}: {{Integrating Text-to-Image AI}} in {{3D Design Workflows}}},
	shorttitle = {{3DALL-E}},
	booktitle = {Proceedings of the 2023 {{ACM Designing Interactive Systems Conference}}},
	author = {Liu, Vivian and Vermeulen, Jo and Fitzmaurice, George and Matejka, Justin},
	year = 2023,
	month = jul,
	pages = {1955--1977},
	publisher = {ACM},
	address = {Pittsburgh PA USA},
	doi = {10.1145/3563657.3596098},
	urldate = {2024-09-13},
	isbn = {978-1-4503-9893-0},
	langid = {english},
}

@article{liuLivingHeritageHistoric2012,
	title = {The Living Heritage of Historic Crises: Curating the {{Bhopal}} Disaster in the Social Media Landscape},
	shorttitle = {The Living Heritage of Historic Crises},
	author = {Liu, Sophia B.},
	year = 2012,
	month = may,
	journal = {Interactions},
	volume = 19,
	number = 3,
	pages = {20--24},
	issn = {1072-5520, 1558-3449},
	doi = {10.1145/2168931.2168938},
	urldate = {2024-09-13},
	langid = {english}
}

@article{luceroMobileCollocatedInteractions2013,
	title = {Mobile Collocated Interactions: Taking an Offline Break Together},
	shorttitle = {Mobile Collocated Interactions},
	author = {Lucero, Andr{\'e}s and Jones, Matt and Jokela, Tero and Robinson, Simon},
	year = 2013,
	month = mar,
	journal = {Interactions},
	volume = 20,
	number = 2,
	pages = {26--32},
	issn = {1072-5520, 1558-3449},
	doi = {10.1145/2427076.2427083},
	urldate = {2024-09-13},
	langid = {english}
}

@inproceedings{lugerHavingReallyBad2016b,
	title = {"{{Like Having}} a {{Really Bad PA}}": {{The Gulf}} between {{User Expectation}} and {{Experience}} of {{Conversational Agents}}},
	shorttitle = {"{{Like Having}} a {{Really Bad PA}}"},
	booktitle = {Proceedings of the 2016 {{CHI Conference}} on {{Human Factors}} in {{Computing Systems}}},
	author = {Luger, Ewa and Sellen, Abigail},
	year = 2016,
	month = may,
	pages = {5286--5297},
	publisher = {ACM},
	address = {San Jose California USA},
	doi = {10.1145/2858036.2858288},
	urldate = {2024-09-13},
	isbn = {978-1-4503-3362-7},
	langid = {english}
}

@inproceedings{luParticipatoryNoticingPhotovoice2023,
	title = {Participatory {{Noticing}} through {{Photovoice}}: {{Engaging Arts-}} and {{Community-Based Approaches}} in {{Design Research}}},
	shorttitle = {Participatory {{Noticing}} through {{Photovoice}}},
	booktitle = {Proceedings of the 2023 {{ACM Designing Interactive Systems Conference}}},
	author = {Lu, Alex Jiahong and Sannon, Shruti and Moy, Cameron and Brewer, Savana and Green, Jaye and Jackson, Kisha N and Reeder, Daivon and Wafer, Camaria and Ackerman, Mark S. and Dillahunt, Tawanna R},
	year = 2023,
	month = jul,
	pages = {2489--2508},
	publisher = {ACM},
	address = {Pittsburgh PA USA},
	doi = {10.1145/3563657.3596041},
	urldate = {2024-09-13},
	isbn = {978-1-4503-9893-0},
	langid = {english},
}

@article{maloneyTalkingVoiceUnderstanding2020,
	title = {"{{Talking}} without a {{Voice}}": {{Understanding Non-verbal Communication}} in {{Social Virtual Reality}}},
	shorttitle = {"{{Talking}} without a {{Voice}}"},
	author = {Maloney, Divine and Freeman, Guo and Wohn, Donghee Yvette},
	year = 2020,
	month = oct,
	journal = {Proceedings of the ACM on Human-Computer Interaction},
	volume = 4,
	number = {CSCW2},
	pages = {1--25},
	issn = {2573-0142},
	doi = {10.1145/3415246},
	urldate = {2024-09-13},
	langid = {english}
}

@inproceedings{marcusTwitinfoAggregatingVisualizing2011,
	title = {Twitinfo: Aggregating and Visualizing Microblogs for Event Exploration},
	shorttitle = {Twitinfo},
	booktitle = {Proceedings of the {{SIGCHI Conference}} on {{Human Factors}} in {{Computing Systems}}},
	author = {Marcus, Adam and Bernstein, Michael S. and Badar, Osama and Karger, David R. and Madden, Samuel and Miller, Robert C.},
	year = 2011,
	month = may,
	pages = {227--236},
	publisher = {ACM},
	address = {Vancouver BC Canada},
	doi = {10.1145/1978942.1978975},
	urldate = {2024-09-13},
	isbn = {978-1-4503-0228-9},
	langid = {english},
}

@inproceedings{markResilienceCollaborationTechnology2008,
	title = {Resilience in Collaboration: Technology as a Resource for New Patterns of Action},
	shorttitle = {Resilience in Collaboration},
	booktitle = {Proceedings of the 2008 {{ACM}} Conference on {{Computer}} Supported Cooperative Work},
	author = {Mark, Gloria and Semaan, Bryan},
	year = 2008,
	month = nov,
	pages = {137--146},
	publisher = {ACM},
	address = {San Diego CA USA},
	doi = {10.1145/1460563.1460585},
	urldate = {2024-09-13},
	isbn = {978-1-60558-007-4},
	langid = {english}
}

@inproceedings{marlowActivityTracesSignals2013,
	title = {Activity Traces and Signals in Software Developer Recruitment and Hiring},
	booktitle = {Proceedings of the 2013 Conference on {{Computer}} Supported Cooperative Work},
	author = {Marlow, Jennifer and Dabbish, Laura},
	year = 2013,
	month = feb,
	pages = {145--156},
	publisher = {ACM},
	address = {San Antonio Texas USA},
	doi = {10.1145/2441776.2441794},
	urldate = {2024-09-13},
	isbn = {978-1-4503-1331-5},
	langid = {english}
}

@inproceedings{matthiesEarFieldSensingNovelEar2017a,
	title = {{{{\emph{EarFieldSensing}}}}: {{A Novel In-Ear Electric Field Sensing}} to {{Enrich Wearable Gesture Input}} through {{Facial Expressions}}},
	shorttitle = {{{\emph{EarFieldSensing}}}},
	booktitle = {Proceedings of the 2017 {{CHI Conference}} on {{Human Factors}} in {{Computing Systems}}},
	author = {Matthies, Denys J. C. and Strecker, Bernhard A. and Urban, Bodo},
	year = 2017,
	month = may,
	pages = {1911--1922},
	publisher = {ACM},
	address = {Denver Colorado USA},
	doi = {10.1145/3025453.3025692},
	urldate = {2024-09-13},
	isbn = {978-1-4503-4655-9},
	langid = {english}
}

@inproceedings{mccarthyContextContentCommunity2008,
	title = {The Context, Content \& Community Collage: Sharing Personal Digital Media in the Physical Workplace},
	shorttitle = {The Context, Content \& Community Collage},
	booktitle = {Proceedings of the 2008 {{ACM}} Conference on {{Computer}} Supported Cooperative Work},
	author = {McCarthy, Joseph F. and Congleton, Ben and Harper, F. Maxwell},
	year = 2008,
	month = nov,
	pages = {97--106},
	publisher = {ACM},
	address = {San Diego CA USA},
	doi = {10.1145/1460563.1460580},
	urldate = {2024-09-13},
	isbn = {978-1-60558-007-4},
	langid = {english}
}

@article{mcgillCreatingAugmentingKeyboards2022,
	title = {Creating and {{Augmenting Keyboards}} for {{Extended Reality}} with the {{K}} Eyboard {{A}} Ugmentation {{T}} Oolkit},
	author = {McGill, Mark and Brewster, Stephen and De Sa Medeiros, Daniel Pires and Bovet, Sidney and Gutierrez, Mario and Kehoe, Aidan},
	year = 2022,
	month = apr,
	journal = {ACM Transactions on Computer-Human Interaction},
	volume = 29,
	number = 2,
	pages = {1--39},
	issn = {1073-0516, 1557-7325},
	doi = {10.1145/3490495},
	urldate = {2024-09-13},
	langid = {english},
}

@inproceedings{mcnallyCodesigningMobileOnline2018,
	title = {Co-Designing {{Mobile Online Safety Applications}} with {{Children}}},
	booktitle = {Proceedings of the 2018 {{CHI Conference}} on {{Human Factors}} in {{Computing Systems}}},
	author = {McNally, Brenna and Kumar, Priya and Hordatt, Chelsea and Mauriello, Matthew Louis and Naik, Shalmali and Norooz, Leyla and Shorter, Alazandra and Golub, Evan and Druin, Allison},
	year = 2018,
	month = apr,
	pages = {1--9},
	publisher = {ACM},
	address = {Montreal QC Canada},
	doi = {10.1145/3173574.3174097},
	urldate = {2024-09-13},
	isbn = {978-1-4503-5620-6},
	langid = {english}
}

@inproceedings{mentisInvisibleEmotionInformation2010a,
	title = {Invisible Emotion: Information and Interaction in an Emergency Room},
	shorttitle = {Invisible Emotion},
	booktitle = {Proceedings of the 2010 {{ACM}} Conference on {{Computer}} Supported Cooperative Work},
	author = {Mentis, Helena M. and Reddy, Madhu and Rosson, Mary Beth},
	year = 2010,
	month = feb,
	pages = {311--320},
	publisher = {ACM},
	address = {Savannah Georgia USA},
	doi = {10.1145/1718918.1718975},
	urldate = {2024-09-13},
	isbn = {978-1-60558-795-0},
	langid = {english}
}

@article{messerschmidtANISMAPrototypingToolkit2022,
	title = {{{ANISMA}}: {{A Prototyping Toolkit}} to {{Explore Haptic Skin Deformation Applications Using Shape-Memory Alloys}}},
	shorttitle = {{ANISMA}},
	author = {Messerschmidt, Moritz Alexander and Muthukumarana, Sachith and Hamdan, Nur Al-Huda and Wagner, Adrian and Zhang, Haimo and Borchers, Jan and Nanayakkara, Suranga Chandima},
	year = 2022,
	month = jun,
	journal = {ACM Transactions on Computer-Human Interaction},
	volume = 29,
	number = 3,
	pages = {1--34},
	issn = {1073-0516, 1557-7325},
	doi = {10.1145/3490497},
	urldate = {2024-09-13},
	langid = {english}
}

@inproceedings{miltonSeeMeHere2023,
	title = {``{{I See Me Here}}'': {{Mental Health Content}}, {{Community}}, and {{Algorithmic Curation}} on {{TikTok}}},
	shorttitle = {``{{I See Me Here}}''},
	booktitle = {Proceedings of the 2023 {{CHI Conference}} on {{Human Factors}} in {{Computing Systems}}},
	author = {Milton, Ashlee and Ajmani, Leah and DeVito, Michael Ann and Chancellor, Stevie},
	year = 2023,
	month = apr,
	pages = {1--17},
	publisher = {ACM},
	address = {Hamburg Germany},
	doi = {10.1145/3544548.3581489},
	urldate = {2024-09-13},
	isbn = {978-1-4503-9421-5},
	langid = {english}
}

@inproceedings{mizrahiDigitalGastronomyMethods2016,
	title = {Digital {{Gastronomy}}: {{Methods}} \& {{Recipes}} for {{Hybrid Cooking}}},
	shorttitle = {Digital {{Gastronomy}}},
	booktitle = {Proceedings of the 29th {{Annual Symposium}} on {{User Interface Software}} and {{Technology}}},
	author = {Mizrahi, Moran and Golan, Amos and Mizrahi, Ariel Bezaleli and Gruber, Rotem and Lachnise, Alexander Zoonder and Zoran, Amit},
	year = 2016,
	month = oct,
	pages = {541--552},
	publisher = {ACM},
	address = {Tokyo Japan},
	doi = {10.1145/2984511.2984528},
	urldate = {2024-09-13},
	isbn = {978-1-4503-4189-9},
	langid = {english}
}

@inproceedings{muellerJoggingDistanceEurope2010,
	title = {Jogging over a Distance between {{Europe}} and {{Australia}}},
	booktitle = {Proceedings of the 23nd Annual {{ACM}} Symposium on {{User}} Interface Software and Technology},
	author = {Mueller, Florian and Vetere, Frank and Gibbs, Martin R. and Edge, Darren and Agamanolis, Stefan and Sheridan, Jennifer G.},
	year = 2010,
	month = oct,
	pages = {189--198},
	publisher = {ACM},
	address = {New York New York USA},
	doi = {10.1145/1866029.1866062},
	urldate = {2024-09-13},
	isbn = {978-1-4503-0271-5},
	langid = {english}
}

@inproceedings{muellerNextStepsHumanComputer2020,
	title = {Next {{Steps}} for {{Human-Computer Integration}}},
	booktitle = {Proceedings of the 2020 {{CHI Conference}} on {{Human Factors}} in {{Computing Systems}}},
	author = {Mueller, Florian Floyd and Lopes, Pedro and Strohmeier, Paul and Ju, Wendy and Seim, Caitlyn and Weigel, Martin and Nanayakkara, Suranga and Obrist, Marianna and Li, Zhuying and Delfa, Joseph and Nishida, Jun and Gerber, Elizabeth M. and Svanaes, Dag and Grudin, Jonathan and Greuter, Stefan and Kunze, Kai and Erickson, Thomas and Greenspan, Steven and Inami, Masahiko and Marshall, Joe and Reiterer, Harald and Wolf, Katrin and Meyer, Jochen and Schiphorst, Thecla and Wang, Dakuo and Maes, Pattie},
	year = 2020,
	month = apr,
	pages = {1--15},
	publisher = {ACM},
	address = {Honolulu HI USA},
	doi = {10.1145/3313831.3376242},
	urldate = {2024-09-13},
	isbn = {978-1-4503-6708-0},
	langid = {english},
}

@inproceedings{mullenbachExploringAffectiveCommunication2014,
	title = {Exploring Affective Communication through Variable-Friction Surface Haptics},
	booktitle = {Proceedings of the {{SIGCHI Conference}} on {{Human Factors}} in {{Computing Systems}}},
	author = {Mullenbach, Joe and Shultz, Craig and Colgate, J. Edward and Piper, Anne Marie},
	year = 2014,
	month = apr,
	pages = {3963--3972},
	publisher = {ACM},
	address = {Toronto Ontario Canada},
	doi = {10.1145/2556288.2557343},
	urldate = {2024-09-13},
	isbn = {978-1-4503-2473-1},
	langid = {english}
}

@inproceedings{mullerLookingGlassField2012,
	title = {Looking Glass: A Field Study on Noticing Interactivity of a Shop Window},
	shorttitle = {Looking Glass},
	booktitle = {Proceedings of the {{SIGCHI Conference}} on {{Human Factors}} in {{Computing Systems}}},
	author = {M{\"u}ller, J{\"o}rg and Walter, Robert and Bailly, Gilles and Nischt, Michael and Alt, Florian},
	year = 2012,
	month = may,
	pages = {297--306},
	publisher = {ACM},
	address = {Austin Texas USA},
	doi = {10.1145/2207676.2207718},
	urldate = {2024-09-13},
	isbn = {978-1-4503-1015-4},
	langid = {english}
}

@article{mulliganThisThingCalled2019,
	title = {This {{Thing Called Fairness}}: {{Disciplinary Confusion Realizing}} a {{Value}} in {{Technology}}},
	shorttitle = {This {{Thing Called Fairness}}},
	author = {Mulligan, Deirdre K. and Kroll, Joshua A. and Kohli, Nitin and Wong, Richmond Y.},
	year = 2019,
	month = nov,
	journal = {Proceedings of the ACM on Human-Computer Interaction},
	volume = 3,
	number = {CSCW},
	pages = {1--36},
	issn = {2573-0142},
	doi = {10.1145/3359221},
	urldate = {2024-09-13},
	langid = {english},
}

@inproceedings{nakagakiChainFORMLinearIntegrated2016,
	title = {{{ChainFORM}}: {{A Linear Integrated Modular Hardware System}} for {{Shape Changing Interfaces}}},
	shorttitle = {{ChainFORM}},
	booktitle = {Proceedings of the 29th {{Annual Symposium}} on {{User Interface Software}} and {{Technology}}},
	author = {Nakagaki, Ken and Dementyev, Artem and Follmer, Sean and Paradiso, Joseph A. and Ishii, Hiroshi},
	year = 2016,
	month = oct,
	pages = {87--96},
	publisher = {ACM},
	address = {Tokyo Japan},
	doi = {10.1145/2984511.2984587},
	urldate = {2024-09-13},
	isbn = {978-1-4503-4189-9},
	langid = {english}
}

@inproceedings{nakajimaReflectingHumanBehavior2008,
	title = {Reflecting Human Behavior to Motivate Desirable Lifestyle},
	booktitle = {Proceedings of the 7th {{ACM}} Conference on {{Designing}} Interactive Systems},
	author = {Nakajima, Tatsuo and Lehdonvirta, Vili and Tokunaga, Eiji and Kimura, Hiroaki},
	year = 2008,
	month = feb,
	pages = {405--414},
	publisher = {ACM},
	address = {Cape Town South Africa},
	doi = {10.1145/1394445.1394489},
	urldate = {2024-09-13},
	isbn = {978-1-60558-002-9},
	langid = {english}
}

@inproceedings{niiyamaPoimoPortableInflatable2020a,
	title = {Poimo: {{Portable}} and {{Inflatable Mobility Devices Customizable}} for {{Personal Physical Characteristics}}},
	shorttitle = {Poimo},
	booktitle = {Proceedings of the 33rd {{Annual ACM Symposium}} on {{User Interface Software}} and {{Technology}}},
	author = {Niiyama, Ryuma and Sato, Hiroki and Tsujimura, Kazzmasa and Narumi, Koya and Seong, Young Ah and Yamamura, Ryosuke and Kakehi, Yasuaki and Kawahara, Yoshihiro},
	year = 2020,
	month = oct,
	pages = {912--923},
	publisher = {ACM},
	address = {Virtual Event USA},
	doi = {10.1145/3379337.3415894},
	urldate = {2024-09-13},
	isbn = {978-1-4503-7514-6},
	langid = {english},
}

@inproceedings{nunesSharingDigitalPhotographs2008,
	title = {Sharing Digital Photographs in the Home through Physical Mementos, Souvenirs, and Keepsakes},
	booktitle = {Proceedings of the 7th {{ACM}} Conference on {{Designing}} Interactive Systems},
	author = {Nunes, Michael and Greenberg, Saul and Neustaedter, Carman},
	year = 2008,
	month = feb,
	pages = {250--260},
	publisher = {ACM},
	address = {Cape Town South Africa},
	doi = {10.1145/1394445.1394472},
	urldate = {2024-09-13},
	isbn = {978-1-60558-002-9},
	langid = {english}
}

@inproceedings{obristOpportunitiesOdorExperiences2014,
	title = {Opportunities for Odor: Experiences with Smell and Implications for Technology},
	shorttitle = {Opportunities for Odor},
	booktitle = {Proceedings of the {{SIGCHI Conference}} on {{Human Factors}} in {{Computing Systems}}},
	author = {Obrist, Marianna and Tuch, Alexandre N. and Hornbaek, Kasper},
	year = 2014,
	month = apr,
	pages = {2843--2852},
	publisher = {ACM},
	address = {Toronto Ontario Canada},
	doi = {10.1145/2556288.2557008},
	urldate = {2024-09-13},
	isbn = {978-1-4503-2473-1},
	langid = {english}
}

@inproceedings{oduorFrustrationsBenefitsMobile2016,
	title = {The {{Frustrations}} and {{Benefits}} of {{Mobile Device Usage}} in the {{Home}} When {{Co-Present}} with {{Family Members}}},
	booktitle = {Proceedings of the 2016 {{ACM Conference}} on {{Designing Interactive Systems}}},
	author = {Oduor, Erick and Neustaedter, Carman and Odom, William and Tang, Anthony and Moallem, Niala and Tory, Melanie and Irani, Pourang},
	year = 2016,
	month = jun,
	pages = {1315--1327},
	publisher = {ACM},
	address = {Brisbane QLD Australia},
	doi = {10.1145/2901790.2901809},
	urldate = {2024-09-13},
	isbn = {978-1-4503-4031-1},
	langid = {english},
}

@inproceedings{oelenCrowdsourcingScholarlyDiscourse2021,
	title = {Crowdsourcing {{Scholarly Discourse Annotations}}},
	booktitle = {26th {{International Conference}} on {{Intelligent User Interfaces}}},
	author = {Oelen, Allard and Stocker, Markus and Auer, S{\"o}ren},
	year = 2021,
	month = apr,
	pages = {464--474},
	publisher = {ACM},
	address = {College Station TX USA},
	doi = {10.1145/3397481.3450685},
	urldate = {2024-09-13},
	isbn = {978-1-4503-8017-1},
	langid = {english},
}

@article{oharaBlendedInteractionSpaces2011,
	title = {Blended Interaction Spaces for Distributed Team Collaboration},
	author = {O'hara, Kenton and Kjeldskov, Jesper and Paay, Jeni},
	year = 2011,
	month = apr,
	journal = {ACM Transactions on Computer-Human Interaction},
	volume = 18,
	number = 1,
	pages = {1--28},
	issn = {1073-0516, 1557-7325},
	doi = {10.1145/1959022.1959025},
	urldate = {2024-09-13},
	langid = {english},
}

@article{olesonTeachingInclusiveDesign2023,
	title = {Teaching {{Inclusive Design Skills}} with the {{CIDER Assumption Elicitation Technique}}},
	author = {Oleson, Alannah and Solomon, Meron and Perdriau, Christopher and Ko, Amy},
	year = 2023,
	month = feb,
	journal = {ACM Transactions on Computer-Human Interaction},
	volume = 30,
	number = 1,
	pages = {1--49},
	issn = {1073-0516, 1557-7325},
	doi = {10.1145/3549074},
	urldate = {2024-09-13},
	langid = {english},
}

@inproceedings{ouCilllia3DPrinted2016,
	title = {Cilllia: {{3D Printed Micro-Pillar Structures}} for {{Surface Texture}}, {{Actuation}} and {{Sensing}}},
	shorttitle = {Cilllia},
	booktitle = {Proceedings of the 2016 {{CHI Conference}} on {{Human Factors}} in {{Computing Systems}}},
	author = {Ou, Jifei and Dublon, Gershon and Cheng, Chin-Yi and Heibeck, Felix and Willis, Karl and Ishii, Hiroshi},
	year = 2016,
	month = may,
	pages = {5753--5764},
	publisher = {ACM},
	address = {San Jose California USA},
	doi = {10.1145/2858036.2858257},
	urldate = {2024-09-13},
	isbn = {978-1-4503-3362-7},
	langid = {english}
}

@inproceedings{parkManifestationDepressionLoneliness2015,
	title = {Manifestation of {{Depression}} and {{Loneliness}} on {{Social Networks}}: {{A Case Study}} of {{Young Adults}} on {{Facebook}}},
	shorttitle = {Manifestation of {{Depression}} and {{Loneliness}} on {{Social Networks}}},
	booktitle = {Proceedings of the 18th {{ACM Conference}} on {{Computer Supported Cooperative Work}} \& {{Social Computing}}},
	author = {Park, Sungkyu and Kim, Inyeop and Lee, Sang Won and Yoo, Jaehyun and Jeong, Bumseok and Cha, Meeyoung},
	year = 2015,
	month = feb,
	pages = {557--570},
	publisher = {ACM},
	address = {Vancouver BC Canada},
	doi = {10.1145/2675133.2675139},
	urldate = {2024-09-13},
	isbn = {978-1-4503-2922-4},
	langid = {english}
}

@inproceedings{parkMetaverseWorkspaceOpportunities2023,
	title = {Towards a {{Metaverse Workspace}}: {{Opportunities}}, {{Challenges}}, and {{Design Implications}}},
	shorttitle = {Towards a {{Metaverse Workspace}}},
	booktitle = {Proceedings of the 2023 {{CHI Conference}} on {{Human Factors}} in {{Computing Systems}}},
	author = {Park, Hyanghee and Ahn, Daehwan and Lee, Joonhwan},
	year = 2023,
	month = apr,
	pages = {1--20},
	publisher = {ACM},
	address = {Hamburg Germany},
	doi = {10.1145/3544548.3581306},
	urldate = {2024-09-13},
	isbn = {978-1-4503-9421-5},
	langid = {english},
}

@inproceedings{peiHandInterfacesUsing2022,
	title = {Hand {{Interfaces}}: {{Using Hands}} to {{Imitate Objects}} in {{AR}}/{{VR}} for {{Expressive Interactions}}},
	shorttitle = {Hand {{Interfaces}}},
	booktitle = {{{CHI Conference}} on {{Human Factors}} in {{Computing Systems}}},
	author = {Pei, Siyou and Chen, Alexander and Lee, Jaewook and Zhang, Yang},
	year = 2022,
	month = apr,
	pages = {1--16},
	publisher = {ACM},
	address = {New Orleans LA USA},
	doi = {10.1145/3491102.3501898},
	urldate = {2024-09-13},
	isbn = {978-1-4503-9157-3},
	langid = {english},
}

@inproceedings{perraultWatchitSimpleGestures2013,
	title = {Watchit: Simple Gestures and Eyes-Free Interaction for Wristwatches and Bracelets},
	shorttitle = {Watchit},
	booktitle = {Proceedings of the {{SIGCHI Conference}} on {{Human Factors}} in {{Computing Systems}}},
	author = {Perrault, Simon T. and Lecolinet, Eric and Eagan, James and Guiard, Yves},
	year = 2013,
	month = apr,
	pages = {1451--1460},
	publisher = {ACM},
	address = {Paris France},
	doi = {10.1145/2470654.2466192},
	urldate = {2024-09-13},
	isbn = {978-1-4503-1899-0},
	langid = {english},
}

@inproceedings{pfeifferCruiseControlPedestrians2015,
	title = {Cruise {{Control}} for {{Pedestrians}}: {{Controlling Walking Direction}} Using {{Electrical Muscle Stimulation}}},
	shorttitle = {Cruise {{Control}} for {{Pedestrians}}},
	booktitle = {Proceedings of the 33rd {{Annual ACM Conference}} on {{Human Factors}} in {{Computing Systems}}},
	author = {Pfeiffer, Max and D{\"u}nte, Tim and Schneegass, Stefan and Alt, Florian and Rohs, Michael},
	year = 2015,
	month = apr,
	pages = {2505--2514},
	publisher = {ACM},
	address = {Seoul Republic of Korea},
	doi = {10.1145/2702123.2702190},
	urldate = {2024-09-13},
	isbn = {978-1-4503-3145-6},
	langid = {english}
}

@inproceedings{pierceInterfaceUserExploratory2018,
	title = {An {{Interface}} without {{A User}}: {{An Exploratory Design Study}} of {{Online Privacy Policies}} and {{Digital Legalese}}},
	shorttitle = {An {{Interface}} without {{A User}}},
	booktitle = {Proceedings of the 2018 {{Designing Interactive Systems Conference}}},
	author = {Pierce, James and Fox, Sarah and Merrill, Nick and Wong, Richmond and DiSalvo, Carl},
	year = 2018,
	month = jun,
	pages = {1345--1358},
	publisher = {ACM},
	address = {Hong Kong China},
	doi = {10.1145/3196709.3196818},
	urldate = {2024-09-13},
	isbn = {978-1-4503-5198-0},
	langid = {english},
}

@inproceedings{porfirioAuthoringVerifyingHumanRobot2018,
	title = {Authoring and {{Verifying Human-Robot Interactions}}},
	booktitle = {Proceedings of the 31st {{Annual ACM Symposium}} on {{User Interface Software}} and {{Technology}}},
	author = {Porfirio, David and Saupp{\'e}, Allison and Albarghouthi, Aws and Mutlu, Bilge},
	year = 2018,
	month = oct,
	pages = {75--86},
	publisher = {ACM},
	address = {Berlin Germany},
	doi = {10.1145/3242587.3242634},
	urldate = {2024-09-13},
	isbn = {978-1-4503-5948-1},
	langid = {english},
}

@inproceedings{prakashAutoDescFacilitatingConvenient2023,
	title = {{{AutoDesc}}: {{Facilitating Convenient Perusal}} of {{Web Data Items}} for {{Blind Users}}},
	shorttitle = {{AutoDesc}},
	booktitle = {Proceedings of the 28th {{International Conference}} on {{Intelligent User Interfaces}}},
	author = {Prakash, Yash and Sunkara, Mohan and Lee, Hae-Na and Jayarathna, Sampath and Ashok, Vikas},
	year = 2023,
	month = mar,
	pages = {32--45},
	publisher = {ACM},
	address = {Sydney NSW Australia},
	doi = {10.1145/3581641.3584049},
	urldate = {2024-09-13},
	isbn = 9798400701061,
	langid = {english},
}

@inproceedings{qianInferringMotionDirection2017a,
	title = {Inferring {{Motion Direction}} Using {{Commodity Wi-Fi}} for {{Interactive Exergames}}},
	booktitle = {Proceedings of the 2017 {{CHI Conference}} on {{Human Factors}} in {{Computing Systems}}},
	author = {Qian, Kun and Wu, Chenshu and Zhou, Zimu and Zheng, Yue and Yang, Zheng and Liu, Yunhao},
	year = 2017,
	month = may,
	pages = {1961--1972},
	publisher = {ACM},
	address = {Denver Colorado USA},
	doi = {10.1145/3025453.3025678},
	urldate = {2024-09-13},
	isbn = {978-1-4503-4655-9},
	langid = {english}
}

@inproceedings{qianScalARAuthoringSemantically2022,
	title = {{{ScalAR}}: {{Authoring Semantically Adaptive Augmented Reality Experiences}} in {{Virtual Reality}}},
	shorttitle = {{ScalAR}},
	booktitle = {{{CHI Conference}} on {{Human Factors}} in {{Computing Systems}}},
	author = {Qian, Xun and He, Fengming and Hu, Xiyun and Wang, Tianyi and Ipsita, Ananya and Ramani, Karthik},
	year = 2022,
	month = apr,
	pages = {1--18},
	publisher = {ACM},
	address = {New Orleans LA USA},
	doi = {10.1145/3491102.3517665},
	urldate = {2024-09-13},
	copyright = {http://www.acm.org/publications/policies/copyright\_policy\#Background},
	isbn = {978-1-4503-9157-3},
	langid = {english},
}

@inproceedings{rajanTaskLoadEstimation2016,
	title = {Task {{Load Estimation}} and {{Mediation Using Psycho-physiological Measures}}},
	booktitle = {Proceedings of the 21st {{International Conference}} on {{Intelligent User Interfaces}}},
	author = {Rajan, Rahul and Selker, Ted and Lane, Ian},
	year = 2016,
	month = mar,
	pages = {48--59},
	publisher = {ACM},
	address = {Sonoma California USA},
	doi = {10.1145/2856767.2856769},
	urldate = {2024-09-13},
	isbn = {978-1-4503-4137-0},
	langid = {english}
}

@article{rappExploringLivedExperience2023,
	title = {Exploring the {{Lived Experience}} of {{Behavior Change Technologies}}: {{Towards}} an {{Existential Model}} of {{Behavior Change}} for {{HCI}}},
	shorttitle = {Exploring the {{Lived Experience}} of {{Behavior Change Technologies}}},
	author = {Rapp, Amon and Boldi, Arianna},
	year = 2023,
	month = dec,
	journal = {ACM Transactions on Computer-Human Interaction},
	volume = 30,
	number = 6,
	pages = {1--50},
	issn = {1073-0516, 1557-7325},
	doi = {10.1145/3603497},
	urldate = {2024-09-13},
	langid = {english},
}

@article{reichertsItsGoodTalk2022,
	title = {It's {{Good}} to {{Talk}}: {{A Comparison}} of {{Using Voice Versus Screen-Based Interactions}} for {{Agent-Assisted Tasks}}},
	shorttitle = {It's {{Good}} to {{Talk}}},
	author = {Reicherts, Leon and Rogers, Yvonne and Capra, Licia and Wood, Ethan and Duong, Tu Dinh and Sebire, Neil},
	year = 2022,
	month = jun,
	journal = {ACM Transactions on Computer-Human Interaction},
	volume = 29,
	number = 3,
	pages = {1--41},
	issn = {1073-0516, 1557-7325},
	doi = {10.1145/3484221},
	urldate = {2024-09-13},
	langid = {english},
}

@inproceedings{retelnyExpertCrowdsourcingFlash2014,
	title = {Expert Crowdsourcing with Flash Teams},
	booktitle = {Proceedings of the 27th Annual {{ACM}} Symposium on {{User}} Interface Software and Technology},
	author = {Retelny, Daniela and Robaszkiewicz, S{\'e}bastien and To, Alexandra and Lasecki, Walter S. and Patel, Jay and Rahmati, Negar and Doshi, Tulsee and Valentine, Melissa and Bernstein, Michael S.},
	year = 2014,
	month = oct,
	pages = {75--85},
	publisher = {ACM},
	address = {Honolulu Hawaii USA},
	doi = {10.1145/2642918.2647409},
	urldate = {2024-09-13},
	isbn = {978-1-4503-3069-5},
	langid = {english}
}

@inproceedings{richardsmaldonadoReaderQuizzerAugmentingResearch2023a,
	title = {{{ReaderQuizzer}}: {{Augmenting Research Papers}} with {{Just-In-Time Learning Questions}} to {{Facilitate Deeper Understanding}}},
	shorttitle = {{ReaderQuizzer}},
	booktitle = {Computer {{Supported Cooperative Work}} and {{Social Computing}}},
	author = {Richards Maldonado, Liam and Abouzied, Azza and Gleason, Nancy W.},
	year = 2023,
	month = oct,
	pages = {391--394},
	publisher = {ACM},
	address = {Minneapolis MN USA},
	doi = {10.1145/3584931.3607494},
	urldate = {2024-09-13},
	isbn = 9798400701290,
	langid = {english}
}

@article{roffarelloAchievingDigitalWellbeing2023,
	title = {Achieving {{Digital Wellbeing Through Digital Self-control Tools}}: {{A Systematic Review}} and {{Meta-analysis}}},
	shorttitle = {Achieving {{Digital Wellbeing Through Digital Self-control Tools}}},
	author = {Roffarello, Alberto Monge and De Russis, Luigi},
	year = 2023,
	month = aug,
	journal = {ACM Transactions on Computer-Human Interaction},
	volume = 30,
	number = 4,
	pages = {1--66},
	issn = {1073-0516, 1557-7325},
	doi = {10.1145/3571810},
	urldate = {2024-09-13},
	langid = {english},
}

@article{ruanComparingSpeechKeyboard2018,
	title = {Comparing {{Speech}} and {{Keyboard Text Entry}} for {{Short Messages}} in {{Two Languages}} on {{Touchscreen Phones}}},
	author = {Ruan, Sherry and Wobbrock, Jacob O. and Liou, Kenny and Ng, Andrew and Landay, James A.},
	year = 2018,
	month = jan,
	journal = {Proceedings of the ACM on Interactive, Mobile, Wearable and Ubiquitous Technologies},
	volume = 1,
	number = 4,
	pages = {1--23},
	issn = {2474-9567},
	doi = {10.1145/3161187},
	urldate = {2024-09-13},
	langid = {english},
}

@article{sahaLanguageLGBTQMinority2019,
	title = {The {{Language}} of {{LGBTQ}}+ {{Minority Stress Experiences}} on {{Social Media}}},
	author = {Saha, Koustuv and Kim, Sang Chan and Reddy, Manikanta D. and Carter, Albert J. and Sharma, Eva and Haimson, Oliver L. and De Choudhury, Munmun},
	year = 2019,
	month = nov,
	journal = {Proceedings of the ACM on Human-Computer Interaction},
	volume = 3,
	number = {CSCW},
	pages = {1--22},
	issn = {2573-0142},
	doi = {10.1145/3361108},
	urldate = {2024-09-13},
	langid = {english},
}

@inproceedings{sanchesMindBodyDesigning2010a,
	title = {Mind the Body!: Designing a Mobile Stress Management Application Encouraging Personal Reflection},
	shorttitle = {Mind the Body!},
	booktitle = {Proceedings of the 8th {{ACM Conference}} on {{Designing Interactive Systems}}},
	author = {Sanches, Pedro and H{\"o}{\"o}k, Kristina and Vaara, Elsa and Weymann, Claus and Bylund, Markus and Ferreira, Pedro and Peira, Nathalie and Sj{\"o}linder, Marie},
	year = 2010,
	month = aug,
	pages = {47--56},
	publisher = {ACM},
	address = {Aarhus Denmark},
	doi = {10.1145/1858171.1858182},
	urldate = {2024-09-13},
	isbn = {978-1-4503-0103-9},
	langid = {english}
}

@article{satriadiMapsMe3D2020,
	title = {Maps {{Around Me}}: {{3D Multiview Layouts}} in {{Immersive Spaces}}},
	shorttitle = {Maps {{Around Me}}},
	author = {Satriadi, Kadek Ananta and Ens, Barrett and Cordeil, Maxime and Czauderna, Tobias and Jenny, Bernhard},
	year = 2020,
	month = nov,
	journal = {Proceedings of the ACM on Human-Computer Interaction},
	volume = 4,
	number = {ISS},
	pages = {1--20},
	issn = {2573-0142},
	doi = {10.1145/3427329},
	urldate = {2024-09-13},
	langid = {english}
}

@article{scheuermanHowComputersSee2019,
	title = {How {{Computers See Gender}}: {{An Evaluation}} of {{Gender Classification}} in {{Commercial Facial Analysis Services}}},
	shorttitle = {How {{Computers See Gender}}},
	author = {Scheuerman, Morgan Klaus and Paul, Jacob M. and Brubaker, Jed R.},
	year = 2019,
	month = nov,
	journal = {Proceedings of the ACM on Human-Computer Interaction},
	volume = 3,
	number = {CSCW},
	pages = {1--33},
	issn = {2573-0142},
	doi = {10.1145/3359246},
	urldate = {2024-09-13},
	langid = {english},
}

@article{scheuermanHowWeveTaught2020,
	title = {How {{We}}'ve {{Taught Algorithms}} to {{See Identity}}: {{Constructing Race}} and {{Gender}} in {{Image Databases}} for {{Facial Analysis}}},
	shorttitle = {How {{We}}'ve {{Taught Algorithms}} to {{See Identity}}},
	author = {Scheuerman, Morgan Klaus and Wade, Kandrea and Lustig, Caitlin and Brubaker, Jed R.},
	year = 2020,
	month = may,
	journal = {Proceedings of the ACM on Human-Computer Interaction},
	volume = 4,
	number = {CSCW1},
	pages = {1--35},
	issn = {2573-0142},
	doi = {10.1145/3392866},
	urldate = {2024-09-13},
	langid = {english}
}

@article{scheuermanSafeSpacesSafe2018a,
	title = {Safe {{Spaces}} and {{Safe Places}}: {{Unpacking Technology-Mediated Experiences}} of {{Safety}} and {{Harm}} with {{Transgender People}}},
	shorttitle = {Safe {{Spaces}} and {{Safe Places}}},
	author = {Scheuerman, Morgan Klaus and Branham, Stacy M. and Hamidi, Foad},
	year = 2018,
	month = nov,
	journal = {Proceedings of the ACM on Human-Computer Interaction},
	volume = 2,
	number = {CSCW},
	pages = {1--27},
	issn = {2573-0142},
	doi = {10.1145/3274424},
	urldate = {2024-09-13},
	langid = {english}
}

@inproceedings{schlesingerLetsTalkRace2018,
	title = {Let's {{Talk About Race}}: {{Identity}}, {{Chatbots}}, and {{AI}}},
	shorttitle = {Let's {{Talk About Race}}},
	booktitle = {Proceedings of the 2018 {{CHI Conference}} on {{Human Factors}} in {{Computing Systems}}},
	author = {Schlesinger, Ari and O'Hara, Kenton P. and Taylor, Alex S.},
	year = 2018,
	month = apr,
	pages = {1--14},
	publisher = {ACM},
	address = {Montreal QC Canada},
	doi = {10.1145/3173574.3173889},
	urldate = {2024-09-13},
	isbn = {978-1-4503-5620-6},
	langid = {english},
}

@inproceedings{schmidtCrossdeviceInteractionStyle2012a,
	title = {A Cross-Device Interaction Style for Mobiles and Surfaces},
	booktitle = {Proceedings of the {{Designing Interactive Systems Conference}}},
	author = {Schmidt, Dominik and Seifert, Julian and Rukzio, Enrico and Gellersen, Hans},
	year = 2012,
	month = jun,
	pages = {318--327},
	publisher = {ACM},
	address = {Newcastle Upon Tyne United Kingdom},
	doi = {10.1145/2317956.2318005},
	urldate = {2024-09-13},
	isbn = {978-1-4503-1210-3},
	langid = {english},
}

@inproceedings{schoenebeckGivingTwitterLent2014,
	title = {Giving up {{Twitter}} for {{Lent}}: How and Why We Take Breaks from Social Media},
	shorttitle = {Giving up {{Twitter}} for {{Lent}}},
	booktitle = {Proceedings of the {{SIGCHI Conference}} on {{Human Factors}} in {{Computing Systems}}},
	author = {Schoenebeck, Sarita Yardi},
	year = 2014,
	month = apr,
	pages = {773--782},
	publisher = {ACM},
	address = {Toronto Ontario Canada},
	doi = {10.1145/2556288.2556983},
	urldate = {2024-09-13},
	isbn = {978-1-4503-2473-1},
	langid = {english}
}

@article{schrillsHowUsersExperience2023,
	title = {How {{Do Users Experience Traceability}} of {{AI Systems}}? {{Examining Subjective Information Processing Awareness}} in {{Automated Insulin Delivery}} ({{AID}}) {{Systems}}},
	shorttitle = {How {{Do Users Experience Traceability}} of {{AI Systems}}?},
	author = {Schrills, Tim and Franke, Thomas},
	year = 2023,
	month = dec,
	journal = {ACM Transactions on Interactive Intelligent Systems},
	volume = 13,
	number = 4,
	pages = {1--34},
	issn = {2160-6455, 2160-6463},
	doi = {10.1145/3588594},
	urldate = {2024-09-13},
	langid = {english},
}

@inproceedings{schroederPocketSkillsConversational2018,
	title = {Pocket {{Skills}}: {{A Conversational Mobile Web App To Support Dialectical Behavioral Therapy}}},
	shorttitle = {Pocket {{Skills}}},
	booktitle = {Proceedings of the 2018 {{CHI Conference}} on {{Human Factors}} in {{Computing Systems}}},
	author = {Schroeder, Jessica and Wilkes, Chelsey and Rowan, Kael and Toledo, Arturo and Paradiso, Ann and Czerwinski, Mary and Mark, Gloria and Linehan, Marsha M.},
	year = 2018,
	month = apr,
	pages = {1--15},
	publisher = {ACM},
	address = {Montreal QC Canada},
	doi = {10.1145/3173574.3173972},
	urldate = {2024-09-13},
	isbn = {978-1-4503-5620-6},
	langid = {english}
}

@article{schweikardtSUSTAINABLYOURSUserCentered2009,
	title = {{{SUSTAINABLY OURSUser}} Centered Is off Center},
	author = {Schweikardt, Eric},
	year = 2009,
	month = may,
	journal = {Interactions},
	volume = 16,
	number = 3,
	pages = {12--15},
	issn = {1072-5520, 1558-3449},
	doi = {10.1145/1516016.1516019},
	urldate = {2024-09-13},
	langid = {english}
}

@article{seboRobotsGroupsTeams2020,
	title = {Robots in {{Groups}} and {{Teams}}: {{A Literature Review}}},
	shorttitle = {Robots in {{Groups}} and {{Teams}}},
	author = {Sebo, Sarah and Stoll, Brett and Scassellati, Brian and Jung, Malte F.},
	year = 2020,
	month = oct,
	journal = {Proceedings of the ACM on Human-Computer Interaction},
	volume = 4,
	number = {CSCW2},
	pages = {1--36},
	issn = {2573-0142},
	doi = {10.1145/3415247},
	urldate = {2024-09-13},
	langid = {english},
}

@inproceedings{shiMarkitTalkitLowBarrier2017,
	title = {Markit and {{Talkit}}: {{A Low-Barrier Toolkit}} to {{Augment 3D Printed Models}} with {{Audio Annotations}}},
	shorttitle = {Markit and {{Talkit}}},
	booktitle = {Proceedings of the 30th {{Annual ACM Symposium}} on {{User Interface Software}} and {{Technology}}},
	author = {Shi, Lei and Zhao, Yuhang and Azenkot, Shiri},
	year = 2017,
	month = oct,
	pages = {493--506},
	publisher = {ACM},
	address = {Qu{\'e}bec City QC Canada},
	doi = {10.1145/3126594.3126650},
	urldate = {2024-09-13},
	isbn = {978-1-4503-4981-9},
	langid = {english}
}

@inproceedings{silvaVehicleDriverRecognition2012a,
	title = {In-Vehicle Driver Recognition Based on Hand {{ECG}} Signals},
	booktitle = {Proceedings of the 2012 {{ACM}} International Conference on {{Intelligent User Interfaces}}},
	author = {Silva, Hugo and Louren{\c c}o, Andr{\'e} and Fred, Ana},
	year = 2012,
	month = feb,
	pages = {25--28},
	publisher = {ACM},
	address = {Lisbon Portugal},
	doi = {10.1145/2166966.2166971},
	urldate = {2024-09-13},
	isbn = {978-1-4503-1048-2},
	langid = {english}
}

@article{sodenInformatingCrisisExpanding2018,
	title = {Informating {{Crisis}}: {{Expanding Critical Perspectives}} in {{Crisis Informatics}}},
	shorttitle = {Informating {{Crisis}}},
	author = {Soden, Robert and Palen, Leysia},
	year = 2018,
	month = nov,
	journal = {Proceedings of the ACM on Human-Computer Interaction},
	volume = 2,
	number = {CSCW},
	pages = {1--22},
	issn = {2573-0142},
	doi = {10.1145/3274431},
	urldate = {2024-09-13},
	langid = {english},
}

@inproceedings{sohnDiaryStudyMobile2008,
	title = {A Diary Study of Mobile Information Needs},
	booktitle = {Proceedings of the {{SIGCHI Conference}} on {{Human Factors}} in {{Computing Systems}}},
	author = {Sohn, Timothy and Li, Kevin A. and Griswold, William G. and Hollan, James D.},
	year = 2008,
	month = apr,
	pages = {433--442},
	publisher = {ACM},
	address = {Florence Italy},
	doi = {10.1145/1357054.1357125},
	urldate = {2024-09-13},
	isbn = {978-1-60558-011-1},
	langid = {english},
}

@inproceedings{steichenUseradaptiveInformationVisualization2013,
	title = {User-Adaptive Information Visualization: Using Eye Gaze Data to Infer Visualization Tasks and User Cognitive Abilities},
	shorttitle = {User-Adaptive Information Visualization},
	booktitle = {Proceedings of the 2013 International Conference on {{Intelligent}} User Interfaces},
	author = {Steichen, Ben and Carenini, Giuseppe and Conati, Cristina},
	year = 2013,
	month = mar,
	pages = {317--328},
	publisher = {ACM},
	address = {Santa Monica California USA},
	doi = {10.1145/2449396.2449439},
	urldate = {2024-09-13},
	isbn = {978-1-4503-1965-2},
	langid = {english}
}

@inproceedings{steimleFlexpadHighlyFlexible2013,
	title = {Flexpad: Highly Flexible Bending Interactions for Projected Handheld Displays},
	shorttitle = {Flexpad},
	booktitle = {Proceedings of the {{SIGCHI Conference}} on {{Human Factors}} in {{Computing Systems}}},
	author = {Steimle, J{\"u}rgen and Jordt, Andreas and Maes, Pattie},
	year = 2013,
	month = apr,
	pages = {237--246},
	publisher = {ACM},
	address = {Paris France},
	doi = {10.1145/2470654.2470688},
	urldate = {2024-09-13},
	isbn = {978-1-4503-1899-0},
	langid = {english}
}

@inproceedings{stewartCharacteristicsPressurebasedInput2010a,
	title = {Characteristics of Pressure-Based Input for Mobile Devices},
	booktitle = {Proceedings of the {{SIGCHI Conference}} on {{Human Factors}} in {{Computing Systems}}},
	author = {Stewart, Craig and Rohs, Michael and Kratz, Sven and Essl, Georg},
	year = 2010,
	month = apr,
	pages = {801--810},
	publisher = {ACM},
	address = {Atlanta Georgia USA},
	doi = {10.1145/1753326.1753444},
	urldate = {2024-09-13},
	isbn = {978-1-60558-929-9},
	langid = {english}
}

@inproceedings{streliHOOVHandOutView2023,
	title = {{{HOOV}}: {{Hand Out-Of-View Tracking}} for {{Proprioceptive Interaction}} Using {{Inertial Sensing}}},
	shorttitle = {{HOOV}},
	booktitle = {Proceedings of the 2023 {{CHI Conference}} on {{Human Factors}} in {{Computing Systems}}},
	author = {Streli, Paul and Armani, Rayan and Cheng, Yi Fei and Holz, Christian},
	year = 2023,
	month = apr,
	pages = {1--16},
	publisher = {ACM},
	address = {Hamburg Germany},
	doi = {10.1145/3544548.3581468},
	urldate = {2024-09-13},
	isbn = {978-1-4503-9421-5},
	langid = {english},
}

@inproceedings{strohmayerTechnologiesSocialJustice2019,
	title = {Technologies for {{Social Justice}}: {{Lessons}} from {{Sex Workers}} on the {{Front Lines}}},
	shorttitle = {Technologies for {{Social Justice}}},
	booktitle = {Proceedings of the 2019 {{CHI Conference}} on {{Human Factors}} in {{Computing Systems}}},
	author = {Strohmayer, Angelika and Clamen, Jenn and Laing, Mary},
	year = 2019,
	month = may,
	pages = {1--14},
	publisher = {ACM},
	address = {Glasgow Scotland Uk},
	doi = {10.1145/3290605.3300882},
	urldate = {2024-09-13},
	isbn = {978-1-4503-5970-2},
	langid = {english},
}

@inproceedings{suDesigningNomadicWork2008a,
	title = {Designing for Nomadic Work},
	booktitle = {Proceedings of the 7th {{ACM}} Conference on {{Designing}} Interactive Systems},
	author = {Su, Norman Makoto and Mark, Gloria},
	year = 2008,
	month = feb,
	pages = {305--314},
	publisher = {ACM},
	address = {Cape Town South Africa},
	doi = {10.1145/1394445.1394478},
	urldate = {2024-09-13},
	isbn = {978-1-60558-002-9},
	langid = {english}
}

@inproceedings{suhAISocialGlue2021,
	title = {{{AI}} as {{Social Glue}}: {{Uncovering}} the {{Roles}} of {{Deep Generative AI}} during {{Social Music Composition}}},
	shorttitle = {{{AI}} as {{Social Glue}}},
	booktitle = {Proceedings of the 2021 {{CHI Conference}} on {{Human Factors}} in {{Computing Systems}}},
	author = {Suh, Minhyang (Mia) and Youngblom, Emily and Terry, Michael and Cai, Carrie J},
	year = 2021,
	month = may,
	pages = {1--11},
	publisher = {ACM},
	address = {Yokohama Japan},
	doi = {10.1145/3411764.3445219},
	urldate = {2024-09-13},
	isbn = {978-1-4503-8096-6},
	langid = {english},
}

@inproceedings{suhSensecapeEnablingMultilevel2023,
	title = {Sensecape: {{Enabling Multilevel Exploration}} and {{Sensemaking}} with {{Large Language Models}}},
	shorttitle = {Sensecape},
	booktitle = {Proceedings of the 36th {{Annual ACM Symposium}} on {{User Interface Software}} and {{Technology}}},
	author = {Suh, Sangho and Min, Bryan and Palani, Srishti and Xia, Haijun},
	year = 2023,
	month = oct,
	pages = {1--18},
	publisher = {ACM},
	address = {San Francisco CA USA},
	doi = {10.1145/3586183.3606756},
	urldate = {2024-09-13},
	isbn = 9798400701320,
	langid = {english},
}

@inproceedings{sunInvestigatingExplainabilityGenerative2022,
	title = {Investigating {{Explainability}} of {{Generative AI}} for {{Code}} through {{Scenario-based Design}}},
	booktitle = {27th {{International Conference}} on {{Intelligent User Interfaces}}},
	author = {Sun, Jiao and Liao, Q. Vera and Muller, Michael and Agarwal, Mayank and Houde, Stephanie and Talamadupula, Kartik and Weisz, Justin D.},
	year = 2022,
	month = mar,
	pages = {212--228},
	publisher = {ACM},
	address = {Helsinki Finland},
	doi = {10.1145/3490099.3511119},
	urldate = {2024-09-13},
	isbn = {978-1-4503-9144-3},
	langid = {english},
}

@inproceedings{suraleTabletInVRExploringDesign2019a,
	title = {{{TabletInVR}}: {{Exploring}} the {{Design Space}} for {{Using}} a {{Multi-Touch Tablet}} in {{Virtual Reality}}},
	shorttitle = {{TabletInVR}},
	booktitle = {Proceedings of the 2019 {{CHI Conference}} on {{Human Factors}} in {{Computing Systems}}},
	author = {Surale, Hemant Bhaskar and Gupta, Aakar and Hancock, Mark and Vogel, Daniel},
	year = 2019,
	month = may,
	pages = {1--13},
	publisher = {ACM},
	address = {Glasgow Scotland Uk},
	doi = {10.1145/3290605.3300243},
	urldate = {2024-09-13},
	isbn = {978-1-4503-5970-2},
	langid = {english}
}

@article{sutherlandGigEconomyInformation2017,
	title = {The {{Gig Economy}} and {{Information Infrastructure}}: {{The Case}} of the {{Digital Nomad Community}}},
	shorttitle = {The {{Gig Economy}} and {{Information Infrastructure}}},
	author = {Sutherland, Will and Jarrahi, Mohammad Hossein},
	year = 2017,
	month = dec,
	journal = {Proceedings of the ACM on Human-Computer Interaction},
	volume = 1,
	number = {CSCW},
	pages = {1--24},
	issn = {2573-0142},
	doi = {10.1145/3134732},
	urldate = {2024-09-13},
	langid = {english}
}

@inproceedings{suzukiAugmentedRealityRobotics2022,
	title = {Augmented {{Reality}} and {{Robotics}}: {{A Survey}} and {{Taxonomy}} for {{AR-enhanced Human-Robot Interaction}} and {{Robotic Interfaces}}},
	shorttitle = {Augmented {{Reality}} and {{Robotics}}},
	booktitle = {{{CHI Conference}} on {{Human Factors}} in {{Computing Systems}}},
	author = {Suzuki, Ryo and Karim, Adnan and Xia, Tian and Hedayati, Hooman and Marquardt, Nicolai},
	year = 2022,
	month = apr,
	pages = {1--33},
	publisher = {ACM},
	address = {New Orleans LA USA},
	doi = {10.1145/3491102.3517719},
	urldate = {2024-09-13},
	isbn = {978-1-4503-9157-3},
	langid = {english},
}

@inproceedings{suzukiShapeBotsShapechangingSwarm2019,
	title = {{{ShapeBots}}: {{Shape-changing Swarm Robots}}},
	shorttitle = {{ShapeBots}},
	booktitle = {Proceedings of the 32nd {{Annual ACM Symposium}} on {{User Interface Software}} and {{Technology}}},
	author = {Suzuki, Ryo and Zheng, Clement and Kakehi, Yasuaki and Yeh, Tom and Do, Ellen Yi-Luen and Gross, Mark D. and Leithinger, Daniel},
	year = 2019,
	month = oct,
	pages = {493--505},
	publisher = {ACM},
	address = {New Orleans LA USA},
	doi = {10.1145/3332165.3347911},
	urldate = {2024-09-13},
	isbn = {978-1-4503-6816-2},
	langid = {english},
}

@inproceedings{teibrichPatchingPhysicalObjects2015,
	title = {Patching {{Physical Objects}}},
	booktitle = {Proceedings of the 28th {{Annual ACM Symposium}} on {{User Interface Software}} \& {{Technology}}},
	author = {Teibrich, Alexander and Mueller, Stefanie and Guimbreti{\`e}re, Fran{\c c}ois and Kovacs, Robert and Neubert, Stefan and Baudisch, Patrick},
	year = 2015,
	month = nov,
	pages = {83--91},
	publisher = {ACM},
	address = {Charlotte NC USA},
	doi = {10.1145/2807442.2807467},
	urldate = {2024-09-13},
	isbn = {978-1-4503-3779-3},
	langid = {english}
}

@inproceedings{tolmieThisHasBe2016,
	title = {``{{This}} Has to Be the Cats'': {{Personal Data Legibility}} in {{Networked Sensing Systems}}},
	shorttitle = {``{{This}} Has to Be the Cats''},
	booktitle = {Proceedings of the 19th {{ACM Conference}} on {{Computer-Supported Cooperative Work}} \& {{Social Computing}}},
	author = {Tolmie, Peter and Crabtree, Andy and Rodden, Tom and Colley, James and Luger, Ewa},
	year = 2016,
	month = feb,
	pages = {491--502},
	publisher = {ACM},
	address = {San Francisco California USA},
	doi = {10.1145/2818048.2819992},
	urldate = {2024-09-13},
	isbn = {978-1-4503-3592-8},
	langid = {english},
}

@article{toTheyJustDont2020,
	title = {"{{They Just Don}}'t {{Get It}}": {{Towards Social Technologies}} for {{Coping}} with {{Interpersonal Racism}}},
	shorttitle = {"{{They Just Don}}'t {{Get It}}"},
	author = {To, Alexandra and Sweeney, Wenxia and Hammer, Jessica and Kaufman, Geoff},
	year = 2020,
	month = may,
	journal = {Proceedings of the ACM on Human-Computer Interaction},
	volume = 4,
	number = {CSCW1},
	pages = {1--29},
	issn = {2573-0142},
	doi = {10.1145/3392828},
	urldate = {2024-09-13},
	langid = {english},
}

@inproceedings{trajkovaAlexaToyExploring2020,
	title = {"{{Alexa}} Is a {{Toy}}": {{Exploring Older Adults}}' {{Reasons}} for {{Using}}, {{Limiting}}, and {{Abandoning Echo}}},
	shorttitle = {"{{Alexa}} Is a {{Toy}}"},
	booktitle = {Proceedings of the 2020 {{CHI Conference}} on {{Human Factors}} in {{Computing Systems}}},
	author = {Trajkova, Milka and {Martin-Hammond}, Aqueasha},
	year = 2020,
	month = apr,
	pages = {1--13},
	publisher = {ACM},
	address = {Honolulu HI USA},
	doi = {10.1145/3313831.3376760},
	urldate = {2024-09-13},
	isbn = {978-1-4503-6708-0},
	langid = {english},
}

@article{trujilloMakeRedditGreat2022,
	title = {Make {{Reddit Great Again}}: {{Assessing Community Effects}} of {{Moderation Interventions}} on r/{{The}}\_{{Donald}}},
	shorttitle = {Make {{Reddit Great Again}}},
	author = {Trujillo, Amaury and Cresci, Stefano},
	year = 2022,
	month = nov,
	journal = {Proceedings of the ACM on Human-Computer Interaction},
	volume = 6,
	number = {CSCW2},
	pages = {1--28},
	issn = {2573-0142},
	doi = {10.1145/3555639},
	urldate = {2024-09-13},
	langid = {english},
}

@inproceedings{tsaknakiExpandingWabiSabiDesign2016a,
	title = {Expanding on {{Wabi-Sabi}} as a {{Design Resource}} in {{HCI}}},
	booktitle = {Proceedings of the 2016 {{CHI Conference}} on {{Human Factors}} in {{Computing Systems}}},
	author = {Tsaknaki, Vasiliki and Fernaeus, Ylva},
	year = 2016,
	month = may,
	pages = {5970--5983},
	publisher = {ACM},
	address = {San Jose California USA},
	doi = {10.1145/2858036.2858459},
	urldate = {2024-09-13},
	isbn = {978-1-4503-3362-7},
	langid = {english},
}

@inproceedings{tuddenhamGraspablesRevisitedMultitouch2010,
	title = {Graspables Revisited: Multi-Touch vs. Tangible Input for Tabletop Displays in Acquisition and Manipulation Tasks},
	shorttitle = {Graspables Revisited},
	booktitle = {Proceedings of the {{SIGCHI Conference}} on {{Human Factors}} in {{Computing Systems}}},
	author = {Tuddenham, Philip and Kirk, David and Izadi, Shahram},
	year = 2010,
	month = apr,
	pages = {2223--2232},
	publisher = {ACM},
	address = {Atlanta Georgia USA},
	doi = {10.1145/1753326.1753662},
	urldate = {2024-09-13},
	isbn = {978-1-60558-929-9},
	langid = {english}
}

@inproceedings{uhlTangibleImmersiveTrauma2023a,
	title = {Tangible {{Immersive Trauma Simulation}}: {{Is Mixed Reality}} the next Level of Medical Skills Training?},
	shorttitle = {Tangible {{Immersive Trauma Simulation}}},
	booktitle = {Proceedings of the 2023 {{CHI Conference}} on {{Human Factors}} in {{Computing Systems}}},
	author = {Uhl, Jakob Carl and {Schrom-Feiertag}, Helmut and Regal, Georg and Gallhuber, Katja and Tscheligi, Manfred},
	year = 2023,
	month = apr,
	pages = {1--17},
	publisher = {ACM},
	address = {Hamburg Germany},
	doi = {10.1145/3544548.3581292},
	urldate = {2024-09-13},
	isbn = {978-1-4503-9421-5},
	langid = {english},
}

@inproceedings{vaishTwitchCrowdsourcingCrowd2014,
	title = {Twitch Crowdsourcing: Crowd Contributions in Short Bursts of Time},
	shorttitle = {Twitch Crowdsourcing},
	booktitle = {Proceedings of the {{SIGCHI Conference}} on {{Human Factors}} in {{Computing Systems}}},
	author = {Vaish, Rajan and Wyngarden, Keith and Chen, Jingshu and Cheung, Brandon and Bernstein, Michael S.},
	year = 2014,
	month = apr,
	pages = {3645--3654},
	publisher = {ACM},
	address = {Toronto Ontario Canada},
	doi = {10.1145/2556288.2556996},
	urldate = {2024-09-13},
	isbn = {978-1-4503-2473-1},
	langid = {english}
}

@inproceedings{valentineFlashOrganizationsCrowdsourcing2017,
	title = {Flash {{Organizations}}: {{Crowdsourcing Complex Work}} by {{Structuring Crowds As Organizations}}},
	shorttitle = {Flash {{Organizations}}},
	booktitle = {Proceedings of the 2017 {{CHI Conference}} on {{Human Factors}} in {{Computing Systems}}},
	author = {Valentine, Melissa A. and Retelny, Daniela and To, Alexandra and Rahmati, Negar and Doshi, Tulsee and Bernstein, Michael S.},
	year = 2017,
	month = may,
	pages = {3523--3537},
	publisher = {ACM},
	address = {Denver Colorado USA},
	doi = {10.1145/3025453.3025811},
	urldate = {2024-09-13},
	isbn = {978-1-4503-4655-9},
	langid = {english},
}

@article{whitneyCOVERSTORYTheCounterfeit2009,
	title = {{{COVER STORYThe}} Counterfeit You},
	author = {Whitney, Hunter},
	year = 2009,
	month = mar,
	journal = {Interactions},
	volume = 16,
	number = 2,
	pages = {37--40},
	issn = {1072-5520, 1558-3449},
	doi = {10.1145/1487632.1487641},
	urldate = {2024-09-13},
	langid = {english}
}

@String{Computing = "Computing" }

@String{Computer = "{IEEE} Computer" }

@String{Chelsea = "Chelsea" }

\end{refsection}

\end{document}